\documentclass[11pt,a4paper]{article}
\pdfoutput=1

\usepackage{jheppub}

\usepackage[utf8]{inputenc}
\usepackage{mathtools}
\usepackage{enumitem}
\usepackage{amsmath}
\usepackage{amsthm} 
\usepackage{amssymb}
\usepackage{amscd}
\usepackage{comment}
\usepackage{xcolor}
\usepackage{array} 
\usepackage{graphicx}
\usepackage{adjustbox}
\usepackage{mathrsfs}
\usepackage[samesize]{cancel}
\usepackage{ytableau}
\usepackage{booktabs}
\usepackage{tikz-cd}
\usepackage{bm}
\usepackage[capitalize]{cleveref}

\usepackage{tikz}
\usetikzlibrary{positioning,fit}
\usetikzlibrary{calc,intersections,through,backgrounds}
\usetikzlibrary{calc,arrows,decorations.markings,shapes.arrows,patterns, decorations.pathmorphing}
\usetikzlibrary{positioning,matrix,arrows,decorations.pathmorphing,decorations.pathreplacing}
\usetikzlibrary{shapes}
\usetikzlibrary{shapes.geometric}
\usetikzlibrary{decorations.text}
\usetikzlibrary{arrows.meta}

\tikzset{
  ->-/.style={decoration={markings, mark=at position 0.5 with {\arrow{to}}},
              postaction={decorate}},}
\tikzset{
  -<-/.style={decoration={markings, mark=at position 0.5 with {\arrow{to reversed}}},
              postaction={decorate}},}
              
\tikzset{
  pics/torus/.style n args={3}{
    code = {
      \providecolor{pgffillcolor}{rgb}{1,1,1}
      \begin{scope}[
          yscale=cos(#3),
          outer torus/.style = {draw,line width/.expanded={\the\dimexpr2\pgflinewidth+#2*2},line join=round},
          inner torus/.style = {draw=pgffillcolor,line width={#2*2}}
        ]
        \draw[outer torus] circle(#1);\draw[inner torus] circle(#1);
        \draw[outer torus] (180:#1) arc (180:360:#1);\draw[inner torus,line cap=round] (180:#1) arc (180:360:#1);
      \end{scope}
    }
  }
}

\newcommand{\tikznode}[2]{\relax
    \ifmmode%
    \tikz[remember picture,baseline=(#1.base),inner sep=0pt] \node (#1) {$#2$};
    \else
    \tikz[remember picture,baseline=(#1.base),inner sep=0pt] \node (#1) {#2};%
    \fi
}
\newcommand{\no}{\nonumber}

\newcommand{\cE}{\mathcal E}

\newcommand{\cJ}{\mathcal J}
\newcommand{\cL}{\mathcal L}

\newcommand{\cO}{\mathcal O}\newcommand{\cP}{\mathcal P}

\newcommand{\sfu}{\mathsf u}\newcommand{\sfv}{\mathsf v}
\newcommand{\sfx}{\mathsf x}\newcommand{\sfy}{\mathsf y}
\newcommand{\sfw}{\mathsf w}\newcommand{\sfz}{\mathsf z}

\newcommand{\la}{\lambda}
\newcommand{\Tr}{{\rm Tr}}

\newcommand{\bmh}{{\bm h}}

\newcommand{\CS}{\textrm{CS}}

\newcommand{\llangle}{\langle\!\langle}
\newcommand{\rrangle}{\rangle\!\rangle}

\newcommand{\bmp}{{\bm j}_{*}}

\graphicspath{{./figures/}}

\allowdisplaybreaks

\begin{document}

\pagenumbering{Alph}
\begin{titlepage}
\begin{flushright}
TTI-MATHPHYS-39
\end{flushright}
 
\vskip 1.5in
\begin{center}
     {\bf\Large{$T\bar{T}$ and root-$T\bar{T}$ deformations }} 
\\
     \bigskip 
     {{\bf\Large in four-dimensional Chern-Simons theory}}
\vskip 1.0cm 
{Jun-ichi Sakamoto$^{1}$, Roberto Tateo$^{2}$, and Masahito Yamazaki$^{3,4,5}$} \vskip 0.05in 
\vskip 0.5cm 
\vskip 0 cm 
\textit{$^1$Mathematical Physics Laboratory, Toyota Technological Institute,\\
Hisakata 2-12-1, Tempaku-ku, Nagoya, Japan 468-8511}
\vskip 0 cm
\textit{$^2$Dipartimento di Fisica, Universit\'a di Torino, INFN Sezione di Torino, \\
Via P. Giuria 1, 10125, Torino, Italy}
\vskip 0 cm 
\textit{$^3$Department of Physics, Graduate School of Science, University of Tokyo, \\
Hongo 7-3-1, Bunkyo-ku, Tokyo 113-0033, Japan}
\vskip 0 cm 
\textit{$^4$Kavli Institute for the Physics and Mathematics of the Universe,\\ UTIAS, University of Tokyo, 5-1-5 Kashiwanoha, Chiba 277-8583, Japan}
\vskip 0 cm 
\textit{$^5$Trans-Scale Quantum Science Institute, University of Tokyo, \\
Hongo 7-3-1, Bunkyo-ku, Tokyo 113-0033, Japan}

\end{center}


\vskip 0.5in
\baselineskip 16pt

\begin{abstract}

The four-dimensional Chern-Simons (CS) theory provides a systematic procedure for realizing two-dimensional integrable field theories. It is therefore a natural question to ask whether integrable deformations of the theories can be realized in the four-dimensional CS theory. In this work, we study $T\bar{T}$ and root-$T\bar{T}$ deformations of two-dimensional integrable field theories, formulated in terms of dynamical coordinate transformations, within the framework of four-dimensional CS theory coupled to disorder defects. We illustrate our procedure in detail for the degenerate $\mathcal{E}$-model, a specific construction that captures and unifies a broad range of integrable systems, including the principal chiral model.

\end{abstract}

\end{titlepage}
\pagenumbering{arabic} 

\tableofcontents
\newpage

\section{Introduction}
  \label{sec:introduction}

The four-dimensional Chern-Simons (CS) theory \cite{nekrassov1996four, Costello:2013zra,Costello:2017dso,Costello:2018gyb,Costello:2019tri} emerged as a novel framework
that explains and unifies many different aspects of integrable models, both in 
integrable lattice models \cite{Costello:2013zra,Costello:2017dso,Costello:2018gyb,Yamazaki:2025yan} and integrable field theories \cite{Costello:2019tri,Vicedo:2019dej,Delduc:2019whp,Lacroix:2021iit}. 

The aim of this paper is to discuss integrable deformations of integrable 
field theories, specifically the $T\bar{T}$ \cite{Smirnov:2016lqw,Cavaglia:2016oda} and root-$T\bar{T}$ \cite{Rodriguez:2021tcz,Babaei-Aghbolagh:2022uij, Conti:2022egv,Ferko:2022cix, Babaei-Aghbolagh:2022leo} deformations,
in the context of the four-dimensional CS theory.
In the string theory interpretation, four-dimensional theory arises from the open sector \cite{Ashwinkumar:2019mtj,Costello:2018txb},
while the integrable deformations are described by the closed-string degrees of freedom, 
i.e.\ gravity.
The crucial observation for us is an incarnation of this philosophy: some integrable $T\bar{T}$-type deformations can be described by suitable dynamical coordinate transformations \cite{Dubovsky:2017cnj, Conti:2018tca}.
Such coordinate transformations can then be incorporated into a suitable coordinate transformation of the four-dimensional setup, which contains defects localized at points on the spectral curve. 

To carry out this procedure, we need to carefully discuss the realization of two-dimensional integrable field theory on curved spaces
in the four-dimensional theory, which is one of the technical results of this paper.
For concreteness, we will discuss these ideas in detail for the example of the so-called degenerate $\mathcal{E}$-model  \cite{Klimcik:1996np}, which contains a large class of integrable field theories such as the principal chiral model (PCM),
and whose realization in the four-dimensional theory has been discussed previously in \cite{Benini:2020skc,Lacroix:2020flf,Liniado:2023uoo}.
We expect, however, that many of our discussions will have even broader applicability.

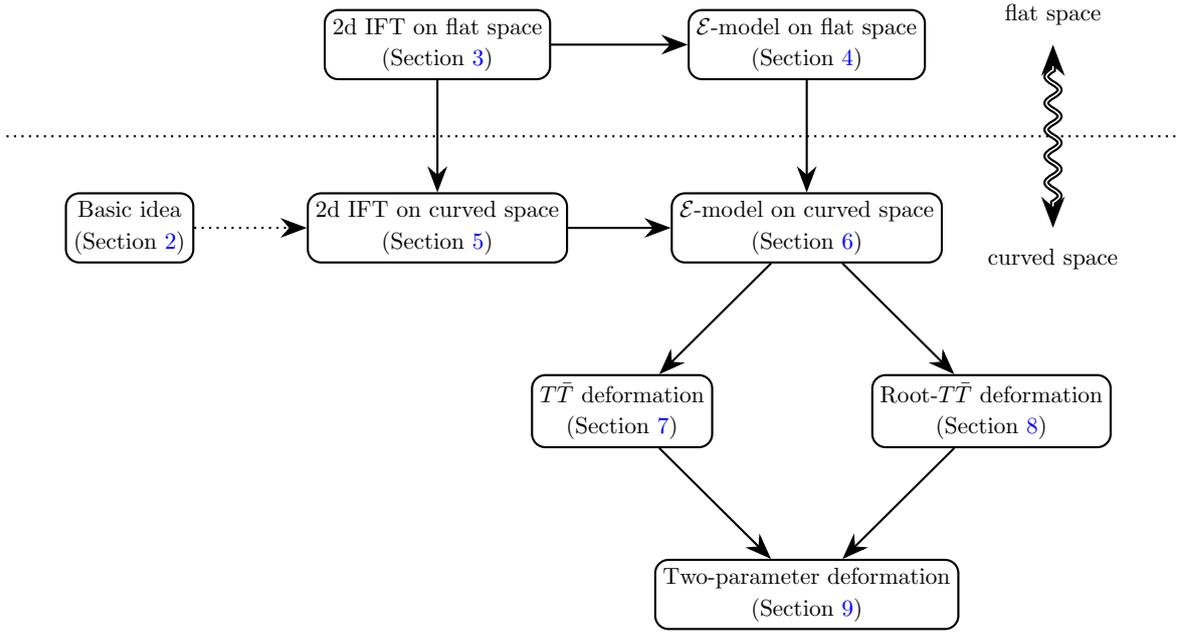
\begin{figure}[htbp]
\centering 
\begin{tikzpicture}[scale=0.81,
every path/.style={thick, >={Stealth[scale=1.5]}},
every node/.style={scale=0.8}]
\tikzset{
   blob/.style={shape= rectangle,rounded corners=5pt,draw,align= center,thick},
   double-blob/.style={shape= rectangle,double,rounded corners=5pt,draw,align= center,thick}
 }

\node[blob] (S2) at (0,-3) 
{Basic idea \\ (\cref{sec:basic})};

\node[blob] (S3) at (5,0) 
{2d IFT  on flat space \\ (\cref{sec:2d_flat})};

\node[blob] (S4) at (11,0) 
{$\mathcal{E}$-model on flat space \\ (\cref{sec:E-model})};

\node[blob] (S5) at (5,-3) 
{2d IFT on curved space \\ (\cref{sec:curved})};

\node[blob] (S6) at (11,-3) 
{$\mathcal{E}$-model on curved space \\ (\cref{sec:E-model_curved})};

\node[blob] (S7) at (8,-6) 
{$T\bar{T}$ deformation \\ (\cref{sec:TT})};

\node[blob] (S8) at (14,-6) 
{Root-$T\bar{T}$ deformation \\ (\cref{sec:root_TT})};

\node[blob] (S9) at (11,-9) 
{Two-parameter deformation \\ (\cref{sec:2_parameter})};

\node[] (flat) at (15,0.5) 
{flat space};

\node[] (curved) at (15,-3.5) 
{curved space};

\draw[dotted] (-2,-1.5) -- (17,-1.5);

\draw[->, dotted] (S2) -- (S5);
\draw[->] (S3) -- (S4);
\draw[->] (S3) -- (S5);
\draw[->] (S4) -- (S6);
\draw[->] (S5) -- (S6);
\draw[->] (S6) -- (S7);
\draw[->] (S6) -- (S8);
\draw[->] (S7) -- (S9);
\draw[->] (S8) -- (S9);
\draw[->,double,decorate,decoration={coil,aspect=0}] (15, -1.5) -- (15, 0);
\draw[->,double,decorate,decoration={coil,aspect=0}]  (15, -1.5) -- (15, -3);

\end{tikzpicture}
\caption{Outline of the structure of this paper.}
\label{fig.flow_chart}
\end{figure}

The rest of this paper is organized as follows.
In \cref{sec:basic} we first outline our basic idea for describing $T\bar{T}$-type deformations. In \cref{sec:2d_flat} we summarize the derivation of two-dimensional integrable field theories on flat spaces from four-dimensional theory.
In \cref{sec:E-model} we specialize to the case of the degenerate $\mathcal{E}$-model.
In \cref{sec:curved} we discuss two-dimensional integrable field theories on curved spaces from four-dimensional theory.
The discussion is then applied to the degenerate $\mathcal{E}$-model in \cref{sec:E-model_curved}.
In \cref{sec:TT} and \cref{sec:root_TT}
we discuss in turn the root-$T\bar{T}$ and $T\bar{T}$ deformations from the four-dimensional theory.
In \cref{sec:2_parameter} we discuss two-parameter deformations of the degenerate $\mathcal{E}$-model.
In \cref{sec:conclusion}, we conclude with a summary and discussions.
Several appendices are included to provide technical details.
The relative dependencies of the sections are summarized in \cref{fig.flow_chart}.

\section{Framework for describing \texorpdfstring{$T\bar{T}$}{T\bar{T}} and root-\texorpdfstring{$T\bar{T}$}{T\bar{T}} deformations in four-dimensional CS theory}
\label{sec:basic}

This work aims to elucidate how the $T\bar{T}$ and root-$T\bar{T}$ deformations of two-dimensional integrable field theories can be described within the 4d CS theory. These deformations are generated by adding specific, irrelevant, and marginal operators constructed from the energy-momentum tensor.
By denoting the original action and the associated energy-momentum tensor by $S^{(0)}[\phi]$ and $T_{\mu\nu}^{(0)}$, the $T\bar{T}$ and the root-$T\bar{T}$ deformed actions $S^{(\la)}\,, S^{(\gamma)}$ at the leading order are
\begin{align}
    \label{eq:TTbar}
    S^{(\la)}&=S^{(0)}[\phi]+\la\int_{\Sigma} {\rm det}(T_{\mu\nu}^{(0)})\,dx^+\wedge dx^-+\cO(\la^2)\,,\\
    \label{eq:root_TTbar}
    S^{(\gamma)}&=S^{(0)}[\phi]+\gamma\int_{\Sigma}\sqrt{{\det}(\widetilde{T}_{\mu\nu}^{(0)})}\,dx^+\wedge dx^-+\cO(\gamma^2)\,,
\end{align}
where $\la\in\mathbb{R}$ and $\gamma\in\mathbb{R}$ are deformation parameters of the $T\bar{T}$ and the root-$T\bar{T}$ deformations, respectively. $\widetilde{T}^{(0)}_{\mu\nu}$ in the root-$T\bar{T}$ operator is the traceless part of $T_{\mu\nu}^{(0)}$.

For our purposes, it is important to note that 
integrability has been observed for many examples of the deformations above.\footnote{See e.g.\ \cite{Borsato:2022tmu} for discussions of integrability for root-$T\bar{T}$ deformations. It is presently unknown if a general root-$T\bar{T}$ deformation preserves integrability. In this paper, we discuss only those deformations preserving integrability.}
This suggests that these integrable deformations may be described within the framework of 4d Chern–Simons theory.

\subsubsection*{Our procedure}

Our strategy for describing $T\bar{T}$-type deformations in the four-dimensional theory is motivated by the fact that the deformed Lax pairs are formally equivalent to the Lax pairs of the original model with a field-dependent metric. For example, in the case of the $T\bar{T}$ deformation, as discussed in \cite{Conti:2018tca}, the deformed Lax pair is obtained by applying a dynamical coordinate transformation, which is characterized by the energy-momentum tensor, to the undeformed Lax pair. The result indicates that the deformed Lax pair can be recast as the Lax pair of the original model with a field-dependent metric. Similarly, for the root-$T\bar{T}$ deformation, it is confirmed in some examples that the deformed Lax pair can be explicitly rewritten as the Lax pair of the original model with a field-dependent metric, as in the case of the PCM.

By employing this fact, we first construct deformed Lax pairs corresponding to the $T\bar{T}$ and the root-$T\bar{T}$ deformations and show that they solve the equations of motion of the four-dimensional theory.
In particular, we highlight the two-dimensional degenerate $\cE$-model \cite{Klimcik:1996np} that describes a broad class of two-dimensional integrable field theories, including the PCM, in a unified manner. 
When verifying that the deformed Lax pairs are solutions to the equations of motion for four-dimensional theory, special attention must be taken in handling the boundary conditions because the deformed Lax pair involves a field-dependent metric in contrast to the undeformed case. This point is discussed in detail in Sections \ref{rootTT-E} and \ref{TT-E}.
Once such a classical solution can be constructed, the corresponding two-dimensional field theory actions are derived according to the standard four-dimensional theory procedure \cite{Costello:2019tri}. We then verify directly that these actions obey $T\bar{T}$-type flow equations.

\subsubsection*{Comments on other approaches}

These two $T\bar{T}$-type deformations admit several other descriptions in the literature. 
For example, $T\bar{T}$ deformations can be described by coupling a two-dimensional field theory with flat space Jackiw-Teitelboim (JT) gravity. 
Similarly, simultaneous $T\bar{T}$ and root-$T\bar{T}$ deformations can also be expressed as the coupling to an extension of flat-space JT gravity \cite{Babaei-Aghbolagh:2024hti}\footnote{Pure root-$T\bar{T}$ deformation is still not known to have a description using a two-dimensional gravity theory.}.

The second description is to consider an infinite class of integrable field theories that describe deformations of the PCM and its non-abelian $T$-dual by coupling to an auxiliary field with certain self-interactions \cite{Ferko:2024ali,Bielli:2024khq}.
The $T\bar{T}$ and root-$T\bar{T}$ deformations can be obtained by integrating out the auxiliary field. 
In addition, the authors of \cite{Fukushima:2024nxm}  derived the PCM with an auxiliary field by considering four-dimensional CS theory coupled to such a field.

Both of these procedures introduce an external field. By contrast, our work investigates whether $T\bar{T}$ and root-$T\bar{T}$ deformations can be described within the framework of the four-dimensional theory without introducing an external field.
Of course, it would be interesting to extend these works to the case of the degenerate $\cE$ model and its realization in the four-dimensional theory, but we leave a detailed study of those questions for future work.

The realization of $T\bar{T}$-deformations within four-dimensional theory was discussed previously in \cite{Py:2022hoa}. In this paper, the $T\bar{T}$-deformation was described similarly to the $J\bar{J}$-deformation in \cite{Costello:2019tri}, where the gauge group was taken to be $SL_2(\mathbb{C})$. 
In addition to obvious differences in the setups, such as the fact that the gauge group is general in our discussion,  
we emphasize that our discussion 
applies also to non-linear orders in the deformation parameter $\lambda$ in 
\eqref{eq:TTbar} and \eqref{eq:root_TTbar}, 
which does not appear to have been incorporated into the discussion in \cite{Py:2022hoa}.

It remains an interesting question to further explore mutual relations between all the different approaches to $T\bar{T}$-type deformations, which we leave for future work.

\section{Two-dimensional integrable field theories on flat space from four-dimensional CS theory }
\label{sec:2d_flat}

We summarize a derivation of two-dimensional integrable field theories on flat space from four-dimensional theory coupled with disorder defects \cite{Costello:2019tri}. We will use the extended four-dimensional action to describe disorder defects considered in \cite{Benini:2020skc}, and largely follow the formalism in \cite{Benini:2020skc,Lacroix:2020flf,Liniado:2023uoo}. 

\subsection{Four-dimensional Chern-Simons action and meromorphic one-form}

Let $G_{\mathbb{C}}$ be a complexified Lie group with a Lie algebra $\mathfrak{g}_{\mathbb{C}}$. We denote the one-dimensional complex space by $C$, identified with the spectral parameter space (spectral curve), and taken as $\mathbb{CP}^1$ in this paper.
The four-dimensional theory is defined on a product space $\Sigma \times C$, and the four-dimensional action is given by \cite{Costello:2019tri}
\begin{align}
    S_{\CS}[A]&=\frac{i}{4\pi}\int_{\Sigma\times C}\omega \wedge \CS(A)\,,\label{CS-action}
\end{align}
where the gauge field $A$ takes values in $\mathfrak{g}_{\mathbb{C}}$\,, and $\CS(A)$ is the CS three-form defined as 
\begin{align}
\CS(A):= \Tr\left( A \wedge dA \, + \, \frac{2}{3}A\wedge A\wedge A\right)\,.
\end{align}

The construction of the associated two-dimensional integrable field theories within four-dimensional theory starts with a meromorphic one-form $\omega$ on $C$:
\begin{align}
    \omega=\varphi(z)dz\,.
\end{align}
Since $\omega$ is proportional to $dz$, the action (\ref{CS-action}) exhibits a $U(1)$ gauge symmetry
\begin{align}\label{u1-gauge}
    A'=A+f\,dz\,,
\end{align}
where $f$ is a $\mathfrak{g}_{\mathbb{C}}$-valued function on $\Sigma\times C$.
For simplicity, we assume that by using the gauge transformation (\ref{u1-gauge}), the gauge field $A$ everywhere on $\Sigma\times C$ can be brought to the form
\begin{align}
    A=A_{+}dx^++A_{-}dx^-+A_{\bar{z}}d\bar{z}\,,
\end{align}
where we introduced the light-cone coordinates on $\Sigma$ as
$x^{\pm}:=t\pm x$.

We denote the sets of poles and zeros of $\omega$ by $\mathfrak{p}$ and $\mathfrak{z}$, and the orders of a pole $x\in\mathfrak{p}$ and a zero $y \in \mathfrak{z}$ by $n_x$ and $m_y$, respectively. 
Then, $\omega$ is expressed as
\begin{align}\label{omega-analytic}
    \omega=\left(\sum_{x\in\mathfrak{p}'}\sum_{p=0}^{n_x-1}\frac{l_{p}^{x}}{(z-x)^{p+1}}-\sum_{p=1}^{n_{\infty}-1} l_p^{\infty} z^{p-1}\right)dz\,,
\end{align}
where $\mathfrak{p}=\mathfrak{p}'\cup \{\infty\}$ and $l_{p=0, \dots, n_x-1}^x\in\mathbb{C}$ are referred to as levels.
We impose a reality condition on the meromorphic function $\varphi(z)$ as
\begin{align}\label{real-omega}
    \overline{\varphi(z)}=\varphi(\overline{z})\,.
\end{align}
Writing $\mathfrak{p}_{r}=\mathfrak{p}_{r}'\sqcup \{\infty\}$ for the real poles, and $\mathfrak{p}_{c}$ and $\mathfrak{p}_{\overline{c}}$ for the poles corresponding to complex numbers whose imaginary part is positive and their complex conjugates, respectively, the set $\mathfrak{p}$ is divided as $\mathfrak{p}=\mathfrak{p}_r\sqcup \mathfrak{p}_c \sqcup \mathfrak{p}_{\overline{c}}$.
The reality condition (\ref{real-omega}) implies that for real poles $x\in \mathfrak{p}_r$, we have $l_p^x\in\mathbb{R}$, and for complex poles $x\in \mathfrak{p}_c$, we have the levels $l_x$ satisfying $\overline{l_p^{x}}=l_{p}^{\overline{x}}$ and $n_{x}=n_{\overline{x}}$.
The Riemann-Roch theorem relates the number of zeros of a meromorphic one form $\omega$ on $C=\mathbb{CP}^1$ to the number of its poles via
\begin{align}\label{z-p}
    \sum_{x\in \mathfrak{p}}n_x=\sum_{x \in \mathfrak{z}}m_x+2\,.
\end{align}

\subsection{Equations of motion and boundary conditions}

The equations of motion
of the four-dimensional theory (\ref{CS-action}) are given by:
\begin{align}\label{CS-eom1}
    \omega \wedge F(A)=0\,,
\end{align}
where $F(A):=dA+A\wedge A$ is the field strength of the gauge field $A$.
The meromorphic one-form is retained because it specifies the analytic structure of the gauge field in the spectral parameter space $C$.

In order to obtain a well-defined variational problem, the boundary term in the variation
\begin{align}
    \int_{\Sigma\times C} d\omega \wedge \Tr\left( A \wedge \delta A\right)
    \label{beom-int}
\end{align}
must vanish. Since $\omega$ is holomorphic, except at the poles of $\omega$ on $C$,
we can write $\omega= dz'$ for a
local holomorphic coordinate $z'$, leading to $d\omega=0$.
This implies that the integral localizes onto the so-called disorder surface defects supported on $\Sigma_x:=\Sigma \times \{x\}\subset \Sigma \times C$ for all $x\in\mathfrak{p}$. For later use, we denote the collection of all surface defects $\Sigma\times \mathfrak{p}$ by $\Sigma_{\mathfrak{p}}$. In this process
the gauge field $A$ localizes to defects. 

At a defect $x\in \mathfrak{p}$, one may expect the gauge field to take values in the 
Lie algebra $\mathfrak{g}$, which can be regarded as the tangent space $T_{e} G$ at the origin $e$ of the Lie group $G$.
For our discussion, however, we need the behaviour of the gauge field beyond the linear order,
and to some polynomial order up to a degree as specified by the order $n_x$ of the pole $x\in \mathfrak{p}$.
This can be mathematically formulated as an element of the $n_x$-tangent space\footnote{This can be formulated as the tangent space to the $n_x$-jet bundle $J^{n_x-1}G$.}---
in practice, this takes values in $\mathfrak{g}\otimes_{\mathbb{R}} \mathfrak{I}_x$
where $\mathfrak{I}_x=\mathbb{R}[\varepsilon_x]/(\varepsilon_x^{n_x})$ for all $x\in \mathfrak{p}_r$ and $\mathfrak{I}_x=\mathbb{C}[\varepsilon_x]/(\varepsilon_x^{n_x})$ for all $x\in \mathfrak{p}_c$.
We can collect this algebra across all the poles $x\in \mathfrak{p}$, and 
define the defect Lie algebra $\mathfrak{d}$ \cite{Benini:2020skc,Lacroix:2020flf,Liniado:2023uoo} 
\begin{align}\label{defect-Liealg} 
    \mathfrak{d}
    :=\prod_{x\in \mathfrak{p}_r}\left(\mathfrak{g}\otimes_{\mathbb{R}}\mathfrak{I}_{x}\right)\times \prod_{x\in \mathfrak{p}_c}\left(\mathfrak{g}_{\mathbb{C}}\otimes_{\mathbb{C}}\mathfrak{I}_{x}\right) \;,
\end{align}
with the commutation relations
\begin{align}
    [\mathsf{u}\otimes \varepsilon^p_x, \mathsf{v}\otimes \varepsilon^q_y]=\delta_{xy}[\mathsf{u},\mathsf{v}]\otimes \varepsilon_x^{p+q}\,,
\end{align}
where $\varepsilon_x^{p+q}=0$ for $p+q\geq n_x$ and $\delta_{xy}$ is the Kronecker delta. This Lie algebra is associated with a Lie group, which we denote by $\bm{D}$.\footnote{The Lie group $\bm{D}$ can be defined in terms of the jet bundle, which as a set is given by a direct product of the groups associated with each pole of a meromorphic one-form $\omega$, as in
\begin{align}\label{defect-Liegroup}
    {\bm D}=\prod_{x\in \mathfrak{p}_r}\left(G\times(\mathfrak{g}\otimes_{\mathbb{R}}\varepsilon_x\mathfrak{I}_{x})\right)
    \times \prod_{x\in \mathfrak{p}_c}\left(G_{\mathbb{C}}\times (\mathfrak{g}_{\mathbb{C}}\otimes_{\mathbb{C}}\varepsilon_x\mathfrak{I}_{x})\right)\,.
\end{align}
}

The localization of the gauge field is translated into the statement that
the gauge field $A$ is pulled back to a field ${\bm j}^*A\in \Omega^1(\Sigma_{\mathfrak{p}},\mathfrak{d})$,
where  the pull-back map ${\bm j}^*:\Omega^1(\Sigma\times C,\mathfrak{g}_{\mathbb{C}}) \to \Omega^1(\Sigma_{\mathfrak{p}},\mathfrak{d})$ is defined as
\begin{align}\label{jet-map-alg}
    {\bm j}^*: A\mapsto {\bm j}^*A=\left(\sum_{p=0}^{n_x-1}\frac{1}{p!}(\partial^{p}_{\xi_x}A)\Bigl\lvert_{z=x}\otimes \,\varepsilon_x^{p}\right)_{x\in \mathfrak{p}}\,.
\end{align}

The defect algebra is canonically equipped with a non-degenerate adjoint-invariant symmetric bilinear form
\begin{align}
    \llangle \cdot , \cdot\rrangle_{\mathfrak{d}}: \mathfrak{d}\times \mathfrak{d}\to \mathbb{R} \,.
\end{align}
This is defined by 
the orthogonal properties
\begin{align}\label{d-bi-orth}
    \llangle \sfu\otimes \varepsilon_{x}^{p}, \sfv \otimes \varepsilon_y^q\rrangle_{\mathfrak{d}}
    :=
    \begin{cases}
            \delta_{xy}\,l_{p+q}^{x}\Tr(\sfu\,\sfv)\qquad x\,,y \in \mathfrak{p}_r\\
            \delta_{xy}\left(l_{p+q}^{x}\Tr(\sfu\,\sfv)+l_{p+q}^{\bar{x}}\Tr(\tau(\sfu)\,\tau(\sfv))\right)\qquad x\,,y \in \mathfrak{p}_c
    \end{cases}\,,
\end{align}
where $\tau$ denotes an anti-linear involution representing the complex conjugation, and the levels $l_{p=0, \dots, n_x-1}^x$ 
were introduced previously
in \eqref{omega-analytic} as parameters for the one-form $\omega$. The relation with the one-form becomes more manifest by noting that 
\begin{align}
\llangle {\bm j}^* f, {\bm j}^* g \rrangle_{\mathfrak{d}}
=
\llangle f, g \rrangle_{\omega}
\end{align}
for $f, g \in R_{\mathfrak{z}}$.
Here $R_{\mathfrak{z}}$ denotes the space of $\mathfrak{g}_{\mathbb{C}}$-valued rational functions with poles in $\mathfrak{z}$,
and $\llangle -, - \rrangle_{\omega}$ is defined by
\begin{align}
\llangle f, g \rrangle_{\omega}
:=\sum_{x\in \mathfrak{p}} \operatorname{Res}_{x}
\operatorname{Tr}( f g) \, \omega \;.
\end{align}

Equation (\ref{beom-int}) reduces to
\begin{align}\label{beom-2}
    \int_{\Sigma_{\mathfrak{p}}}\llangle {\bm j}^*A, {\bm j}^*\delta A\rrangle_{\mathfrak{d}}=0\,,
\end{align}
which imposes boundary conditions for the gauge field $A$ (and its variation $\delta A$) on the surface defects $\Sigma_{\mathfrak{p}}$.
We can satisfy the condition (\ref{beom-2}) by finding a Lie subalgebra $\mathfrak{k}\subset\mathfrak{d}$ that is maximally isotropic with respect to the bilinear form
\begin{align}
    \label{eq:isotropic}
    \llangle \mathsf{x},\mathsf{y}\rrangle_{\mathfrak{d}}=0,
    \qquad \mathsf{x}\,, \mathsf{y}\in \mathfrak{k}\,,
\end{align}
and imposing the boundary condition 
\begin{align}\label{bc-A-gauge}
    {\bm j}^*A \in \Omega^1(\Sigma_{\mathfrak{p}},\mathfrak{k}) \;.
\end{align} 
For later use, we denote the complement (as a vector space) of $\mathfrak{k}$ within $\mathfrak{d}$ as $\tilde{\mathfrak{k}}$, and the corresponding groups of $\mathfrak{k}\,, \tilde{\mathfrak{k}}$ by ${\bm K}$ and $\tilde{{\bm K}}$, respectively.\footnote{The groups 
 ${\bm K}$ and $\tilde{{\bm K}}$ play complementary roles in the discussion of the Poisson-Lie T-duality.}
 
To summarize this section: we have introduced a defect Lie algebra $\mathfrak{d}$, equipped with a symmetric bilinear pairing $\llangle \cdot, \cdot \rrangle_{\mathfrak{d}}$, together with a maximal isotropic subalgebra $\mathfrak{k}$. 
We will now turn to the discussion of the gauge symmetry of the theory.

\subsection{Gauge symmetry and edge modes}
\label{subsec:edge}

The boundary condition (\ref{bc-A-gauge}) on the surface defects $\Sigma_{\mathfrak{p}}$ breaks the gauge symmetry of the four-dimensional theory:
the gauge transformation
\begin{align}\label{gauge-tr}
    A\mapsto {}^uA :=uAu^{-1}-du u^{-1}\,, \qquad u \in C^{\infty}(\Sigma\times C,G_{\mathbb{C}})\,,
\end{align}
is valid only if the gauge parameter $u$ satisfies the corresponding boundary condition
\begin{align}
    {\bm j}^*u \in C^{\infty}(\Sigma_{\mathfrak{p}},{\bm K})\,,
\end{align}
where, similar to (\ref{jet-map-alg}) for one-forms, we introduce a pull-back map ${\bm j}^*: C^{\infty}(\Sigma\times C, G_{\mathbb{C}})\to C^{\infty}(\Sigma_{\mathfrak{p}}, {\bm D})$ for $C^{\infty}$-functions as
\begin{align}
    {\bm j}^*: g\mapsto {\bm j}^*g=\left(g\lvert_{z=x},\sum_{p=0}^{n_x-1}\frac{1}{p!}\partial^{p-1}_{\xi_x}(\partial_{\xi_x} g g^{-1})\Bigl\lvert_{z=x}\otimes \,\varepsilon_x^{p}\right)_{x\in \mathfrak{p}}\,.
\end{align}

There is another description of the theory that makes the gauge symmetry more manifest.
In this approach, 
we introduce an additional field (an edge mode) $\bmh\in C^{\infty}(\Sigma_{\mathfrak{p}},{\bm D})$ supported on a collection of surface defects $\Sigma_{\mathfrak{p}}$.\footnote{In general, we can describe disorder defects by coupling the four-dimensional theory to a field theory with additional degrees of freedom localized at the defect. For example, a 't Hooft operator in the four-dimensional theory, which is a one-dimensional disordered defect, has been proposed to be described by a Gukov-Witten-type surface defect \cite{Gukov:2006jk} with one-dimensional boundary \cite{Costello:2021zcl}. For other disorder defects in the four-dimensional theory, see \cite{Khan:2022vrx}.}

The two-dimensional disorder defects discussed above are described by the four-dimensional CS action coupled to two-dimensional defects \cite{Benini:2020skc}
\begin{align}
    S_{\text{ext}}[A,\bmh]&=S_{\CS}[A]+S_{\text{defect}}[A,\bmh]\,,\label{4dCS-defect}\\
    S_{\text{defect}}[A,\bmh]&=-\frac{1}{2}\int_{\Sigma_{\mathfrak{p}}}\llangle \bmh^{-1}d\bmh, {\bm j}^*A\rrangle_{\mathfrak{d}}-\frac{1}{2}S_{\mathfrak{d}}^{\text{WZ}}[\bmh]\,. \label{defect-action}
\end{align}
The second term in (\ref{defect-action}) is the Wess-Zumino term defined by
\begin{align}
    S_{\mathfrak{d}}^{\text{WZ}}[\bmh]=-\frac{1}{3!}\int_{\Sigma_{\mathfrak{p}}\times I}\llangle \widehat{\bmh}^{-1} d\widehat{\bmh}, [\widehat{\bmh}^{-1} d\widehat{\bmh}, \widehat{\bmh}^{-1}d\widehat{\bmh}] \rrangle_{\mathfrak{d}}\,,
\end{align}
where $I=[0,1]$ and $\widehat{\bmh}$ is a $G_{\mathbb{C}}$-valued function on $\Sigma_{\mathfrak{p}}\times I$ with $\widehat{\bmh}=\bmh$ at $\Sigma_{\mathfrak{p}}\times \{0\}\subset \Sigma_{\mathfrak{p}}\times I$ and $\widehat{\bmh}=\mathrm{Id}$ at $\Sigma_{\mathfrak{p}}\times \{1\} \subset \Sigma_{\mathfrak{p}}\times I$.
Now the boundary condition \eqref{bc-A-gauge} of the gauge field $A$ is slightly modified and takes the form
\begin{align}\label{bc-gauge}
    {}^{\bmh}({\bm j}^*A)\in \Omega^1(\Sigma_{\mathfrak{p}}, \mathfrak{k})\,.
\end{align}
This modification is necessary to preserve the equation of motion (\ref{CS-eom1}) for the gauge field. Indeed, the variation of $S_{\text{ext}}[A,\bmh]$ with respect to the gauge field $A$ is
\begin{align}
    \delta S_{\text{ext}}[A,\bmh]=\frac{i}{2\pi}\int_{\Sigma\times C}\omega\wedge\Tr(\delta A\wedge F)-\frac{1}{2}\int_{\Sigma_{\mathfrak{p}}}\llangle {\bm j}^*\delta A,({\bm j}^*A-\bmh^{-1}d\bmh)\rrangle_{\mathfrak{d}}\,.
\end{align}
Since ${\bm j}^*A-\bmh^{-1}d\bmh \in \mathrm{Ad}_{\bmh^{-1}}\mathfrak{k}$ and ${\bm j}^*\delta A\in \mathrm{Ad}_{\bmh^{-1}}\mathfrak{k}$, the second term vanishes due to \eqref{eq:isotropic} and the 
$\mathrm{Ad}_{\bmh^{-1}}$-invariance of the inner product. Hence, the equations of motion (\ref{CS-eom1}) remain unchanged.

Thanks to the additional fields, the four-dimensional/two-dimensional action (\ref{4dCS-defect}) is invariant under a gauge transformation \footnote{This can be shown by using the Polyakov-Wiegmann identity \cite{Polyakov:1983tt}
\begin{align}
      S_{\mathfrak{d}}^{\mathrm{WZ}}[\bmh {\bm k}^{-1}]=S_{\mathfrak{d}}^{\mathrm{WZ}}[\bmh ]-S_{\mathfrak{d}}^{\mathrm{WZ}}[{\bm k}]+\int_{\Sigma_{\mathfrak{p}}}\llangle \bmh^{-1} d \bmh, {\bm k}^{-1}d {\bm k}\rrangle_{\mathfrak{d}}\,.
\end{align}
}
\begin{align}\label{gauge-tr2}
    A\mapsto {}^uA:= uAu^{-1}-du u^{-1}\,,
    \qquad 
    \bmh\mapsto {}^u\bmh: =\bmh ({\bm j}^*u)^{-1}\,,
\end{align}
for an arbitrary $u \in C^{\infty}(\Sigma\times C,G_{\mathbb{C}})$.
The boundary condition \eqref{bc-gauge}
is also preserved under an additional symmetry
\begin{align}\label{gauge-tr3}
    \bmh\mapsto {\bm k} \bmh \,,
\end{align}
where ${\bm k} \in C^{\infty}(\Sigma_{\mathfrak{p}}, {\bm K})$.
Due to the latter symmetry, we may regard $\bmh$ as a field taking values in
the left coset $\bm{K}\backslash \bm{D}$. After fixing the gauge symmetry \eqref{gauge-tr3}, the associated Lie algebra is $\tilde{\bm k}$.

\subsection{Ansatz for Lax pair}\label{sec:Lax}

To derive a two-dimensional integrable field theory, we need to solve for the gauge field 
$A_{\pm}$ on the spectral curve space $C$.
We can then equate the gauge field $A_{\pm}$ to a Lax pair $\cL_{\pm}$, by a suitable gauge transformation (\ref{gauge-tr2}) to choose a gauge $A_{\bar{z}}=0$ \cite{Costello:2019tri} (see also \cite{Delduc:2019whp}) \footnote{For simplicity, we assume that the bundle of the gauge field $A$ on $C$ is trivial so that the gauge choice $A_{\bar{z}}=0$ can be taken.}.
Since the equations of motion (\ref{CS-eom1}) are invariant under the gauge transformation, $\cL$ still satisfies 
\begin{align}\label{lax-flat}
    \partial_+\cL_--\partial_-\cL_++[\cL_+,\cL_-]=0\,.
\end{align}
The equation (\ref{lax-flat}) shows that $\cL_{\pm}$ can be identified with a Lax pair for a classical integrable field theory.

In order to obtain an explicit form of $\cL$, we need to solve the $(x^{\pm},\bar{z})$-components of the equations of motion (\ref{CS-eom1}) which takes the form  
\begin{align}
    \omega\wedge \partial_{\bar{z}}\cL=0\,,\label{Lax-eom2}
\end{align}
subject to the boundary condition \eqref{bc-gauge}
\begin{align}\label{Lax-con1}
     {}^{\bmh}({\bm j}^*\cL(\bmh))\in \Omega^1(\Sigma_{\mathfrak{p}},\mathfrak{k})\,.
\end{align}
Note that when we explicitly solve for $\mathcal{L}$
the result depends on the edge-mode field $\bmh$ introduced in \eqref{subsec:edge}, and 
we have explicitly shown the dependence of $\cL$
on $\bmh$. 

We further require consistency with the gauge symmetries
of the theory. The symmetry \eqref{gauge-tr3}
stays exactly the same, where
\begin{align}
    \bmh\mapsto {\bm k} \bmh \,,
\end{align}
and ${\bm k} \in C^{\infty}(\Sigma_{\mathfrak{p}}, {\bm K})$.
For \eqref{gauge-tr2}, 
since we have integrated the theory along $C$, and we no longer have
a dependence on the $C$ coordinates, $u \in C^{\infty}(\Sigma \times C,G_{\mathbb{C}})$
is replaced by $f \in C^{\infty}(\Sigma,G)$ 
and the action on $\bmh$ is given by
\begin{align}
   \bmh \to \bmh \Delta(f)^{-1}\,,
\end{align}
where  $\Delta:G\to G^{\times |\mathfrak{p}|}$ is the diagonal embedding, and $\Delta(f)$ denotes the composition $\Delta\circ f\in C^{\infty}(\Sigma , G^{\times |\mathfrak{p}|})$.
By combining the two, we obtain the requirement that the Lax pair obeys the following transformation rule:
\begin{align}\label{Lax-con2}
    {\bm j}^*\cL({\bm k}\,\bmh \Delta(f)^{-1})={}^{\Delta(f)}\left({\bm j}^*\cL(\bmh)\right)\,.
\end{align}
As we will see later, this condition will be slightly relaxed in the case of $T\bar{T}$ and root-$T\bar{T}$ deformations.

From the analytic structure of $\omega$ described in (\ref{omega-analytic}), a general solution to the equations (\ref{Lax-eom2}) takes the form
\begin{align}
    \cL_{\pm}=\sum_{x\in \mathfrak{z}'}\sum_{q=0}^{m_x-1}\frac{\cL_{\pm}^{(x,q)}}{(z-x)^{q+1}}
    +\sum_{q=0}^{m_{\infty}-1}\cL_{\pm}^{(\infty,q)}z^{q+1}
    +\cL_{\pm}^{c}\,,
\end{align}
where $\cL_{\pm}^{(x,q)}\,, \cL_{\pm}^{c}\in C^{\infty}(\Sigma,\mathfrak{g}_{\mathbb{C}})$.
In addition, it must be possible to integrate the four-dimensional Lagrangian in the local neighborhood of $\Sigma\times \mathfrak{p}$ to obtain a finite two-dimensional action. 
For this reason, we further restrict the above ansatz for the Lax pair \cite{Liniado:2023uoo}.

Let us assume that 
\begin{itemize}
\item we can choose a decomposition $\mathfrak{z}=\mathfrak{z}_+\sqcup\mathfrak{z}_-$ of the set of  zeros,
such that 
\begin{align}
\label{eq:m_equal}
	\sum_{x\in \mathfrak{z}_+}m_x=\sum_{x\in \mathfrak{z}_-}m_x \,.
\end{align}
\end{itemize}
This condition will be assumed throughout the remainder of this paper.
Note that this implies that the number of poles, 
including the multiplicities, of $\omega$ must be even, i.e.
\begin{align}
	\label{eq:assumption_even}
    \sum_{x \in \mathfrak{p}}n_x\in 2\mathbb{Z}_{>0}\,,
\end{align}
which implies the number of zeros (again counted with multiplicities) is also even.

We then consider a solution
in which $\cL_{\pm}$ has poles only in $\mathfrak{z}_{\pm}$. For example,  if we assume that the zero at $z=\infty$ is contained in $\mathfrak{z}_+$, our ansatz is 
\begin{align}\label{Lax-ansatz0}
\begin{split}
    \cL_{+}&=\sum_{x\in \mathfrak{z}'_+}\sum_{q=0}^{m_x-1}\frac{\cL_{+}^{(x,q)}}{(z-x)^{q+1}}
    +\sum_{q=0}^{m_{\infty}-1}\cL_{+}^{(\infty,q)}z^{q+1}
    +\cL_{+}^{c}\,,\\
    \cL_-&=\sum_{x\in \mathfrak{z}'_-}\sum_{q=0}^{m_x-1}\frac{\cL_{-}^{(x,q)}}{(z-x)^{q+1}}
    +\cL_{-}^{c}\,.
\end{split}
\end{align}
Such an ansatz resolves the divergence arising from the cross term between $A_{+}$ and $A_{-}$ in the 
four-dimensional action.

Each coefficient $\cL_{\pm}^{(x,q)}$ and $\cL_{\pm}^c$ is determined by solving the constraint equation (\ref{Lax-con1}).
The number of undetermined functions in this ansatz is $\sum_{x\in \mathfrak{z}}m_x+2$, while the number of constraints coming from the boundary condition (\ref{Lax-con1}) is $\sum_{x\in\mathfrak{p}}n_x$, as can be seen from the definition (\ref{jet-map-alg}) of the map ${\bm j}^*$. Since we have the equality (\ref{z-p}), the free parameters in the ansatz (\ref{Lax-ansatz0}) should be uniquely determined. 

We should note that the ansatz (\ref{Lax-ansatz0}) is not the most general ansatz that gives the corresponding finite two-dimensional action; a solution is called admissible if it yields
a finite two-dimensional action, and a most general admissible solution may not fit into the ansatz (\ref{Lax-ansatz0}).
We will return to more general admissible solutions later, in our discussion of curved spaces in Section \ref{sec:curved}.

\section{Degenerate \texorpdfstring{$\mathcal{E}$-model}{E-model} on flat space}
\label{sec:E-model}

While the discussion of Section \ref{sec:2d_flat} is general and in principle yields two-dimensional integrable field theories,
it is often useful to obtain more explicit expressions for the Lax pair, and hence the two-dimensional action, for the theories.
This is particularly the case for our discussion on $T\bar{T}$-type deformations in later sections, 
where the integrability of the theory is not guaranteed a priori and requires separate verification.

In this section, we discuss  a class of solutions to the equations of motion (\ref{Lax-eom2})
found in \cite{Liniado:2023uoo}, satisfying the boundary condition (\ref{Lax-con1})  and the transformation property (\ref{Lax-con2}).
The resulting two-dimensional theory is identified with the Lax pair for the degenerate $\cE$-model \cite{Klimcik:1996np} on flat space,
as we will now explain.

\subsection{Solution of Lax pair}\label{sec:Lax_solution}

Let $R_{\mathfrak{z}_{\pm}}$ denote the spaces of $\mathfrak{g}_{\mathbb{C}}$-valued rational functions with poles in $\mathfrak{z}_{\pm}$,
and $R'_{\mathfrak{z}_{\pm}}$ denote its subspace without constant pieces.
In other words, given elements $f_{\pm}(z)\in R_{\mathfrak{z}_{\pm}}$, we can expand
\begin{align}
   f_{\pm}(z)&=\sum_{x\in \mathfrak{z}'_{\pm}}\sum_{q=0}^{m_x-1}\frac{f_{\pm}^{(x,q)}}{(z-x)^{q+1}}+\sum_{q=0}^{m_{\infty}-1} f_{\pm}^{(\infty,q)}z^{q+1} + f_{\pm}^c \,,
\end{align}
where $f_{\pm}^{(x,q)}\,,f_{\pm}^{(\infty, q)} \,, f_{\pm}^c \in \mathfrak{g}$ for $x\in \mathfrak{z}_{r}$, and $f_{\pm}^{(x,q)}\in \mathfrak{g}_{\mathbb{C}}$ for $x\in \mathfrak{z}_{c}$;
we have moreover $f_{\pm}^c=0$ for $f_{\pm}(z)\in R'_{\mathfrak{z}_{\pm}}$.
The space $R_{\mathfrak{z}_{\pm}}$ parametrizes all possible ans\"{a}tze 
for the flat-space Lax operator $\mathcal{L}_{\pm}$ discussed previously in
\eqref{Lax-ansatz0}. 
The problem is thus to determine the elements of $R_{\mathfrak{z}_{\pm}}$ corresponding to a given defect field $\bmh$.

By Taylor-expanding a function up to the first $n_x$ terms around each 
pole $x\in \mathfrak{p}$ of the one-form $\omega$, we define a linear map 
$\bm{j}_{\mathfrak{p}}: R'_{\mathfrak{z}_{\pm}}  \to \mathfrak{d}$ as
\begin{align}
f \mapsto \left(\sum_{p=0}^{n_x-1}\frac{1}{p!} (\partial^p_{\zeta_x} f ) \big|_x \otimes \varepsilon_x^p\right)_{x\in \mathfrak{p}}\,.
\end{align} 
The image of this map lies in the defect Lie algebra $\mathfrak{d}$, and can be realized by the ansatz \eqref{Lax-ansatz0}:
\begin{align}
    \mathfrak{r}_{\pm}:=
    {\bm j}_{\mathfrak{p}}(R'_{ \mathfrak{z}_{\pm}})\,.
\end{align}
Since we have imposed the condition \eqref{eq:m_equal}, both $\mathfrak{r}_{\pm}$ have the same dimension, $\mathrm{dim}(\mathfrak{d})/2-\mathrm{dim}(\mathfrak{g})$.
If we include all the zeros $\mathfrak{z}$
 (as opposed to $\mathfrak{z}_{\pm}$), we obtain a larger space
 \begin{align}
    \mathfrak{r}
    := {\bm j}_{\mathfrak{p}}\left(R'_{ \mathfrak{z}}\right)
    = \mathfrak{r}_{+} \dot{+} \mathfrak{r}_{-}\,,
\end{align}
of $\mathrm{dim}(\mathfrak{d})-2 \mathrm{dim}(\mathfrak{g})$,
and the choice of splitting $\mathfrak{r}$ into $\mathfrak{r}_{\pm}$
is the algebraic counterpart of the choices of the splitting of $\mathfrak{z}$ into $\mathfrak{z}_{\pm} \subset \mathfrak{z}$. 

The constant piece is described by the diagonal subalgebra $\mathfrak{f}$ of $\mathfrak{d}$,
defined as the image of the diagonal embedding 
$\Delta: \mathfrak{g}\to \mathfrak{g}^{\times |\mathfrak{p}|}\subset \mathfrak{d}$:
\begin{align}\label{diag-map}
    \Delta: \sfu \mapsto (\sfu \otimes \varepsilon_{x}^0)_{x\in\mathfrak{p}}\qquad \sfu\in\mathfrak{g}\,.
\end{align}
In other words,
\begin{align}
\label{eq:f_def}
    \mathfrak{f}:=\{(\sfu\otimes \varepsilon^{0}_{x})_{x\in \mathfrak{p}}~\lvert~ \sfu \in \mathfrak{g}\}\,.
\end{align}
If we include constant pieces, we have
\begin{align}
	{\bm j}_{\mathfrak{p}}\left(R_{ \mathfrak{z}_{\pm}}\right)
	= \mathfrak{f}\dot{+} \mathfrak{r}_{\pm} \,,
	\quad
	{\bm j}_{\mathfrak{p}}\left(R_{ \mathfrak{z}}\right)
	= \mathfrak{f}\dot{+} \mathfrak{r} \,.
\end{align}

\begin{figure}[htbp]
\centering
\includegraphics[scale=0.35]{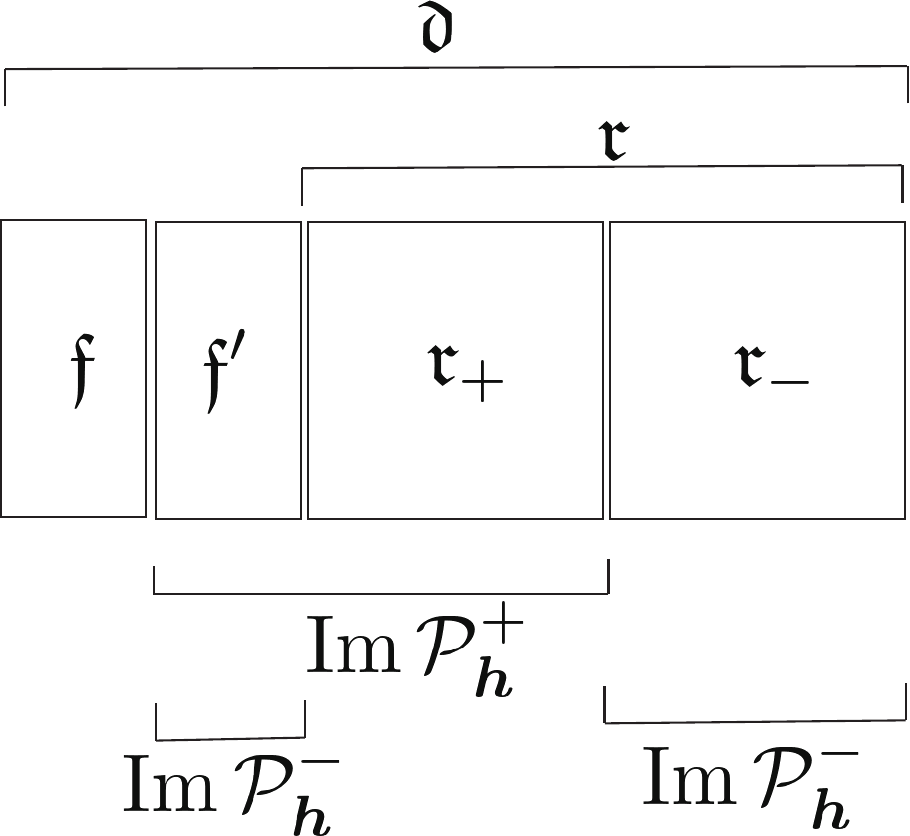}
\caption{Decomposition of $\mathfrak{d}$ into subspaces. We have a decomposition 
$\mathfrak{d}=\mathfrak{f} \dot{+} \mathfrak{f}' \dot{+}  \mathfrak{r}_+ \dot{+}  \mathfrak{r}_{-}$, with 
$\mathrm{dim}\,\mathfrak{f}  =\mathrm{dim}\, \mathfrak{f}' = \mathrm{dim}\,\mathfrak{g}$.}
\label{fig:spaces}
\end{figure}

Since the inner product $\llangle \cdot , \cdot \rrangle_{\mathfrak{d}}$ is defined as in \eqref{d-bi-orth}, we 
obtain, by the residue theorem,
\begin{align}
\label{orthogonality}
	 \llangle \mathfrak{f} , \mathfrak{f} \rrangle_{\mathfrak{d}}
	 = \llangle \mathfrak{f} , \mathfrak{r}_{\pm} \rrangle_{\mathfrak{d}} 
	 = \llangle \mathfrak{r}_{+} , \mathfrak{r}_{-} \rrangle_{\mathfrak{d}} =0 \,.
\end{align}
Since the inner product is non-degenerate, we have a decomposition 
$\mathfrak{d}=\mathfrak{f} \dot{+} \mathfrak{f}' \dot{+}  \mathfrak{r}_+ \dot{+}  \mathfrak{r}_{-}$, with 
$\mathrm{dim}\,\mathfrak{f}  =\mathrm{dim}\, \mathfrak{f}' = \mathrm{dim}\,\mathfrak{g}$
and $\mathrm{dim}\,\mathfrak{r}_{+}  =\mathrm{dim}\, \mathfrak{r}_{-} = \mathrm{dim}\,\mathfrak{d}/2-\mathrm{dim}\,\mathfrak{g}$,
so that $\mathfrak{f}$ and $\mathfrak{f}'$ are mutually non-degenerate with respect to the bilinear form (see Fig.~\ref{fig:spaces} for a summary).

Let us note that $\mathfrak{f}\dot{+} \mathfrak{r}_{\pm}$ has dimension $\mathrm{dim}\,\mathfrak{d}/2$, 
and is maximally isotropic in $\mathfrak{d}$. We have previously encountered another maximally isotropic
subspace $\mathfrak{k}$ in  $\mathfrak{d}$, which remains maximally isotropic
when conjugated to $\mathrm{Ad}_{\bmh^{-1}}\mathfrak{k}$.
Since they are both isotropic, provided that they intersect transversally,
we can introduce projection operators $\mathcal{P}_{\bmh}^{\pm}:\mathfrak{d} \to \mathfrak{d}$ \cite{Klimcik:2019kkf, Klimcik:2021bqm} 
implementing a ``canonical transformation'' between the two subspaces, so that we have
\begin{align}\label{def-W}
    \operatorname{Ker}\,\mathcal{P}_{\bmh}^{\pm}=\operatorname{Ad}_{\bmh^{-1}}\mathfrak{k}\,,\qquad \operatorname{Im}\,\mathcal{P}_{\bmh}^{\pm}=\mathfrak{f}\oplus \mathfrak{r}_{\pm}\,.
\end{align} 
Note that whether the intersection is transverse depends on the value of $\bmh$---this
is satisfied for a generic value of $\bmh$, except for certain non-generic values, which we
interpret as the singular points of the target manifold of the field $\bmh$. 
The images of $\mathcal{P}^{\pm}_{\bmh}$
are orthogonal as a consequence of \eqref{orthogonality}:
\begin{align}
    \llangle \textrm{Im}\, \mathcal{P}^{+}_{\bmh} , \textrm{Im}\, \mathcal{P}^{-}_{\bmh} \rrangle_{\mathfrak{d}}=0 \,.
\end{align}

Now the constraint (\ref{Lax-con1}) can be written as 
${\bm j}^*\cL(\bmh)=\bmh^{-1}k \bmh+\bmh^{-1}d\bmh$ for $k\in \Omega^1(\Sigma_{\mathfrak{p}},\mathfrak{k})$.
By acting with the projection operators $\mathcal{P}_{\bmh}^{\pm}$ on both sides, we obtain
\begin{align}\label{Lax-sol-E}
     {\bm j}^*\cL_{\pm}(\bmh) 
     =\mathcal{P}_{\bmh}^{\pm}(\bmh^{-1}\partial_{\pm}\bmh)\,,
\end{align}
and this satisfies the transformation rule (\ref{Lax-con2}) as a consequence of \eqref{W-tr}.

We can now show ${\bm j}_{\mathfrak{p}}$
 is surjective onto $\mathfrak{f}\dot{+} \mathfrak{r}$ and is invertible \cite[Proposition 3.4]{Liniado:2023uoo}---this is a variant of the statement that a holomorphic function on $\mathbb{CP}^1$ is determined 
 by its poles and residues, apart from the constant pieces taken care of by $\mathfrak{f}$.
 This means that we can define an inverse map\footnote{Moreover, we have $\mathfrak{f}^{\bot}=\mathfrak{f}\dot{+} \mathfrak{r}$,
where the algebra $\mathfrak{f}^{\bot}\subset \mathfrak{d}$ is the subspace orthogonal to $\mathfrak{f}$ (defined previously in \eqref{eq:f_def}) with respect to the non-degenerate bilinear form $\llangle\cdot,\cdot \rrangle_{\mathfrak{d}}$.}
\begin{align}
	\bmp: 
	\mathfrak{f}\dot{+} \mathfrak{r}\to R_{\mathfrak{z}}\,.
\end{align}
Its restriction gives inverse maps
$\mathfrak{f}\dot{+} \mathfrak{r}_{\pm} \to R_{\mathfrak{z}_{\pm}}$, which we also denote by $\bmp$.
Using this map $\bm{j}_{*}$, we can rewrite \eqref{Lax-sol-E}
and the Lax pair $\cL$ for the degenerate $\cE$-model takes a simple form
\begin{align}\label{E-Lax}
    \cL_{\pm}=\bmp\left(\mathcal{P}_{\bmh}^{\pm}(\bmh^{-1}\partial_{\pm}\bmh)\right)\,.
\end{align}

\subsection{Reduction to two-dimensional action on flat space}\label{sec:reduction}

Once a Lax pair is obtained, one can derive the corresponding two-dimensional action from the four-dimensional action by substituting the Lax pair (\ref{E-Lax}) into the two-dimensional-4d action (\ref{4dCS-defect}) \cite{Costello:2019tri}.
We can easily see that the four-dimensional action with $A=\cL$ vanishes due to the equations of motion (\ref{Lax-eom2}).
Hence, the resulting two-dimensional action is given by \cite{Delduc:2019whp} 
\begin{align}\label{2d-formula-1}
    S_{\rm 2d}[\bmh]=-\frac{1}{2}\int_{\Sigma_{\mathfrak{p}}}\llangle\bmh^{-1}d\bmh , {\bm j}^*\cL\rrangle_{\mathfrak{d}}-\frac{1}{2}S_{\mathfrak{d}}^{\text{WZ}}[\bmh]\,.
\end{align}
Substituting the \eqref{E-Lax} for the Lax pair, we obtain the action of the flat-space $\mathcal{E}$-model \cite{Klimcik:1996np,Klimcik:2021bqm}:
\begin{align}\label{dE-action}
    S_{\rm 2d}[\bmh]&=\frac{1}{2}\int_{\Sigma_{\mathfrak{p}}}\biggl(\llangle \bmh^{-1}\partial_-\bmh, \mathcal{P}_{\bmh}^{+}(\bmh^{-1}\partial_+\bmh)\rrangle_{\mathfrak{d}}
    -\llangle \bmh^{-1}\partial_+\bmh, \mathcal{P}_{\bmh}^{-}(\bmh^{-1}\partial_-\bmh)\rrangle_{\mathfrak{d}}\biggr)dx^+\wedge dx^-\no\\
    &\qquad \qquad-\frac{1}{2}S_{\mathfrak{d}}^{\text{WZ}}[\bmh]\,.
\end{align}

Owing to the transformation property (\ref{Lax-con2}), the two-dimensional action (\ref{2d-formula-1}) is invariant under 
\begin{align}\label{2dgauge-sym}
    \bmh\to {\bm k}\,\bmh\Delta(f)^{-1}\,,
\end{align}
where ${\bm k} \in C^{\infty}(\Sigma_{\mathfrak{p}},{\bm K})$ and $f \in C^{\infty}(\Sigma,G)$.
Therefore, the ${\bm D}$-valued field ${\bm h}$ is valued in the double coset
\begin{align}
\label{eq:double_coset}
	\bm{K} \backslash \bm{D} /\bm{F} ,
\end{align}
where $\bm{F}$ is the image of the diagonal embedding $\Delta:G\to G^{\times |\mathfrak{p}|}$.
The Lie algebra associated with the group $\bm{F}$ is the diagonal subalgebra $\mathfrak{f}$ of $\mathfrak{d}$
introduced previously in \eqref{eq:f_def}.
Note that we can use the $\bm{K}$ symmetry to
 regard ${\bm h}$ as taking values in $\tilde{{\bm K}}\subset {\bm D}$.

In the literature of the $\mathcal{E}$-model, sigma models
associated with the right coset $\bm{K} \backslash \bm{D}$
are called non-degenerate $\mathcal{E}$-models \cite{Klimcik:1995ux,Klimcik:1996nq,Klimcik:1995dy}, 
while those associated with the double coset $\bm{K} \backslash \bm{D} /\bm{F}$
are called a degenerate $\mathcal{E}$-models \cite{Klimcik:1996np,Klimcik:2019kkf,Klimcik:2021bqm},
where the latter can be regarded as a gauging of the former.
The double coset $\bm{K} \backslash \bm{D} /\bm{F}$ is not necessarily smooth, 
and may have singularities.
 
 In our discussion to this point, the defining data for the two-dimensional theory are
 given by
 \begin{enumerate}
 \item the defect Lie algebra $\mathfrak{d}$, 
 \item the symmetric bilinear pairing  $\llangle \cdot , \cdot\rrangle_{\mathfrak{d}}$
 \item the maximal isotropic algebra $\mathfrak{k}$,
 \item the isotropic algebra $\mathfrak{f}$, and
 \item the decomposition $\mathfrak{r}= \mathfrak{r}_{+} \dot{+} \mathfrak{r}_{-}$.
 \end{enumerate}
The last decomposition defines an involution $\mathcal{E}_{\mathfrak{r}}$ acting as $\pm 1$ on $\mathfrak{r}_{\pm}$. This involution extends to an involution $\mathcal{E}$ on the whole of $\mathfrak{d}$,
which is discussed in the literature under the same name and is the origin of the name ``$\mathcal{E}$-model.''

\subsubsection*{Classical integrability}

For the two-dimensional system (\ref{2d-formula-1}) thus constructed to be classically integrable, its equations of motion must be equivalent to the flatness condition of the Lax pair. In order to establish this equivalence, we need to impose an additional constraint on a variation of the Lax pair.

To see this, consider a variation of the Lax pair and the edge mode
\begin{align}\label{var}
    \cL'=\cL+\epsilon \, l\,,\qquad \bmh'=\bmh \, e^{\epsilon \sfu}\,,
\end{align}
with $l \in \Omega^1(\Sigma,\mathfrak{g})$ and $\sfu \in C^{\infty}(\Sigma_{\mathfrak{p}},\mathfrak{d})$.
We require that, under the variation, the pair $(\cL',\bmh')$ still satisfies the boundary condition \eqref{Lax-con1}:
\begin{align}\label{bc-var}
    {}^{\bmh'}({\bm j}^*\cL')\in\Omega^1(\Sigma_{\mathfrak{p}},\mathfrak{k})\,.
\end{align}
This condition leads to the constraint
\begin{align}\label{const-dh}
    \mathrm{Ad}_{\bmh}(d_{\Sigma}\sfu+[{\bm j}^*\cL,\sfu]-{\bm j}^* l)\in \Omega^{1}(\Sigma_{\mathfrak{p}},\mathfrak{k})\,.
\end{align}
The variation of the action (\ref{4dCS-defect}) with the constraint (\ref{const-dh}) yields
\cite{Benini:2020skc} 
\begin{align}\label{24action-val}
    \delta S_{\text{ext}}[\cL,\bmh]&=\frac{i \epsilon}{2\pi}\int_{\Sigma\times C}\omega\wedge\Tr( l\wedge \bar{\partial}\cL)
   +\epsilon\int_{\Sigma_{\mathfrak{p}}}
    \llangle \sfu,d_{\Sigma}({\bm j}^*\cL)+\frac{1}{2}[{\bm j}^*\cL,{\bm j}^*\cL]\rrangle_{\mathfrak{d}}\no\\
    &\quad+\frac{\epsilon}{2}\int_{\Sigma_{\mathfrak{p}}}\llangle \mathrm{Ad}_{\bmh}(d_{\Sigma}\sfu+[{\bm j}^*\cL,\sfu]-{\bm j}^*l), {}^\bmh({\bm j}^*\cL)\rrangle_{\mathfrak{d}}\no\\
    &=\frac{i \epsilon}{2\pi}\int_{\Sigma\times C}\omega\wedge\Tr( l\wedge \bar{\partial}\cL)
    +\epsilon\int_{\Sigma_{\mathfrak{p}}}
    \llangle \sfu,d_{\Sigma}({\bm j}^*\cL)+\frac{1}{2}[{\bm j}^*\cL,{\bm j}^*\cL]\rrangle_{\mathfrak{d}}\,.
\end{align}
The first term vanishes as a consequence of the equations of motion (\ref{CS-eom1}) with the gauge choice $\cL_{\bar{z}}=0$. Hence, the equations of motion for the four-dimensional/two-dimensional action (\ref{4dCS-defect}) are equivalent to the flatness condition of the Lax pair
\begin{align}
    d_{\Sigma}({\bm j}^*\cL)+\frac{1}{2}[{\bm j}^*\cL,{\bm j}^*\cL]=0\,.
\end{align}
The extended action (\ref{4dCS-defect}) thus describes a two-dimensional integrable field theory.

\section{Two-dimensional integrable field theories on curved space from four-dimensional CS theory}
\label{sec:curved}

We now turn to two-dimensional integrable field theories on curved space $\Sigma$.\footnote{In Costello-Stefa\'nski's procedure \cite{Costello:2020lpi}, the metric on $\Sigma$ is introduced by extending the four-dimensional Chern–Simons action to include couplings of the Beltrami differentials with the gauge field $A$.
In the present work, however, we follow a different approach: we first construct a Lax pair which is covariant under a general coordinate transformation on the surface defects $\Sigma_{\mathfrak{p}}$, and then introduce the metric into the four-dimensional action.}
The starting point of our discussion is the four-dimensional/two-dimensional system (\ref{4dCS-defect}) with the boundary condition (\ref{bc-gauge}).

As in the flat space case, the two-dimensional integrable field theory is obtained by integrating the four-dimensional/two-dimensional system along the spectral curve $C$.
At first glance, the original four-dimensional action (\ref{CS-action}) is written in a coordinate-independent way, and there appears to be no room to introduce metric dependence on $\Sigma$. However, two-dimensional integrable field theories are realized on surface defects $\Sigma_{\mathfrak{p}}$ using solutions (Lax pairs) that break diffeomorphism invariance on $\Sigma$ due to the existence of poles on $C$, as we will see below.
Thus, the insertion of the defects into the four-dimensional theory makes it possible to introduce metric dependence on each surface defect.

\subsubsection*{Coordinate transformation}

Let us recall that our boundary condition (\ref{Lax-con1}) is expressed in a coordinate-independent manner\footnote{Not all boundary conditions in the four-dimensional theory are diffeomorphism invariant. For example, chiral/anti-chiral Dirichlet boundary conditions $A_+=0$ (or $A_{-}=0)$ at each simple pole $x\in \mathfrak{p}$ explicitly break the diffeomorphism symmetry.}.
This means that the metric dependence in this condition appears only when $\cL$ is written down in the edge mode field $\bmh$.
Therefore, our task is to construct a Lax pair that transforms covariantly under general coordinate transformations.

To this end, it is convenient to introduce a notation in which the coordinates on the flat space $\Sigma$ are denoted with primes, i.e. $(x^{+’}, x^{-’})$ for the light-cone coordinates, and the metric is
\begin{align}\label{2d-metric-t}
    ds^2=-dx^{+'}dx^{-'}\,.
\end{align}
We modify the curvature on a surface defect $\Sigma_x$ by deforming the complex structure of the metric (\ref{2d-metric-t}) such that
\begin{align}\label{comp-def}
    dx^{+'}=e^{\phi}(dx^++\mu_{x} dx^-)\,,\qquad dx^{-'}=e^{\phi}(dx^-+\bar{\mu}_{x} dx^+)\,,
\end{align}
where $\mu_x, \bar{\mu}_x$ are the Beltrami differentials and the subscript indicates the position of the surface defect $\Sigma_x$ on $C$. (In the following, we sometimes denote them simply as
$\mu$ and $\bar{\mu}$.)
This yields the Beltrami parameterization of the metric on the surface defects $\Sigma_x$
\begin{align}
    ds^2=-e^{2\phi}(dx^++\mu dx^-)(dx^-+\bar{\mu}dx^+)\,.
\end{align}
The local invertibility of the transformation (\ref{comp-def}) requires that the Beltrami differentials and the conformal factor satisfy the condition
\begin{align}
    (\partial_--\mu\partial_+)e^{\phi}&=(\partial_+ \mu)e^{\phi}\,,\qquad
     (\partial_+-\bar{\mu} \partial_-)e^{\phi}=(\partial_- \bar{\mu})e^{\phi}\,.
\end{align}

\subsubsection*{Lax pair}

To construct a covariant Lax pair with the components $\cL_{\pm}$, we require the Lax connection $\cL$ to be invariant under the deformation. This leads to the identity
\begin{align}\label{Lax-id}
    \cL=\cL_{+}dx^++\cL_{-}dx^-=\cL_{+'}dx^{+'}+\cL_{-'}dx^{-'}\,,
\end{align}
where $\cL_{\pm'}$ denotes a Lax pair satisfying the ansatz (\ref{Lax-ansatz0}) in the flat-space case.
The covariant Lax pair $\cL_{\pm}$ continues to satisfy the transformation property (\ref{Lax-con2}).
We temporarily assume the coefficient functions are covariant under this deformation, i.e.,
\begin{align}
\begin{split}\label{coeff-diff}
    \cL_{+'}^{(x,q)}&=\frac{1}{1-\mu\bar{\mu}} \cL_{+}^{(x,q)}-\frac{\bar{\mu}}{1-\mu\bar{\mu}}\cL_{-}^{(x,q)}\,,\\ 
   \cL_{-'}^{(x,q)}&=\frac{1}{1-\mu\bar{\mu}}\cL_{-}^{(x,q)}-\frac{\mu}{1-\mu\bar{\mu}}\cL_{+}^{(x,q)}\,.
\end{split}
\end{align}
As we will see in the following subsection, this holds for the degenerate $\cE$-model, which provides a unified description of 2d integrable field theories associated with the boundary condition (\ref{Lax-con1}).

From the relations (\ref{Lax-id}) and (\ref{coeff-diff}), we can straightforwardly write down the Lax pair on a curved space. To write its concrete form, it is convenient to introduce the world-sheet projection operator
\begin{align}\label{sheet-proj}
    P^{(\pm)\mu\nu}:=\frac{\gamma^{\mu\nu}\pm \epsilon^{\mu\nu}}{2}\,,
\end{align}
where $\gamma^{\mu\nu}=\sqrt{-g}g^{\mu\nu}$ denotes the Weyl-invariant metric (see appendix \ref{Weyl-metric} for details), and the antisymmetric tensor $\epsilon^{\mu\nu}$ is normalized as $\epsilon^{+-}=1$.
By using the relation (\ref{proj-bel}), each component of the Lax pair on a curved space can be expressed as  
\begin{align}\label{curve-Lax}
    \cL&=\left(\sum_{x\in \mathfrak{z}'_+}\frac{P_{+}^{(+)\mu}\cL_{\mu}^{(x,0)}}{z-x}+\sum_{x\in \mathfrak{z}'_-}\frac{P_{+}^{(-)\mu}\cL_{\mu}^{(x,0)}}{z-x}+\sum_{q=0}^{m_{\infty}-1}P_{+}^{(+)\mu}\cL_{\mu}^{(\infty,q)}z^{q+1}+\cL_{+}^{c}\right)dx^+\no\\
    &\quad+\left(\sum_{x\in \mathfrak{z}'_-}\frac{P_{-}^{(-)\mu}\cL_{\mu}^{(x,0)}}{z-x}+\sum_{x\in \mathfrak{z}'_+}\frac{P_{-}^{(+)\mu}\cL_{\mu}^{(x,0)}}{z-x}+\sum_{q=0}^{m_{\infty}-1}P_{-}^{(+)\mu}\cL_{\mu}^{(\infty,q)}z^{q+1}+\cL_{-}^{c}\right)dx^-\,,
\end{align}
where we defined
\begin{align}
    P_{\mu}^{(\pm)\nu} :=\gamma_{\mu\rho}P^{(\pm)\rho\nu}=\frac{1}{2}(\delta_{\mu}^{\nu}\pm \gamma_{\mu\rho}\epsilon^{\rho\nu})\,.
\end{align}
Since, by definition, the constant terms $\cL_{\pm}^{c}$ do not have poles on $C$, the diffeomorphism invariance on $\Sigma$ is preserved, and the metric does not appear in this term.
For consistency, one can verify that in flat-space limit, the projection operators $P_{\mu}^{(\pm)\nu}$ becomes $P_{\pm}^{(\pm)\mu}=\delta_{\pm}^{\mu}$, $P_{\mp}^{(\pm)\mu}=0$, and the Lax pair (\ref{curve-Lax}) reduces to the admissible ansatz (\ref{Lax-ansatz0}) in flat space.

\subsubsection*{On the finiteness of the four-dimensional CS action and reduction to two-dimensional action}
\label{subsec:finiteness}

In contrast to the flat-space solutions \eqref{Lax-ansatz0} considered in Section \ref{sec:Lax},
each component of the ansatz (\ref{curve-Lax}) simultaneously contains poles associated with both $\mathfrak{z}_+$ and $\mathfrak{z}_-$. 
This raises the question of whether the ansatz is admissible, i.e.\ if the four-dimensional action remains finite after integration.
In fact, the Lagrangian with the gauge choice $\cL_{\bar{z}}=0$ has potentially divergent terms, given by
\begin{align}
    \omega \wedge \CS(\cL)&\supset\frac{1}{2}\omega \wedge \biggl(P_{+}^{(+)[\mu}P_{-}^{(+)\nu]}\Tr(\cL_{\mu}^{(+)}\partial_{\bar{z}}\cL_{\nu}^{(+)})\no\\
    &\qquad\qquad+P_{+}^{(-)[\mu}P_{-}^{(-)\nu]}\Tr(\cL_{\mu}^{(-)}\partial_{\bar{z}}\cL_{\nu}^{(-)})\biggr)d\bar{z}\wedge dx^+\wedge dx^-\,,\label{div-Lag}
\end{align}
where the indices $\mu$ and $\nu$ are anti-symmetrized, and $\cL^{(+)}_{\mu}(\cL^{(-)}_{\mu})$ denotes the set of terms containing the poles $1/(z-x)$ for $x \in \mathfrak{z}_+(\mathfrak{z}_-)$ in the Lax pair (\ref{curve-Lax}).
Each contribution diverges due to the presence of $1/(z-x)\delta^{(2)}(z-x)$.
To avoid the divergence of the action, we regularize the delta function $\delta^{(2)}(z-x)$, which arises from the derivative $\partial_{\bar{z}}$ of the poles $1/(z-x)$ for $x \in \mathfrak{z}$, by a smooth function.
The right hand side of (\ref{div-Lag}) then vanishes since the regularized terms $\Tr(\cL_{\mu}^{(\pm)}\partial_{\bar{z}}\cL_{\nu}^{(\pm)})$ are symmetric with respect to $\mu$ and $\nu$.
Also for the Lagrangian $\omega \wedge \CS(A)$ before fixing the gauge $A_{\bar{z}}=0$, we can verify that potentially divergent terms cancel out after performing a similar regularization of the delta function.
Thus, after proper regularization, the four-dimensional action remains finite even for the ansatz (\ref{curve-Lax}). The Lax pair (\ref{curve-Lax}) is an admissible solution in this sense.

Similarly, the argument concerning the classical integrability of the effective two-dimensional theory discussed in Section \ref{sec:reduction} requires a modification. In the flat-space case, we used the relation 
\begin{align}
    \frac{i}{4\pi}\int_{\Sigma\times C}d\omega\wedge \Tr(l\wedge \cL)=-\frac{1}{2}\int_{\Sigma_{\mathfrak{p}}}\llangle {\bm j}^*l,{\bm j}^*\cL\rrangle_{\mathfrak{d}}
\end{align}
obtained by integrating over $C$. In the present case, the term $l\wedge \cL$ contains poles of higher order than the zeros of $\omega$, and after regularizations as above, 
the variation (\ref{24action-val}) of the extended four-dimensional/two-dimensional action is modified as
\begin{align}\label{m4d2d-var}
    \delta S_{\text{ext}}[\cL,\bmh]
    &=\frac{i \epsilon}{2\pi}\int_{\Sigma\times C}\omega\wedge\Tr( l\wedge \bar{\partial}\cL)\lvert_{\text{reg}}
    +\epsilon\int_{\Sigma_{\mathfrak{p}}}
    \llangle \sfu,d_{\Sigma}({\bm j}^*\cL)+\frac{1}{2}[{\bm j}^*\cL,{\bm j}^*\cL]\rrangle_{\mathfrak{d}}\,,
\end{align}
where $( l\wedge \bar{\partial}\cL)\lvert_{\text{reg}}$ denotes the regularized part of $l\wedge \bar{\partial}\cL$, removing singular terms such as $1/(z-x)\delta^{(2)}(z-x)$.  
One can then show that the first term of (\ref{m4d2d-var}) vanishes again under the equations of motion $\omega \wedge \bar{\partial}\cL=0$, and the effective two-dimensional action is classically integrable.

\section{Degenerate \texorpdfstring{$\cE$}{\mathcal{E}}-model on curved space}
\label{sec:E-model_curved}

To proceed further, we require more explicit solutions of the Lax pair.
For this reason, we derive the action and the Lax pair for the degenerate $\cE$-model on a curved space $\Sigma$,
whose flat-space counterpart was discussed in Section \ref{sec:E-model}.
Since the general formalism in Section \ref{curve-E} is technically involved, 
it may be useful to consult the example of PCM in Section \ref{ex:PCM} while reading through the general discussion
in Section \ref{curve-E}.

\subsection{Solution of Lax pair}\label{curve-E}

Since we have already solved for the Lax pair in flat space (Section \ref{sec:Lax_solution}) and 
derived the transformation property under a coordinate transformation
(Section \ref{sec:curved}), we are now ready to discuss the case of a curved space.
Under the complex structure deformation (\ref{comp-def}), the left-invariant current $\bmh^{-1}\partial_{\pm'}\bmh$ transforms according to the same rule as (\ref{coeff-diff}),
\begin{align}
\begin{split}\label{hdh-diff}
    \bmh^{-1}\partial_{+'}\bmh&=\frac{1}{1-\mu\bar{\mu}} \bmh^{-1}\partial_{+}\bmh-\frac{\bar{\mu}}{1-\mu\bar{\mu}}\bmh^{-1}\partial_{-}\bmh\,,\\ 
    \bmh^{-1}\partial_{-'}\bmh&=\frac{1}{1-\mu\bar{\mu}}\bmh^{-1}\partial_{-}\bmh-\frac{\mu}{1-\mu\bar{\mu}}\bmh^{-1}\partial_{+}\bmh\,.
\end{split}
\end{align}
This leads to the Lax pair for the degenerate $\cE$-model on curved space:
\begin{align}
\begin{split}\label{Lax-curve-E}
    \cL_{+}&=P_{+}^{(+)\mu}\,\bmp\left(\mathcal{P}_{\bmh}^{+}({\bm h}^{-1}\partial_{\mu}{\bm h})\right)+P_{+}^{(-)\mu}\,\bmp\left(\mathcal{P}_{\bmh}^{-}({\bm h}^{-1}\partial_{\mu}{\bm h})\right)\,,\\
    \cL_{-}&=P_{-}^{(-)\mu}\,\bmp\left(\mathcal{P}_{\bmh}^{-}({\bm h}^{-1}\partial_{\mu}{\bm h})\right)+P_{-}^{(+)\mu}\,\bmp\left(\mathcal{P}_{\bmh}^{+}({\bm h}^{-1}\partial_{\mu}{\bm h})\right)\,,
\end{split}
\end{align}
or equivalently
\begin{align}\label{Lax-curve-E1}
    \cL_{\pm}&=\frac{1}{2}\bmp\left((\mathcal{P}_{\bmh}^{+}+\mathcal{P}_{\bmh}^{-})({\bm h}^{-1}\partial_{\pm}{\bm h})\right)+\frac{1}{2}\gamma_{\pm\rho}\epsilon^{\rho\mu}\,\bmp\left((\mathcal{P}_{\bmh}^{+}-\mathcal{P}_{\bmh}^{-})({\bm h}^{-1}\partial_{\mu}{\bm h})\right)\,.
\end{align}
This solution satisfies the boundary condition:
\begin{align}
\begin{split}
    {}^{\bmh}({\bm j}^*\cL_{\pm})&=P_{\pm}^{(+)\mu}\mathrm{Ad}_{\bmh}((\mathcal{P}_{\bmh}^{+}-\mathrm{Id})({\bm h}^{-1}\partial_{\mu}{\bm h}))\\
    &\quad+P_{\pm}^{(-)\mu}\mathrm{Ad}_{\bmh}((\mathcal{P}_{\bmh}^{-}-\mathrm{Id})({\bm h}^{-1}\partial_{\mu}{\bm h}))\in\Omega^1(\Sigma_{\mathfrak{p}},\mathfrak{k})\,.
\end{split}
\end{align}
In the following, we will give a more explicit form of the Lax pair (\ref{Lax-curve-E}) for specific choices of boundary conditions.

\subsection{Reduction to two-dimensional action on curved space}\label{sec:reduction_curved}

Next, we derive the action of the degenerate $\cE$-model on the curved surface $\Sigma$.
Substituting (\ref{Lax-curve-E}) into (\ref{2d-formula-1}) leads to
\begin{align}\label{cE-action2}
\begin{split}
    S_{\rm 2d}[\bmh]&=\frac{1}{4}\int_{\Sigma_{\mathfrak{p}}}\biggl(\gamma^{\mu\nu}\llangle \bmh^{-1}\partial_{\mu}\bmh, (\mathcal{P}_{\bmh}^{+}-\mathcal{P}_{\bmh}^{-})(\bmh^{-1}\partial_{\nu}\bmh)\rrangle_{\mathfrak{d}}
   \\
    &\qquad\qquad
    +\varepsilon^{\mu\nu}\llangle \bmh^{-1}\partial_{\mu}\bmh, (\mathcal{P}_{\bmh}^{+}+\mathcal{P}_{\bmh}^{-})(\bmh^{-1}\partial_{\nu}\bmh)\rrangle_{\mathfrak{d}}\biggr)dx^+\wedge dx^-
    -\frac{1}{2}S_{\mathfrak{d}}^{\text{WZ}}[\bmh]\,.
\end{split}
\end{align}
The projection operators $\mathcal{P}^{\pm}_{\bmh}$ satisfy the relation
\begin{align}
 \llangle  \sfu,\mathcal{P}^{+}_{\bmh}(\sfv)\rrangle_{\mathfrak{d}}=\llangle (\mathrm{Id}- \mathcal{P}^{-}_{\bmh})(\sfu),\sfv\rrangle_{\mathfrak{d}}\qquad \text{for}\quad \sfu\,,\sfv\in\mathfrak{d}\,.
\end{align}
In the flat-space limit, this reduces to the action of the degenerate $\cE$-model on flat space 
in \eqref{dE-action}. 
The equations of motion are 
\begin{align}\label{e-eom}
\begin{split}
    &-\partial_{\mu}(P^{(-)\mu\nu}\mathcal{P}^{-}_{\bmh}(\bmh^{-1}\partial_{\nu}\bmh))+\partial_{\mu}(P^{(+)\mu\nu}\mathcal{P}^{+}_{\bmh}(\bmh^{-1}\partial_{\nu}\bmh))\\
    &\quad-[P^{(+)\mu\rho}\mathcal{P}^{+}_{\bmh}(\bmh^{-1}\partial_{\rho}\bmh),P_{\mu}^{(-)\sigma}\mathcal{P}^{-}_{\bmh}(\bmh^{-1}\partial_{\sigma}\bmh)]=0\,.
\end{split}
\end{align}
For any $\sfu_{\pm}\in \mathfrak{f}\oplus \mathfrak{r}_{\pm}=\operatorname{Im}\,\mathcal{P}_{\bmh}^{\pm}$\,, the inverse map $\bmp$ satisfies
\begin{align}
    \bmp\left([\sfu_+,\sfu_-]\right)=[\bmp(\sfu_+),\bmp(\sfu_-)]\,,
\end{align}
which follows since ${\bm j}_{*}$ is the inverse of ${\bm j}^*$,
the latter being compatible with the commutation relations by definition.
By virtue of this property, the equations of motion for the Lax pair (\ref{Lax-curve-E}) are equivalent to (\ref{E-eom}).
The degenerate $\cE$-model is therefore classically integrable.

Finally, we present the energy-momentum tensor,
which will be required for the discussion of  the $T\bar{T}$ and root-$T\bar{T}$ deformations of the degenerate $\cE$-model in a later section.
By varying the action (\ref{cE-action2}) with respect to the metric, we obtain the energy-momentum tensor
\begin{align}\label{em-dE}
    T_{\pm\pm}^{(0)}=-\frac{1}{2}\llangle \bmh^{-1}\partial_{\pm}\bmh ,(\mathcal{P}_{\bmh}^{-}-\mathcal{P}_{\bmh}^{+})(\bmh^{-1}\partial_{\pm}\bmh) \rrangle_{\mathfrak{d}}\,,\qquad T_{+-}^{(0)}=0\,.
\end{align}
The energy-momentum tensor is invariant under the gauge transformation (\ref{2dgauge-sym}), i.e.\
\begin{align}\label{em-tr}
     T_{\pm\pm}^{(0)}({\bm k}\,\bmh\Delta(f)^{-1})&= T_{\pm\pm}^{(0)}(\bmh)\,,
\end{align}
as shown in Appendix \ref{gaugetr-em}.
As we will see, this implies that both the $T\bar{T}$- and root-$T\bar{T}$-deformed actions are invariant under the gauge transformation (\ref{2dgauge-sym}).

\subsubsection*{On classical integrability}

We now explicitly check that the variation of the Lax pair (\ref{Lax-curve-E1}) for the degenerate $\cE$-model on curved space satisfies the boundary condition (\ref{bc-var}).
Under the variation (\ref{var}), the left-invariant current for $\bmh$ and its projections in $\mathcal{P}_{\bmh}^{\pm}$ are 
\begin{align}
 \bmh'^{-1}d_{\Sigma}\bmh'&=\bmh^{-1}d_{\Sigma}\bmh+\epsilon \left(d_{\Sigma}\sfu+[\bmh^{-1}d_{\Sigma}\bmh,\sfu]\right)\,,\\
    \mathcal{P}_{\bmh'}^{\pm}(\bmh^{'-1}d_{\Sigma}\bmh')&=\mathcal{P}_{\bmh}^{\pm}(\bmh^{-1}d_{\Sigma}\bmh)+\epsilon\left(\mathcal{P}_{\bmh}^{\pm}(d_{\Sigma}\sfu)+[\mathcal{P}_{\bmh}^{\pm}(\bmh^{-1}d_{\Sigma}\bmh),\sfu]\right)\,,
\end{align}
where in the second equality, we have used the transformation rule (\ref{W-tr}) of the $\mathcal{P}_{\bmh}^{\pm}$.
The Lax pair (\ref{Lax-curve-E1}) transforms as
\begin{align}
    {\bm j}^*\cL_{\pm}'&={\bm j}^*\cL_{\pm}+\epsilon\,[{\bm j}^*\cL_{\pm},\sfu]\no\\
    &\quad+\frac{1}{2}\epsilon\,[(\mathcal{P}_{\bmh}^{+}+\mathcal{P}_{\bmh}^{-})(\partial_{\pm}\sfu)+\gamma_{\pm\rho}\varepsilon^{\rho\mu}(\mathcal{P}_{\bmh}^{+}-\mathcal{P}_{\bmh}^{-})(\partial_{\mu}\sfu)]\no\\
    &={\bm j}^*\cL_{\pm}+\epsilon\left([{\bm j}^*\cL_{\pm},\sfu]
    +(P_{\pm}^{(+)\mu}\mathcal{P}_{\bmh}^{+}(\partial_{\mu}\sfu)+P_{\pm}^{(-)\mu}\mathcal{P}_{\bmh}^{-}(\partial_{\mu}\sfu))\right)\,,
\end{align}
and the variation $l_{\pm}$ takes the form
\begin{align}\label{E-model-var}
    {\bm j}^*l_{\pm}=[{\bm j}^*\cL_{\pm},\sfu]
    +(P_{\pm}^{(+)\mu}\mathcal{P}_{\bmh}^{+}(\partial_{\mu}\sfu)+P_{\pm}^{(-)\mu}\mathcal{P}_{\bmh}^{-}(\partial_{\mu}\sfu))\,.
\end{align}
The constraint (\ref{const-dh}) can be rewritten as
\begin{align}
    & \mathrm{Ad}_{\bmh}\left(d_{\Sigma}\sfu+[{\bm j}^*\cL,\sfu]-{\bm j}^* l\right)\no\\
    &=\mathrm{Ad}_{\bmh}\left(P_{+}^{(+)\mu}(\mathrm{Id}-\mathcal{P}_{\bmh}^{+})\partial_{\mu}\sfu+P_{+}^{(-)\mu}(\mathrm{Id}-\mathcal{P}_{\bmh}^{-})\partial_{\mu}\sfu\right)dx^+\no\\
    &\quad+\mathrm{Ad}_{\bmh}\left(P_{-}^{(+)\mu}(\mathrm{Id}-\mathcal{P}_{\bmh}^{+})\partial_{\mu}\sfu+P_{-}^{(-)\mu}(\mathrm{Id}-\mathcal{P}_{\bmh}^{-})\partial_{\mu}\sfu\right)dx^-\,.
\end{align}
From the definition (\ref{def-W}) of $\mathcal{P}_{\bmh}^{\pm}$, we have $\mathrm{Ad}_{\bmh}(\mathrm{Id}-\mathcal{P}_{\bmh}^{\pm})\partial_{\pm}\sfu\in \Omega^1(\Sigma_{\mathfrak{p}}, \mathfrak{k})$.

\subsection{Example: PCM on curved space}\label{ex:PCM}

The method for constructing Lax pairs and actions described above is abstract but can be easily extended to cases where $\omega$ contains higher-order poles. Here, we illustrate how to compute the Lax pair and the action using this method for the simplest example, the principal chiral model (PCM) for a Lie group $G$ on a curved space. Readers who do not need details of the calculations can ignore this section.

The PCM can be derived from the four-dimensional theory with the meromorphic one-form \cite{Costello:2019tri}
\begin{align}
    \omega=T\frac{1-z^2}{z^2}dz\qquad T: \, \text{constant}\,.\label{omega-pcm}
\end{align}
The meromorphic one-form $\omega$ exhibits the following double poles and simple zeros:
\begin{align}
    \mathfrak{p}=\{0\,, \infty\}\,,\qquad \mathfrak{z}=\{\pm 1\}\,,
\end{align}
and the levels $l_{p=0,1}^{x \in \mathfrak{p}}$ of $\omega$ are given by
\begin{align}
    l^{0}_{0}=0\,,\quad l^{0}_{1}=T\,,\quad l^{\infty}_{0}=0\,,\quad l^{\infty}_{1}=T\,.
\end{align}
The defect Lie algebra $\mathfrak{d}$ \eqref{defect-Liealg} in this example takes the form
\begin{align}
    \mathfrak{d}=\left(\mathfrak{g}\otimes\mathbb{R}[\varepsilon_0]/(\varepsilon_0^2)\right)\times \left(\mathfrak{g}\otimes\mathbb{R}[\varepsilon_{\infty}]/(\varepsilon_{\infty}^2)\right)
\end{align}
with the commutation relations
\begin{align}\label{comm-semiab}
    [\sfu^0,\sfv^0]=[\sfu,\sfv]^0\,,\qquad [\sfu^0,\sfv^1]=[\sfu^1,\sfv^0]=[\sfu,\sfv]^1\,,\qquad [\sfu^1,\sfv^1]=0\,.
\end{align}
This shows that $\mathfrak{g}\otimes\mathbb{R}[\varepsilon]/(\varepsilon^2)\simeq\mathfrak{g}\ltimes \mathfrak{g}_{\rm ab}$\,, and the subalgebras $\mathfrak{g}\oplus \{0\}$ and $ \{0\}\oplus \mathfrak{g}_{\rm ab}$ of $\mathfrak{g}\ltimes \mathfrak{g}_{\rm ab}$ are generated by elements of $\mathfrak{g}\otimes\mathbb{R}[\varepsilon]/(\varepsilon^2)$
\begin{align}
   \mathfrak{g}\oplus \{0\}=\{\sfu^0\,\lvert\,\sfu\in \mathfrak{g}\}\,,\qquad \{0\}\oplus \mathfrak{g}_{\rm ab}=\{\sfu^1\,\lvert\,\sfu\in \mathfrak{g}\}\,,
\end{align}
where $\mathfrak{g}_{\rm ab} = \mathfrak{g} / [\mathfrak{g}, \mathfrak{g}]$ is the abelianization of the Lie algebra $\mathfrak{g}$, and we introduce the notation $\sfu^p:=\sfu\,\otimes\,\varepsilon^p$.
The bilinear form on $\mathfrak{d}$ is non-vanishing on each factor of 
\begin{align}
\begin{split}
    \llangle \sfu^0, \sfv^1\rrangle_{\mathfrak{d}}&=T\,\Tr(\sfu\,\sfv)\,,\qquad
     \llangle \sfu^1, \sfv^{0}\rrangle_{\mathfrak{d}}=T\,\Tr(\sfu\,\sfv)\,,
\end{split}
\end{align}
while all the others vanish.
The defect Lie group ${\bm D}$ can be identified with the product of two copies of the tangent bundles $TG$, each isomorphic to $G\ltimes \mathfrak{g}$.
The group multiplication of $TG$ is given by
\begin{align}
    (g,\sfu)(f,\sfv)=(gf,\sfu+\mathrm{Ad}_g\sfv)\,,\qquad (g,\sfu)^{-1}=(g^{-1},-\mathrm{Ad}_g^{-1}\sfu)\,,
\end{align}
where $f\,, g \in G$ and $\sfu\,,\sfv \in\mathfrak{g}$.

\subsubsection*{Boundary conditions}

To specify the boundary conditions of the gauge field $A$, we consider a decomposition of $\mathfrak{d}$ into two maximal isotropic subalgebras:
\begin{align}
    \mathfrak{d}=\mathfrak{k}\,\dot{+}\,\tilde{\mathfrak{k}}\,.
\end{align}
For our purpose, we choose $\mathfrak{k}$ as 
\begin{align}\label{pcm-al}
    \mathfrak{k}=\left(\{0\}\oplus \mathfrak{g}_{\rm ab}\right)\times \left(\{0\}\oplus \mathfrak{g}_{\rm ab}\right)\,,
\end{align}
and this choice corresponds to the Dirichlet boundary condition
\begin{align}
A_{\pm}\lvert_{z=0}&=0\,,\qquad 
A_{\pm}\lvert_{\xi_{\infty}=0}=0 \,.\label{bcPCM1}
\end{align}
Indeed, this choice automatically satisfies the boundary equations of motion (\ref{beom-int}) for (\ref{omega-pcm}) 
\begin{align}
    \llangle A,\delta A\rrangle_{\mathfrak{d}}=T\, \sum_{x\in\mathfrak{p}}\partial_{\xi_x}\Tr\left( A\wedge \delta A\right)\lvert_{z=x}=0\,.\label{beom-int2}
\end{align}

\subsubsection*{Lax pair}

To construct a Lax pair, we decompose $\mathfrak{z}$ into $ \mathfrak{z}_+=\{+1\}$ and $\mathfrak{z}_-=\{-1\}$, leading to the following admissible ansatz:
\begin{align}
    \cL&=\left(\frac{P_{+}^{(+)\mu}\cL_{\mu}^{(1,0)}}{z-1}+\frac{P_{+}^{(-)\mu}\cL_{\mu}^{(-1,0)}}{z+1}+\cL_{+}^{c}\right)dx^+\no\\
    &\quad+\left(\frac{P_{-}^{(-)\mu}\cL_{\mu}^{(-1,0)}}{z+1}+\frac{P_{-}^{(+)\mu}\cL_{\mu}^{(1,0)}}{z-1}+\cL_{-}^{c}\right)dx^-
\end{align}
satisfying the boundary condition
     ${}^{\bmh}({\bm j}^*\cL(\bmh))\in \Omega^1(\Sigma_{\mathfrak{p}},\mathfrak{k})$.
Using the gauge transformation (\ref{2dgauge-sym}), we choose the edge field as $\bmh=(g,\tilde{g}) \in C^{\infty}(\Sigma_{\mathfrak{p}},\tilde{K})$. The corresponding left-invariant current for $\bmh$ is
\begin{align}\label{c-edge-pcm}
    \bmh^{-1}d_{\Sigma}\bmh=(j^0,\tilde{j}^{0})\,,\qquad j=g^{-1}d_{\Sigma}g\,,\qquad \tilde{j}=\tilde{g}^{-1}d_{\Sigma}\tilde{g}\,,
\end{align}
where $d_{\Sigma}$ denotes the exterior derivative on $\Sigma$.
We now specify the kernel and the image of the map $\mathcal{P}_{\bmh}^{\pm}$ defined by (\ref{def-W}).
Since $[\mathfrak{k},\tilde{\mathfrak{k}}]\subset \mathfrak{k}$ by (\ref{comm-semiab}), $\mathrm{Ad}_h^{-1}\mathfrak{k}=\mathfrak{k}$. Hence, the kernel of the projection operators $\operatorname{Ker}\,\mathcal{P}_{\bmh}^{\pm}$ is
\begin{align}
    \operatorname{Ker}\,\mathcal{P}_{\bmh}^{\pm}&=\mathfrak{k}=\left(\mathfrak{g}\otimes\varepsilon_0\right)\times \left(\mathfrak{g}\otimes\varepsilon_{\infty}\right)=\{(\sfy^1,\sfz^1)\,\lvert\,\sfy\,,\sfz\in\mathfrak{g}\}\,.
\end{align}
The diagonal subalgebra $\mathfrak{f}$ takes the form
\begin{align}
    \mathfrak{f}=\{(\sfw^0,\sfw^0)\,\lvert\,\sfw\in\mathfrak{g}\}\,,
\end{align}
and the vector spaces $\mathfrak{r}_{\pm}$ are given by
\begin{align}
    \mathfrak{r}_+&={\bm j}(R'_{\mathfrak{z}_{+}})=\{(-\sfx^0-\sfx^1,\sfx^1)\lvert \sfx \in \mathfrak{g}\}\,,\\
    \mathfrak{r}_-&={\bm j}(R'_{\mathfrak{z}_{-}})=\{(\sfx^0+\sfx^1,\sfx^1)\lvert \sfx \in \mathfrak{g}\}\,.
\end{align}
In terms of these vector spaces, the left-invariant current can be decomposed as
\begin{align}\label{cedge-pcm-dec}
    \bmh^{-1}\partial_{\mu}\bmh=
    \begin{cases}
        [(\sfy^1,\sfz^1)]+[(\sfw^0,\sfw^0)+(-\sfx^0-\sfx^1,\sfx^1)]\qquad \text{for}\quad [\mathfrak{k}]\oplus [\mathfrak{f}\oplus \mathfrak{r}_{+}]\\
         [(\sfy^1,\sfz^1)]+[(\sfw^0,\sfw^0)+(\sfx^0-\sfx^1,\sfx^1)]\qquad \text{for}\quad [\mathfrak{k}]\oplus[\mathfrak{f}\oplus \mathfrak{r}_{-}]
    \end{cases}
\,,
\end{align}
where the first and second terms grouped by the bracket $[\,\cdot\,]$ on the right-hand side correspond to $ \operatorname{Ker}\,\mathcal{P}_{\bmh}^{\pm}$ and $ \text{Im}\,\mathcal{P}_{\bmh}^{\pm}$, respectively.
By comparing (\ref{c-edge-pcm}) with (\ref{cedge-pcm-dec}), we can determine the actions of $ \mathcal{P}_{\bmh}^{\pm}$ on $\bmh^{-1}\partial_{\mu}\bmh$ as being 
\begin{align}
    \mathcal{P}_{\bmh}^{+}(\bmh^{-1}\partial_{\mu}\bmh)&=(\tilde{j}_{\mu}^{0},\tilde{j}_{\mu}^{0})+\left((j_{\mu}-\tilde{j}_{\mu})^0+(j_{\mu}-\tilde{j}_{\mu})^1,-(j_{\mu}-\tilde{j}_{\mu})^1\right)\,,\label{W-leftp}\\
      \mathcal{P}_{\bmh}^{-}(\bmh^{-1}\partial_{\mu}\bmh)&=(\tilde{j}_{\mu}^0,\tilde{j}_{\mu}^0)+\left((j_{\mu}-\tilde{j}_{\mu})^0-(j_{\mu}-\tilde{j}_{\mu})^1,(j_{\mu}-\tilde{j}_{\mu})^1\right)\,.\label{W-leftm}
\end{align}
The actions of the map $\bmp$ on $\mathcal{P}_{\bmh}^{\pm}(\bmh^{-1}\partial_{\mu}\bmh)$ are given by
\begin{align}
    \bmp(\mathcal{P}_{\bmh}^{+}(\bmh^{-1}\partial_{\mu}\bmh))&=\tilde{j}_{\mu}-\frac{j_{\mu}-\tilde{j}_{\mu}}{z-1}\,,\qquad
     \bmp( \mathcal{P}_{\bmh}^{-}(\bmh^{-1}\partial_{\mu}\bmh))=\tilde{j}_{\mu}+\frac{j_{\mu}-\tilde{j}_{\mu}}{z+1}\,,
\end{align}
and substituting this into (\ref{Lax-curve-E}) yields 
\begin{align}\label{Lax-cpcm}
    \cL&=\left(-\frac{P_{+}^{(+)\mu}(j_{\mu}-\tilde{j}_{\mu})}{z-1}+\frac{P_{+}^{(-)\mu}(j_{\mu}-\tilde{j}_{\mu})}{z+1}+\tilde{j}_{+}\right)dx^+\no\\
    &\quad+\left(\frac{P_{-}^{(-)\mu}(j_{\mu}-\tilde{j}_{\mu})}{z+1}-\frac{P_{-}^{(+)\mu}(j_{\mu}-\tilde{j}_{\mu})}{z-1}+\tilde{j}_{-}\right)dx^-\,.
\end{align}
This is the Lax pair of the (gauged) PCM on curved space.
When we choose the gauge $\tilde{g}=1$, the Lax pair (\ref{Lax-cpcm}) takes the same form as the one presented in \cite{Arutyunov:2004yx}.

\subsubsection*{2d action}

We now derive the two-dimensional action corresponding to the Lax pair (\ref{Lax-cpcm}) by substituting it into the general formula (\ref{2d-formula-1}) or (\ref{cE-action2}). From (\ref{W-leftp}) and (\ref{W-leftm}), we compute
\begin{align}
    \llangle \bmh^{-1}\partial_{\mu}\bmh, \mathcal{P}_{\bmh}^{+}(\bmh^{-1}\partial_{\nu}\bmh)\rrangle_{\mathfrak{d}}&=T\,\Tr(j_{\mu}(j_{\nu}-\tilde{j}_{\nu}))\lvert_{z=0}-T\,\Tr(\tilde{j}_{\mu}(j_{\nu}-\tilde{j}_{\nu}))\lvert_{z=\infty}\,,\\
    &=-\llangle \bmh^{-1}\partial_{\mu}\bmh, \mathcal{P}_{\bmh}^{-}(\bmh^{-1}\partial_{\nu}\bmh)\rrangle_{\mathfrak{d}}\,.
\end{align}
This yields the action of the (gauged) PCM
\begin{align}\label{2d-action-C0}
\begin{split}
   S_{\text{2d}}[g,\tilde{g}]&=\frac{T}{2}\,\int_{\Sigma_{0}}\gamma^{\mu\nu}\Tr(j_{\mu}(j_{\nu}-\tilde{j}_{\nu}))\,dx^+\wedge dx^-\\
    &\quad-\frac{T}{2}\,\int_{\Sigma_{\infty}}\gamma^{\mu\nu}\,\Tr(\tilde{j}_{\mu}(j_{\nu}-\tilde{j}_{\nu}))\,dx^+\wedge dx^-\,.
\end{split}
\end{align}
The WZ term vanishes since $\text{res}_{0}\,\omega=0$ and $\text{res}_{\infty}\,\omega=0$.
If we disregard the difference in positions of the surface defects on $C$, the action (\ref{2d-action-C0}) reduces to
\begin{align}\label{2d-action-C3}
   S_{\text{2d}}[g,\tilde{g}]&=\frac{T}{2}\,\int_{\Sigma}\gamma^{\mu\nu}\Tr((j_{\mu}-\tilde{j}_{\mu})(j_{\nu}-\tilde{j}_{\nu}))\,dx^+\wedge dx^-
\end{align}
and the energy-momentum tensor becomes
\begin{align}
    T_{++}=\Tr\left((j_{+}-\tilde{j}_{+})(j_{+}-\tilde{j}_{+})\right)\,,\qquad T_{--}=\Tr\left((j_{-}-\tilde{j}_{-})(j_{-}-\tilde{j}_{-})\right)\,.
\end{align}
Using the remaining two-dimensional gauge symmetry $\bmh \to \bmh\Delta(\tilde{g})^{-1}$ of (\ref{2dgauge-sym}), we can fix the gauge by setting $\tilde{g}=1$.
Under this gauge, the action (\ref{2d-action-C3}) reduces to the standard expression for the PCM action.

\section{\texorpdfstring{$T\bar{T}$}{T\bar{T}} deformation}
\label{sec:TT}

In this section, we discuss a derivation of $T\bar{T}$ deformations of the degenerate $\cE$-model from four-dimensional theory.

\subsection{\texorpdfstring{$T\bar{T}$}{T\bar{T}} deformation of the PCM}

Before considering the degenerate $\cE$-model, we briefly review the $T\bar{T}$ deformation of the PCM and its derivation by using dynamical coordinate transformations \cite{Conti:2018tca}.

\subsubsection*{$T\bar{T}$ deformed action and Lax pair}

The $T\bar{T}$ deformation of the PCM is defined by adding an irrelevant operator, the so-called $T\bar{T}$ operator, to the original action:
\begin{align}
    S_{\text{PCM}}^{(\la)}&=\int_{\mathbb{R}^2} L_{\text{PCM}}^{(\la)}\,dx^+\wedge dx^-\no\\
    &=S_{\text{PCM}}+\la \int_{\mathbb{R}^2}\text{det}\,(T_{\mu\nu}^{(0)})\,dx^+\wedge dx^-+\cO(\la^2)\,,
\end{align}
where $\la$ is a deformation parameter with a negative mass dimension, and the overall constant $T$ in (\ref{omega-pcm}) is set to $1$.
The deformed Lagrangian $L_{\text{PCM}}^{(\la)}$ satisfies the classical flow equation,
\begin{align}\label{flow-TTbar}
    \frac{d}{d\la} L_{\text{PCM}}^{(\la)}=\text{det}(T_{\mu\nu}^{(\la)})\,,
\end{align}
where $T_{\mu\nu}^{(\la)}$ is the energy-momentum tensor for the deformed model.
Solving the flow equation (\ref{flow-TTbar}) yields the $T\bar{T}$-deformed action \cite{Chen:2021aid}
\begin{align}\label{TTpcm}
 S_{\text{PCM}}^{(\la)}
 &=-\frac{1}{2\la}\int_{\mathbb{R}^2} \left(\sqrt{1+2(2\la) x_1+2(2\la)^2\left(x_1^2-x_2\right)}-1\right)\,dx^+\wedge dx^-\,,
\end{align}
where we define 
\begin{align}
    x_1=-\Tr(j_+j_-)\,,\qquad x_2=\frac{1}{2}\left(\Tr(j_+j_+)\Tr(j_-j_-)+\left(\Tr(j_+j_-)\right)^2\right)\,.
\end{align}
From the deformed action (\ref{TTpcm}), the deformed energy-momentum tensor is given by
\begin{align}
    T_{\pm\pm}^{(\la)}(x)&=\frac{\Tr(j_{\pm}j_{\pm}) }{\sqrt{1+2(2\la) x_1+2(2\la)^2\left(x_1^2-x_2\right)}}\,,\\
    T_{+-}^{(\la)}(x)&=\frac{1+2\la x_1-\sqrt{1+2(2\la) x_1+2(2\la)^2\left(x_1^2-x_2\right)}}{2\la\sqrt{1+2(2\la) x_1+2(2\la)^2\left(x_1^2-x_2\right)}}\,,
\end{align}
and one can verify that the deformed Lagrangian (\ref{TTpcm}) satisfies the classical flow equation (\ref{flow-TTbar}).
Note that the $T\bar{T}$ operator coincides with the trace component of the energy-momentum tensor 
\begin{align}\label{detT-Tpm}
    \text{det}(T_{\mu\nu}^{(\la)})=\frac{1}{\la}T_{+-}^{(\la)}\,.
\end{align}
The identity (\ref{detT-Tpm}) follows from the classical scale-invariance of the original model.

\subsubsection*{Classically integrability}

One of the remarkable features of $T\bar{T}$ deformations is that the deformation preserves the classical integrability of the original model.
The Lax pair can be written as \cite{Chen:2021aid} (see also \cite{Borsato:2022tmu})
\begin{align}
    \cL^{(\la)}=\frac{j_++z\,\mathfrak{J}_+^{(\la)}}{1-z^2}dx^{+}+\frac{j_--z\,\mathfrak{J}_-^{(\la)} }{1-z^2}dx^{-}\label{TT-Lax}
\end{align}
in terms of the current $\mathfrak{J}^{(\la)}$ associated with the $G$-symmetry of the deformed action (\ref{TTpcm}), which is given by
\begin{align}\label{dJ-tt}
   \mathfrak{J}_{\pm}^{(\la)}&=\frac{(1+2\la x_1)j_{\pm}+2\la\,\Tr(j_{\pm}j_{\pm}) j_{\mp}}{ \sqrt{1+2(2\la) x_1+2(2\la)^2\left(x_1^2-x_2\right)}}\,.
\end{align}
The deformed current $\mathfrak{J}_{\pm}^{(\la)}$ is conserved on shell but does not satisfy the flatness condition \cite{Borsato:2022tmu}
\begin{align}\label{dcurrent-tt}
\partial_+\mathfrak{J}_{-}^{(\la)}+\partial_-\mathfrak{J}_{+}^{(\la)}=0\,,\qquad \partial_+\mathfrak{J}_-^{(\la)}-\partial_-\mathfrak{J}_+^{(\la)}+[\mathfrak{J}_+^{(\la)},\mathfrak{J}_-^{(\la)}]\neq 0\,.
\end{align}
The first equation coincides with the equations of motion for the action (\ref{TTpcm}).
Using the relation $[\mathfrak{J}_{+}^{(\la)},\mathfrak{J}_{-}^{(\la)}]=[j_+,j_-]$, one can verify that the flatness condition of (\ref{TT-Lax}) is equivalent to the equations of motion for the $T\bar{T}$ deformed PCM.

To understand the relationship with four-dimensional theory, we simplify (\ref{TTpcm}) by using a conserved current for the $G$ symmetry of the deformed action (\ref{TTpcm}), as in the root-$T\bar{T}$ deformation.
In terms of the deformed current (\ref{dcurrent-tt}), the deformed action (\ref{TTpcm}) can be expressed in a simple form
\begin{align}
\begin{split}\label{TT-action2}
    S_{\text{PCM}}^{(\la)}&=\int_{\mathbb{R}^2} \left(\Tr\left(j_+\mathfrak{J}_-^{(\la)}\right)-\la\,\text{det}(T^{(\la)}_{\mu\nu}) \right)\,dx^+\wedge dx^-\\
    &=\int_{\mathbb{R}^2} \left(\Tr\left(j_-\mathfrak{J}_+^{(\la)}\right)-\la\,\text{det}(T^{(\la)}_{\mu\nu}) \right)\,dx^+\wedge dx^-\,.
\end{split}
\end{align}
The first term has a similar form to that in the root-$T\bar{T}$ deformation, while the second term in the $T\bar{T}$ deformed case breaks the classical scale symmetry of the original model. This is consistent with the fact that the $T\bar{T}$ deformation is a perturbation corresponding to an irrelevant operator.

\subsubsection*{Dynamical coordinate transformation}\label{sec:dycoord}

As observed in \cite{Conti:2018tca} (see also \cite{Chen:2021aid}), the $T\bar{T}$-deformed action and the associated Lax pair can be obtained through a dynamical coordinate transformation, which is specified by the energy-momentum tensor.
For simplicity, we consider $T\bar{T}$ deformations of two-dimensional integrable field theories, which are classically scale invariant. 

Let $x'=(x^{+'},x^{-'})$ and $x=(x^{+},x^{-})$ denote the undeformed and deformed coordinates, respectively.
We denote the energy-momentum tensor of the original model in the coordinates $\{x^{\mu'}\}$ by $T^{(0)}_{\pm'\pm'}(x')$.
Assume that these two coordinates are related by a dynamical coordinate transformation
\begin{align}\label{dcoordtr}
    \begin{pmatrix}
        dx^{+'}\\
        dx^{-'}
    \end{pmatrix}
    =\begin{pmatrix}
        \frac{\partial x^{+'}}{\partial x^{+}}& \frac{\partial x^{+'}}{\partial x^{-}}\\
         \frac{\partial x^{-'}}{\partial x^{+}}& \frac{\partial x^{-'}}{\partial x^{-}}
    \end{pmatrix} 
    \begin{pmatrix}
        dx^{+}\\
        dx^{-}
    \end{pmatrix}
    =\frac{1}{\Delta(x')}\begin{pmatrix}
       1& -\la T_{+'+'}^{(0)}(x')\\
          -\la T_{-'-'}^{(0)}(x')&1
    \end{pmatrix} \
    \begin{pmatrix}
         dx^{+}\\
        dx^{-}
    \end{pmatrix}
    \,,
\end{align}
where $\Delta(x')$ is defined as
\begin{align}
    \Delta(x')=1-\la^2\,T_{+'+'}^{(0)}(x')T_{-'-'}^{(0)}(x')\,.
\end{align}
Thus, the $T\bar{T}$ deformed action $S^{(\la)}$ can be derived from the identity \cite{Conti:2018tca,Conti:2019dxg,Coleman:2019dvf}
\begin{align}\label{TT-action}
    S^{(\la)}&=\int_{\Sigma} L^{(\la)}(x)\,dx^+\wedge dx^- \no\\
    &=\int_{\Sigma}\left(L^{(0)}(x')-\la\,\textrm{det}(T_{\nu'}^{(0)\mu'} (x'))\right)\,dx^{+'}\wedge dx^{-'} \,.
\end{align}
The deformed energy-momentum tensor $T_{\mu\nu}^{(\la)}$ is related to the undeformed one by the relation
\begin{align}\label{dT-T0}
    T_{\pm'\pm'}^{(0)}(x')=\frac{T_{\pm\pm}^{(\la)}(x)}{1+\la T_{+-}^{(\la)}(x)}\,,\qquad    \la T_{+'+'}^{(0)}(x')T_{-'-'}^{(0)}(x')=\frac{T_{+-}^{(\la)}(x)}{1+\la T_{+-}^{(\la)}(x)}\,,
\end{align}
and its inverse relation is
\begin{align}\label{dT-T0-inv}
   T_{\pm\pm}^{(\la)}(x)= \frac{T_{\pm'\pm'}^{(0)}(x')}{1-\la^2 T_{+'+'}^{(0)}(x')T_{-'-'}^{(0)}(x')}\,,\qquad   T_{+-}^{(\la)}(x)=\frac{\la T_{+'+'}^{(0)}(x')T_{-'-'}^{(0)}(x')}{1-\la^2 T_{+'+'}^{(0)}(x')T_{-'-'}^{(0)}(x')}\,.
\end{align}
For a derivation of this relation, see, e.g., \cite{Hirano:2024eab}. 
Using the deformed energy-momentum tensor $T_{\mu\nu}^{(\la)}(x)$, the dynamical coordinate transformation (\ref{dcoordtr}) is rewritten as
\begin{align}\label{dcoordtr1}
    \begin{pmatrix}
        dx^{+'}\\
        dx^{-'}
    \end{pmatrix}
    =\begin{pmatrix}
       1+\la T_{+-}^{(\la)}(x)& -\la T_{--}^{(\la)}(x)\\
          -\la T_{++}^{(\la)}(x)&1+\la T_{+-}^{(\la)}(x)
    \end{pmatrix} \
    \begin{pmatrix}
         dx^{+}\\
        dx^{-}
    \end{pmatrix}
    \,.
\end{align}

As stated above, the deformed model is also classically integrable whenever the undeformed model is classically integrable.
The deformed Lax pair $\cL_{\pm}^{(\la)}(x)$ is obtained by requiring that under the dynamical coordinate transformation (\ref{dcoordtr}) the Lax form remain invariant 
\begin{align} \label{Lax-id2}
    \cL&=\cL_+^{(\la)}(x)dx^{+}+\cL_-^{(\la)}(x)dx^{-}=\cL_{+'}^{(0)}(x')dx^{'+}+\cL_{-'}^{(0)}(x')dx^{'-}\,.
\end{align}
where $\cL_{\pm'}^{(0)}(x')$ denotes the undeformed Lax pair defined on the coordinate system $\{x^{'\mu}\}$.
From the identity (\ref{Lax-id2}), the deformed Lax pair is given by
\begin{align} \label{Lax-id3}
\begin{split}
        \cL_+^{(\la)}(x)&=\cL_{+'}^{(0)}(x'(x))\frac{\partial x^{+'}}{\partial x^{+}}+\cL_{-'}^{(0)}(x'(x))\frac{\partial x^{-'}}{\partial x^{+}}\,,\\
    \cL_-^{(\la)}(x)&=\cL_{+'}^{(0)}(x'(x))\frac{\partial x^{+'}}{\partial x^{-}}+\cL_{-'}^{(0)}(x'(x))\frac{\partial x^{-'}}{\partial x^{-}}\,.
\end{split}
\end{align}
It is not a priori guaranteed whether the Lax pair constructed by this procedure is always the Lax pair of the $T\bar{T}$ deformed action (\ref{TT-action}).
Nevertheless, the Lax pair of the $T\bar{T}$ deformed PCM can indeed be constructed in this approach \cite{Chen:2021aid}.

\subsubsection*{Relation to the PCM with the field-dependent metric}

To conclude this section, we remark that the Lax pair of the $T\bar{T}$ deformed PCM can be identified with that of the PCM on curved space with a field-dependent metric.
Indeed, from the relation (\ref{Lax-id3}), the deformed current $\mathfrak{J}^{(\la)}_{\pm}$ can be expressed as
\begin{align}\label{cJ-TT-j}
    \mathfrak{J}^{(\la)}_{\pm}=\pm\gamma_{\pm\rho}^{*(\la)}\epsilon^{\rho\sigma}j_{\sigma}\,,
\end{align}
where the field-dependent Weyl-invariant metric takes the form 
\begin{align}\label{tt-metric-sol2}
\begin{split}
    \gamma_{\pm\pm}^{*(\la)}&=\frac{2\la T_{\pm\pm}^{(\la)}(1+\la T_{+-}^{(\la)}) }{1+2\la T_{+-}^{(\la)} +\la^2 \left((T_{+-}^{(\la)})^2-T_{++}^{(\la)}T_{--}^{(\la)}\right) }\,,\\
    \gamma_{+-}^{*(\la)}=\gamma_{-+}^{*(\la)}&=-\frac{1+2\la T_{+-}^{(\la)} +\la^2 \left((T_{+-}^{(\la)})^2+T_{++}^{(\la)}T_{--}^{(\la)}\right) }{1+2\la T_{+-}^{(\la)} +\la^2 \left((T_{+-}^{(\la)})^2-T_{++}^{(\la)}T_{--}^{(\la)}\right) }\,.
\end{split}
\end{align}
Using the relation (\ref{dT-T0}) and the expression of the energy-momentum tensor for the original model, one can show that (\ref{cJ-TT-j}) together with (\ref{tt-metric-sol2}) coincides with (\ref{dJ-tt}).
In terms of Beltrami differential and conformal factor, the field-dependent metric can be rewritten as\footnote{Since the associated two-dimensional integrable field action and Lax pair only depend on the Weyl invariant metric, they do not depend on the conformal factor $e^{\phi}$, at least at the classical level.}
\begin{align}
    \mu^{*(\la)}=\frac{-\la T_{--}^{(\la)}(x)}{1+\la\,T_{+-}^{(\la)}(x)}\,,\qquad \bar{\mu}^{*(\la)}=\frac{-\la T_{++}^{(\la)}(x)}{1+\la\,T_{+-}^{(\la)}(x)}\,,\qquad e^{\phi^{*(\la)}}=1+\la\,T_{+-}^{(\la)}(x)\,.
\end{align}
For the dynamical coordinate transformation to be locally invertible, the Beltrami differential and the conformal factor must satisfy the following condition
\begin{align}
    (\partial_--\mu^{*(\la)} \partial_+)e^{\phi^{*(\la)}}&=(\partial_+ \mu^{*(\la)})e^{\phi^{*(\la)}}\,,\\
     (\partial_+-\bar{\mu}^{*(\la)} \partial_-)e^{\phi^{*(\la)}}&=(\partial_- \bar{\mu}^{*(\la)})e^{\phi^{*(\la)}}\,,
\end{align}
which is fulfilled on shell due to the conservation law of the deformed energy-momentum tensor $\partial^{\mu}T_{\mu\nu}^{(\la)}=0$.

Using the field-dependent metric (\ref{tt-metric-sol2}), the $T\bar{T}$ deformed PCM action (\ref{TT-action2}) can be rewritten in a  simplified form
\begin{align}\label{TT-pcm-2}
    S_{\text{PCM}}^{(\la)}
    =-\frac{1}{2}\int_{\mathbb{R}^2} \left(\gamma^{*(\la)\mu\nu}\Tr\left(j_{\mu}j_{\nu}\right)+2\la\,\text{det}(T_{\mu\nu}^{(\la)}) \right)\,dx^+\wedge dx^-\,.
\end{align}
Under a variation of the field, one finds an identity
\begin{align}\label{var-id-tt}
    \delta\gamma^{*(\la)\mu\nu}\Tr(j_{\mu}j_{\nu})+2\la\,\delta\left(\text{det}(T_{\mu\nu}^{(\la)})\right)=0\,.
\end{align}
By this identity, one can readily show that the flatness condition of the Lax pair (\ref{TT-Lax}) is equivalent to the equations of motion for the deformed model (\ref{TT-action2}).
Similar to (\ref{var-id-tt}), we also have
\begin{align}
    \frac{d}{d\la}\gamma^{*(\la)\mu\nu}\Tr(j_{\mu}j_{\nu})+\frac{2}{\la} \,\frac{d}{d\la}\left(\la^2\,\text{det}(T_{\mu\nu}^{(\la)})\right) =0\,.
\end{align}
This shows that the deformed action (\ref{TT-pcm-2}) indeed satisfies the classical flow equation (\ref{flow-TTbar}).

\subsection{\texorpdfstring{$T\bar{T}$}{T\bar{T}} deformation of the degenerate \texorpdfstring{$\cE$}{\mathcal{E}}-model from four-dimensional CS theory}\label{TT-E}

In this section, we explain how we derive the $T\bar{T}$ deformed degenerate $\cE$-model from four-dimensional theory, and then we show the classical integrability of the deformed model.
As will be discussed later, treating the $T\bar{T}$ deformation in the context of the four-dimensional theory requires a more involved procedure than in the case of the root-$T\bar{T}$ deformation.

\subsubsection*{Lax pair}

We first construct the Lax pair of the $T\bar{T}$ deformed degenerate $\cE$-model.
As in the PCM case, we find that the deformed Lax pair can be constructed by applying a dynamical coordinate transformation.
We then verify that it solves the equations of motion (\ref{Lax-eom2}) of the four-dimensional action with the same boundary condition (\ref{Lax-con1}) as the original model and satisfies the transformation rule (\ref{2dgauge-sym}).

We apply the dynamical coordinate transformation (\ref{dcoordtr1}) to the Lax pair (\ref{E-Lax}) for the degenerate $\cE$-model. By using the identity (\ref{Lax-id2}), we can write down the deformed Lax pair as
\begin{align}\label{tt-lax}
    \cL_{\pm}^{(\la)}=\frac{1}{2}\bmp\left((\mathcal{P}_{\bmh}^{+}+\mathcal{P}_{\bmh}^{-})({\bm h}^{-1}\partial_{\pm}{\bm h})\right)\pm\frac{1}{2}\,\bmp\left({\bm\cJ}_{\pm}^{(\la)}\right)\,,
\end{align}
where the deformed current ${\bm\cJ}_{\pm}^{(\la)}$ can be expressed as
\begin{align}\label{cJ-jmet}
    {\bm\cJ}_{\pm}^{(\la)}=\pm\gamma_{\pm\rho}^{*(\la)}\epsilon^{\rho\mu}\,(\mathcal{P}_{\bmh}^{+}-\mathcal{P}_{\bmh}^{-})(\bmh^{-1}\partial_{\mu}\bmh)
\end{align}
in terms of the same field-dependent metric $\gamma_{\mu\nu}^{*(\la)}$ as in the PCM case (\ref{tt-metric-sol2}).
The deformed energy-momentum tensor $T_{\pm\pm}^{(\la)}$ is computed explicitly in Appendix \ref{sec:TT-em}. 

We now examine the Lax pair (\ref{tt-lax}) from the viewpoint of the four-dimensional theory. 
Since the deformed Lax pair (\ref{tt-lax}) takes the same form as the Lax pair (\ref{Lax-curve-E1}) for the degenerate $\cE$-model on curved space, and thus constitutes an admissible solution of the equations of motion (\ref{Lax-eom2}) with the same boundary condition 
\begin{align}
     {}^{\bmh}({\bm j}^*\cL^{(\la)}(\bmh))\in \Omega^1(\Sigma_{\mathfrak{p}},\mathfrak{k})
\end{align}
as the original model, and it obeys the transformation law
\begin{align}
    {\bm j}^*\cL^{(\la)}({\bm k}\,\bmh \Delta(f)^{-1})={}^{\Delta(f)}\left({\bm j}^*\cL^{(\la)}(\bmh)\right)
\end{align}
under the gauge transformation (\ref{2dgauge-sym}) since the undeformed energy-momentum tensor is gauge invariant.
Thus, the Lax pair (\ref{tt-lax}) appears to be a solution to the 4d CS theory belonging to the same class as the undeformed one. However, the boundary condition for the variation needs to be carefully examined, as in the case of the root-$T\bar{T}$ deformation.

\subsubsection*{On classical integrability and additional boundary term}\label{sec:bc-4d2d}

Before deriving the corresponding two-dimensional action, we examine whether the four-dimensional/two-dimensional system (\ref{4dCS-defect}) with the Lax pair (\ref{tt-lax}) can describe an integrable field theory.
As discussed in Section \ref{sec:Lax_solution},
variation of the Lax pair has to satisfy the condition (\ref{const-dh}), which is not satisfied for
the $T\bar{T}$-deformed case. Indeed, we have
\begin{align}\label{24-var-curve-tt}
    \delta S_{\text{ext}}[\cL^{(\la)}(\bmh),\bmh]
    &=\frac{1}{4}\int_{\Sigma_{\mathfrak{p}}} \delta\gamma^{*(\la)\mu\nu}\llangle \bmh^{-1}\partial_{\mu}\bmh, (\mathcal{P}_{{\bm h}}^{+}-\mathcal{P}_{{\bm h}}^{-})({\bm h}^{-1}\partial_{\nu}{\bm h})\rrangle_{\mathfrak{d}}\,dx^+\wedge dx^-\no\\
    &\quad+\int_{\Sigma_{\mathfrak{p}}}
    \llangle \sfu,d({\bm j}^*\cL^{(\la)})+\frac{1}{2}[{\bm j}^*\cL^{(\la)},{\bm j}^*\cL^{(\la)}]\rrangle_{\mathfrak{d}}\,.
\end{align}
In the $T\bar{T}$-deformed case, the first line in (\ref{24-var-curve-tt}) does not vanish in contrast to the root-$T\bar{T}$ deformation; it can be shown to be a variation of the determinant of the deformed energy-momentum tensor (For detailed calculation, see appendix \ref{sec:TT-id})
\begin{align}\label{tt-id}
    \delta\gamma^{*(\la)\mu\nu}\llangle \bmh^{-1}\partial_{\mu}\bmh, (\mathcal{P}_{{\bm h}}^{-}-\mathcal{P}_{{\bm h}}^{+})({\bm h}^{-1}\partial_{\nu}{\bm h})\rrangle_{\mathfrak{d}}=4\la\, \delta \left({\rm det}(T^{(\la)}_{\mu\nu})\right)\,.
\end{align}
This generates an extra boundary term 
\begin{align}\label{24-var-curve-tt-2}
    \delta S_{\text{ext}}[\cL^{(\la)}(\bmh),\bmh]
    &=\lambda \int_{\Sigma_{\mathfrak{p}}} \delta\left({\rm det}(T^{(\la)}_{\mu\nu})\right)\,dx^+\wedge dx^-
    \,.
\end{align}
To cancel this new contribution, it is necessary to add a new boundary term to the surface defect action (\ref{defect-action}):
\begin{align}\label{ex-defect}
   & S_{\text{defect}}[A,\bmh]=-\frac{1}{2}\int_{\Sigma_{\mathfrak{p}}}\llangle \bmh^{-1}d\bmh, {\bm j}^*A\rrangle_{\mathfrak{d}}-\frac{1}{2}S_{\mathfrak{d}}^{\text{WZ}}[\bmh]\no\\
    &\longrightarrow  S_{\text{defect}}^{(\la)}[A,\bmh]= S_{\text{defect}}[A,\bmh]-\la \int_{\Sigma_{\mathfrak{p}}}{\rm det}(T^{(\la)}_{\mu\nu})\,dx^+\wedge dx^-\,.
\end{align}
Note that the extra boundary term is meaningful only after specifying the boundary conditions and poles of the gauge field $A$.
This prescription is similar to the derivation of the two-dimensional WZW model from the double 3d CS theory discussed in \cite{Coussaert:1995zp}. 
In this way, the modified four-dimensional/two-dimensional system 
\begin{align}\label{TT-42}
    S_{\text{ext}}^{(\la)}[\cL^{(\la)}(\bmh),\bmh]
    =  S_{\CS}[\cL^{(\la)}(\bmh)]+S_{\text{defect}}^{(\la)}[\cL^{(\la)}(\bmh),\bmh]
\end{align}
describes the two-dimensional integrable field theory, that is, the variation for the defect field $\bmh$ gives 
\begin{align}
    \delta S_{\text{ext}}^{(\la)}[\cL^{(\la)}(\bmh),\bmh]
    &= \delta_{\bmh} S_{\CS}[\cL^{(\la)}(\bmh)]+\delta S_{\text{defect}}^{(\la)}[\cL^{(\la)}(\bmh),\bmh]\no\\
    &=\int_{\Sigma_{\mathfrak{p}}}
    \llangle \sfu,d({\bm j}^*\cL^{(\la)})+\frac{1}{2}[{\bm j}^*\cL^{(\la)},{\bm j}^*\cL^{(\la)}]\rrangle_{\mathfrak{d}}\,,
\end{align}
which is precisely equivalent to the flatness condition of the Lax pair (\ref{tt-lax}).

\subsubsection*{$T\bar{T}$ deformed action}

We now derive the $T\bar{T}$-deformed degenerate $\cE$-model action.
By substituting the Lax pair (\ref{tt-lax}) into the modified four-dimensional/two-dimensional action (\ref{TT-42}), we can obtain 
\begin{align}\label{ttaction-ansatz}
    S_{\rm 2d}^{(\la)}[\bmh]&=\frac{1}{2}\int_{\Sigma_{\mathfrak{p}}}\biggl(\llangle \bmh^{-1}\partial_{+}\bmh, {\bm\cJ}_{-}^{(\la)}\rrangle_{\mathfrak{d}}
    +\frac{1}{2}\varepsilon^{\mu\nu}\llangle \bmh^{-1}\partial_{\mu}\bmh, (\mathcal{P}_{\bmh}^{+}+\mathcal{P}_{\bmh}^{-})(\bmh^{-1}\partial_{\nu}\bmh)\rrangle_{\mathfrak{d}}\biggr)dx^+\wedge dx^-\no\\
    &\qquad-\frac{1}{2}S_{\mathfrak{d}}^{\text{WZ}}[\bmh]-\la \int_{\Sigma_{\mathfrak{p}}}{\rm det}(T^{(\la)}_{\mu\nu}) dx^+\wedge dx^-\,.
\end{align}
where we have used the identity $\llangle \bmh^{-1}\partial_{+}\bmh, {\bm\cJ}_{-}^{(\la)}\rrangle_{\mathfrak{d}}=\llangle \bmh^{-1}\partial_{-}\bmh, {\bm\cJ}_{+}^{(\la)}\rrangle_{\mathfrak{d}}$.
In the undeformed limit $\la\to 0$, the action (\ref{ttaction-ansatz}) reduces to the degenerate $\cE$-model action (\ref{dE-action}).
The remaining task is to show that the deformed action satisfies the classical flow equation for the $T\bar{T}$ deformation.

For this purpose, it is convenient to rewrite the action (\ref{ttaction-ansatz}) in terms of the field-dependent metric $\gamma^{*(\la)\mu\nu}$ using the relation (\ref{cJ-jmet}), which takes the form
\begin{align}
    S_{\rm 2d}^{(\la)}[\bmh]&=\frac{1}{4}\int_{\Sigma_{\mathfrak{p}}}\biggl(\gamma^{*(\la)\mu\nu}\llangle \bmh^{-1}\partial_{\mu}\bmh, (\mathcal{P}_{\bmh}^{+}-\mathcal{P}_{\bmh}^{-})(\bmh^{-1}\partial_{\nu}\bmh)\rrangle_{\mathfrak{d}}
   \no\\
    &\qquad\qquad
    +\varepsilon^{\mu\nu}\llangle \bmh^{-1}\partial_{\mu}\bmh, (\mathcal{P}_{\bmh}^{+}+\mathcal{P}_{\bmh}^{-})(\bmh^{-1}\partial_{\nu}\bmh)\rrangle_{\mathfrak{d}}\biggr)dx^+\wedge dx^-\no\\
    &\quad-\frac{1}{2}S_{\mathfrak{d}}^{\text{WZ}}[\bmh]-\la \int_{\Sigma_{\mathfrak{p}}}{\rm det}(T^{(\la)}_{\mu\nu}) dx^+\wedge dx^-\,.
\end{align}
Using the identity
\begin{align}\label{TT-id-flow}
      \frac{d}{d\la}\gamma^{*(\la)\mu\nu}\llangle \bmh^{-1}\partial_{\mu}\bmh, (\mathcal{P}_{\bmh}^{+}-\mathcal{P}_{\bmh}^{-})({\bm h}^{-1}\partial_{\nu}{\bm h})\rrangle_{\mathfrak{d}}-\frac{4}{\la}\frac{d}{d\la}\left(\la^2\, {\rm det}(T^{(\la)}_{\mu\nu})\right)=0
\end{align}
similar to the relation (\ref{tt-id}), one sees that the deformed action (\ref{ttaction-ansatz}) satisfies the classical flow equation
\begin{align}
    \frac{d}{d\la} S_{\rm 2d}^{(\la)}[\bmh]=\textrm{det}(T_{\mu\nu}^{(\la)})\,.
\end{align}
Thus, the action (\ref{ttaction-ansatz}) represents the $T\bar{T}$-deformation of the degenerate $\cE$-model and is classically integrable.
The degenerate $\cE$-model is known to describe integrable deformations such as Yang-Baxter \cite{Delduc:2013fga}, homogeneous Yang-Baxter \cite{Kawaguchi:2014qwa}, and $\la$ deformations \cite{Hollowood:2014rla}. It immediately follows that the $T\bar{T}$ deformation also commutes with these integrable deformations.

\section{Root-\texorpdfstring{$T\bar{T}$}{T\bar{T}} deformation}
\label{sec:root_TT}

In this section, we derive the root-$T\bar{T}$ deformation of the degenerate $\cE$-model from four-dimensional theory.

\subsection{Root-\texorpdfstring{$T\bar{T}$}{T\bar{T}} deformed PCM and Lax pair}

We briefly review the root-$T\bar{T}$ deformation of the PCM before considering a derivation of the root-$T\bar{T}$ deformation of the degenerate $\cE$-model from four-dimensional theory.

The root-$T\bar{T}$ deformation \cite{Rodriguez:2021tcz,Babaei-Aghbolagh:2022uij, Conti:2022egv,Ferko:2022cix, Babaei-Aghbolagh:2022leo}  of a given two-dimensional field theory is generated by adding the root-$T\bar{T}$ operator, which is a classically marginal operator constructed from the energy-momentum tensor $T_{\mu\nu}^{(0)}$ of the original model:f
\begin{align}
   S^{(\gamma)}_{\text{PCM}}= S_{\text{PCM}}+\gamma \int_{\mathbb{R}^2} \sqrt{T_{++}^{(0)}T_{--}^{(0)}}\,dx^+\wedge dx^-+\cO(\gamma^2)\,.
\end{align}
The deformed Lagrangian $L^{(\gamma)}_{\text{PCM}}$ satisfies the classical flow equation
\begin{align}
    \frac{d}{d\gamma} L^{(\gamma)}_{\text{PCM}}=\sqrt{T_{++}^{(\gamma)}T_{--}^{(\gamma)}}\,,\label{root-flow}
\end{align}
where $T^{(\gamma)}_{\pm\pm}$ denotes the energy-momentum tensor for the deformed model.

By solving the flow equation (\ref{root-flow}), one obtains the action of the root-$T\bar{T}$ deformed PCM (For details, see  \cite{Borsato:2022tmu} for example)
\begin{align}\label{rootTT-PCM}
    S^{(\gamma)}_{\text{PCM}}=\int_{\mathbb{R}^2} \Tr\left(j_+\mathfrak{J}_-^{(\gamma)}\right)\,dx^+\wedge dx^-=\int_{\mathbb{R}^2} \Tr\left(j_-\mathfrak{J}_+^{(\gamma)}\right)\,dx^+\wedge dx^-\,,
\end{align}
where the deformed current $\mathfrak{J}_{\pm}^{(\gamma)}$ is expressed as a linear combination of the left-invariant current $j_{\pm}$ given by
\begin{align}\label{dc-pcm}
    \mathfrak{J}_{\pm}^{(\gamma)}=\cosh\gamma\,j_{\pm}+\sinh\gamma\,\sqrt{\frac{\Tr[j_{\pm}j_{\pm}]}{\Tr[j_{\mp}j_{\mp}]}}j_{\mp}\,.
\end{align}
The deformed current $\mathfrak{J}_{\pm}^{(\gamma)}$ is conserved but not flat:
\begin{align}\label{eom-root}
\partial_+\mathfrak{J}_{-}^{(\gamma)}+\partial_-\mathfrak{J}_{+}^{(\gamma)}=0\,,\qquad \partial_+\mathfrak{J}_-^{(\gamma)}-\partial_-\mathfrak{J}_+^{(\gamma)}+[\mathfrak{J}_+^{(\gamma)},\mathfrak{J}_-^{(\gamma)}]\neq 0\,.
\end{align}
The deformed energy-momentum tensor is
\begin{align}
    T_{++}^{(\gamma)}=\Tr(j_+\mathfrak{J}_+^{(\gamma)})\,,\qquad 
     T_{--}^{(\gamma)}=\Tr(j_-\mathfrak{J}_-^{(\gamma)})\,,\qquad T_{+-}^{(\gamma)}=0\,,
\end{align}
and one can verify that the deformed Lagrangian obeys the flow equation (\ref{root-flow}).

Importantly, this deformation preserves the classical integrability of the original model.
The Lax pair for the root-$T\bar{T}$ deformed PCM (\ref{rootTT-PCM}) has been constructed in \cite{Borsato:2022tmu}, and it takes the form
\begin{align}
    \cL^{(\gamma)}&=\left(\frac{j_{+}+z\,\mathfrak{J}_+^{(\gamma)}}{1-z^2}\right)dx^++\left(\frac{j_{-}-z\,\mathfrak{J}_-^{(\gamma)}}{1-z^2}\right)dx^-\,.\label{root-Lax}
\end{align}
Thanks to the identity $[\mathfrak{J}_{+}^{(\gamma)},\mathfrak{J}_-^{(\gamma)}]=[j_+,j_-]$, the flatness condition of the Lax pair (\ref{root-Lax}) can be shown to be equivalent to the equations of motion (\ref{eom-root}).

\subsubsection*{Relation to the PCM with a field-dependent metric}

One may formally regard the root-$T\bar{T}$ deformed PCM as the PCM with a field-dependent metric. 
To see this, we compare the Lax pairs (\ref{root-Lax}) and (\ref{Lax-cpcm}) of these two integrable field theories, and find that the deformed current $\mathfrak{J}_{\pm}^{(\gamma)}$ can be expressed as
\begin{align}
    \mathfrak{J}_{\pm}^{(\gamma)}=\pm\gamma_{\pm\rho}^{*(\gamma)}\epsilon^{\rho\mu}\,j_{\mu}\,,\label{J-t-root}
\end{align}
where the field-dependent metric $\gamma_{\mu\nu}^{*(\gamma)}$ is given by
\begin{align}
\begin{split}\label{TTbar-gamma}
 \gamma^{*(\gamma)++}&=-\sinh\gamma \sqrt{\frac{\Tr j_-j_-}{\Tr j_+j_+}}\,,\qquad \gamma^{*(\gamma)--}=-\sinh\gamma \sqrt{\frac{\Tr j_+j_+}{\Tr j_-j_-}}\,,\\
  \gamma^{*(\gamma)+-}&=\gamma^{*(\gamma)-+}=-\cosh\gamma\,,
\end{split}
\end{align}
and still satisfies the normalization of the Weyl-invariant metric
\begin{align}
    {\rm det}(\gamma^{*(\gamma)}_{\mu\nu})=-1\,.
\end{align}
In terms of the Beltrami differential, this yields
\begin{align}\label{TTbar-mu}
    \mu^{*(\gamma)}=\tanh\left(\frac{\gamma}{2}\right) \sqrt{\frac{\Tr j_-j_-}{\Tr j_+j_+}}\,,\qquad \bar{\mu}^{*(\gamma)}=\tanh\left(\frac{\gamma}{2}\right)\sqrt{\frac{\Tr j_+j_+}{\Tr j_-j_-}}\,.
\end{align}
In the next section, we use the relation (\ref{J-t-root}) to construct the Lax pair for the root-$T\bar{T}$ deformed degenerate $\cE$-model and derive the associated deformed action by employing four-dimensional theory.

\subsection{Root-\texorpdfstring{$T\bar{T}$}{T\bar{T}} deformed degenerate \texorpdfstring{$\cE$}{\mathcal{E}}-model from four-dimensional CS theory }\label{rootTT-E}

We now derive the root-$T\bar{T}$ deformed degenerate $\cE$-model from four-dimensional theory.
In the previous section, we observed that the root-$T\bar{T}$ deformed PCM may formally be obtained from the PCM with a field-dependent metric.
Thus, we assume that the root-$T\bar{T}$ deformed degenerate $\cE$-model can also be derived from the four-dimensional action with the meromorphic one-form (\ref{omega-analytic}) and the boundary condition (\ref{Lax-con1}) corresponding to the degenerate $\cE$-model.

\subsubsection*{Lax pair}

We begin by constructing the Lax pair for the root-$T\bar{T}$ deformed degenerate $\cE$-model.
As in the PCM case, we find that the deformed Lax pair can be equivalent to the Lax pair (\ref{Lax-curve-E1}) of the original model with a field-dependent metric. 
The deformed Lax pair should take the form
\begin{align}\label{rootTT-flat}
    \cL_{\pm}^{(\gamma)}=\frac{1}{2}\bmp\left((\mathcal{P}_{\bmh}^{+}+\mathcal{P}_{\bmh}^{-})({\bm h}^{-1}\partial_{\pm}{\bm h})\right)\pm\frac{1}{2}\bmp\left({\bm\cJ}_{\pm}^{(\gamma)}\right)\,,
\end{align}
and the deformed current ${\bm\cJ}_{\pm}^{(\gamma)}$ can be obtained by 
\begin{align}
    {\bm\cJ}_{\pm}^{(\gamma)}=\pm\gamma_{\pm\rho}^{*(\gamma)}\epsilon^{\rho\mu}\,(\mathcal{P}_{\bmh}^{+}-\mathcal{P}_{\bmh}^{-})(\bmh^{-1}\partial_{\mu}\bmh)\,,
\end{align}
where the field-dependent Weyl-invariant metric is specified by the undeformed energy-momentum tensor (\ref{em-dE}) like
\begin{align}
\begin{split}\label{root-gamma}
 \gamma^{*(\gamma)++}&=-\sinh\gamma\,\sqrt{\frac{T_{--}^{(0)}}{T_{++}^{(0)}}}\,,\qquad
 \gamma^{*(\gamma)--}=-\sinh\gamma\,\sqrt{\frac{T_{++}^{(0)}}{T_{--}^{(0)}}}\,,\\
  \gamma^{*(\gamma)+-}&=\gamma^{*(\gamma)-+}=-\cosh\gamma\,.
\end{split}
\end{align}
The configuration (\ref{root-gamma}) still holds the identity
\begin{align}
    {\rm det}(\gamma^{*(\gamma)}_{\mu\nu})=-1\,.
\end{align}
Recall that the Lax pair (\ref{Lax-curve-E1}) of the degenerate $\cE$-model with a nontrivial metric is an admissible solution with the boundary condition (\ref{Lax-con1}) and satisfying the transformation rule (\ref{Lax-con2}). This implies that the Lax pair constructed by the above procedure also satisfies the equations of motion (\ref{Lax-eom2}) of the four-dimensional theory with the boundary condition (\ref{Lax-con1}) and obeys the transformation law (\ref{Lax-con1}).
However, as we see below, since the Lax pair (\ref{rootTT-flat}) includes the field-dependent metric, its variation does not necessarily satisfy the boundary condition (\ref{Lax-con1}).
This raises the question of whether the four-dimensional/two-dimensional system (\ref{4dCS-defect}) automatically describes a classically integrable field theory.

\subsubsection*{On classical integrability }

To see this, we consider a variation of the Lax pair (\ref{rootTT-flat}) with respect to the defect field $\bmh$, which gives
\begin{align}\label{Lax-root-var}
    {\bm j}^*\cL_{\pm}^{(\gamma)'}&={\bm j}^*\cL_{\pm}^{(\gamma)}+\epsilon\, {\bm j}^*\bar{l}_{\pm}+\frac{1}{2}\delta \gamma_{\pm\rho}^{*(\gamma)}\epsilon^{\rho\mu}\,(\mathcal{P}_{\bmh}^{+}-\mathcal{P}_{\bmh}^{-})({\bm h}^{-1}\partial_{\mu}{\bm h})\,,
\end{align}
where we renamed the variation ${\bm j}^*l_{\pm}$ in (\ref{E-model-var}) as
\begin{align}
     {\bm j}^*\bar{l}_{\pm}=[{\bm j}^*\cL_{\pm}^{(\gamma)},\sfu] +[P_{+}^{(+)\mu}\mathcal{P}_{\bmh}^{+}(\partial_{\mu}\sfu)+P_{+}^{(-)\mu}\mathcal{P}_{\bmh}^{-}(\partial_{\mu}\sfu)]\,.
\end{align}
As shown in Section \ref{curve-E}, a component of the variation $\bar{l}$ satisfies
\begin{align}
    & \mathrm{Ad}_{\bmh}(d_{\Sigma}\sfu+[{\bm j}^*\cL^{(\gamma)},\sfu]-{\bm j}^* \bar{l})\in \Omega^1(\Sigma_{\mathfrak{p}},\mathfrak{k})\,,
\end{align}
and therefore it does not contribute to the variation of the four-dimensional/two-dimensional action (\ref{4dCS-defect}).
The remaining term in (\ref{Lax-root-var}), proportional to the variation of the field-dependent metric, results in a modification of the variation (\ref{24action-val}) of the four-dimensional/two-dimensional action:
\begin{align}\label{24-var-curve}
    \delta S_{\text{ext}}[\cL^{(\gamma)},\bmh]
    &=\frac{1}{2}\epsilon\int_{\Sigma_{\mathfrak{p}}} \left(\delta\gamma^{*(\gamma)++}\,T^{(0)}_{++}+\delta\gamma^{*(\gamma)--}\,T^{(0)}_{--}\right)dx^+\wedge dx^-\no\\
    &\quad-\epsilon\int_{\Sigma_{\mathfrak{p}}}
    \llangle \sfu,d_{\Sigma}({\bm j}^*\cL^{(\gamma)})+\frac{1}{2}[{\bm j}^*\cL^{(\gamma)},{\bm j}^*\cL^{(\gamma)}]\rrangle_{\mathfrak{d}}\,.
\end{align}
The explicit computation of the variation (\ref{24-var-curve}) is given in Appendix \ref{sec:42-var}.
The existence of this additional term would lead to a violation of classical integrability in the four-dimensional/two-dimensional system (\ref{4dCS-defect}). Fortunately, since the variation of the field-dependent metric (\ref{root-gamma}) is \begin{align}
    \delta\gamma^{*(\gamma)\pm\pm}=\frac{1}{2}\sinh\gamma\left(\sqrt{\frac{T_{\mp\mp}^{(0)}}{T_{\pm\pm}^{(0)}}}\frac{\delta T_{\pm\pm}^{(0)}}{T_{\pm\pm}^{(0)}}-\frac{\delta T_{\mp\mp}^{(0)}}{\sqrt{T_{++}^{(0)}T_{--}^{(0)}}}\right)\,,
\end{align}
we have the identity $\delta\gamma^{*(\gamma)++}\,T^{(0)}_{++}+\delta\gamma^{*(\gamma)--}\,T^{(0)}_{--}=0$, resulting in the first line of (\ref{24-var-curve}) vanishing.
Hence, the four-dimensional/two-dimensional system (\ref{4dCS-defect}) with the Lax pair (\ref{rootTT-flat}) is classically integrable.

\subsubsection*{Root-$T\bar{T}$ deformed $\cE$-model action}

Finally, we derive the root-$T\bar{T}$ deformation of the degenerate $\cE$-model action from the four-dimensional/two-dimensional system (\ref{4dCS-defect}).
After imposing the gauge-fixing condition $A_{\bar{z}}=0$, substituting (\ref{rootTT-flat}) into the effective 2d action (\ref{2d-formula-1}) leads to 
\begin{align}\label{rootaction-ansatz}
    S_{\rm 2d}^{(\gamma)}[\bmh]&=\frac{1}{2}\int_{\Sigma_{\mathfrak{p}}}\biggl(\llangle \bmh^{-1}\partial_{+}\bmh, {\bm\cJ}_{-}^{(\gamma)}\rrangle_{\mathfrak{d}}
    -\frac{1}{2}\varepsilon^{\mu\nu}\llangle \bmh^{-1}\partial_{\mu}\bmh, (\mathcal{P}_{\bmh}^{+}+\mathcal{P}_{\bmh}^{-})(\bmh^{-1}\partial_{\nu}\bmh)\rrangle_{\mathfrak{d}}\biggr)dx^+\wedge dx^-\no\\
    &\qquad-\frac{1}{2}S_{\mathfrak{d}}^{\text{WZ}}[\bmh]\,.
\end{align}
where we have used the fact that, as in the PCM case, the deformed current ${\bm \cJ}^{(\gamma)}_{\pm}$ satisfies
\begin{align}
    \llangle\bmh^{-1}\partial_+\bmh , {\bm \cJ}^{(\gamma)}_{-}\rrangle_{\mathfrak{d}}=\llangle\bmh^{-1}\partial_-\bmh , {\bm \cJ}^{(\gamma)}_{+}\rrangle_{\mathfrak{d}}\,.
\end{align}
From the deformed action, the energy-momentum tensor is given by
\begin{align}\label{em-root}
    T_{\pm\pm}^{(\gamma)}=\frac{1}{2}\llangle \bmh^{-1}\partial_{\pm}\bmh ,{\bm \cJ}_{\pm}^{(\gamma)}\rrangle_{\mathfrak{d}}\,,\qquad T_{+-}^{(\gamma)}=0\,.
\end{align}
In the undeformed limit, the energy-momentum tensor (\ref{em-root}) should reduce to (\ref{em-dE}).
It is straightforward to show that the deformed action (\ref{rootaction-ansatz}) satisfies the classical flow equation 
\begin{align}
    &\frac{d}{d\gamma}S_{\rm 2d}^{(\gamma)}[\bmh]=\int_{\Sigma_{\mathfrak{p}}}\sqrt{T_{++}^{(\gamma)}T_{--}^{(\gamma)}}\,dx^+\wedge dx^-\,.
\end{align}
As a result, the four-dimensional theory with the Lax pair (\ref{rootTT-flat}) describes the root-$T\bar{T}$ deformation of the degenerate $\cE$-model with the action (\ref{rootaction-ansatz}) and demonstrates that the deformed degenerate $\cE$-model is also classically integrable.

\section{Two-parameter deformation of the degenerated \texorpdfstring{$\cE$}{\mathcal{E}}-model}
\label{sec:2_parameter}

An important feature of the root-$T\bar{T}$ deformation is that it commutes with the $T\bar{T}$ deformation \cite{Babaei-Aghbolagh:2022uij, Conti:2022egv,Ferko:2022cix, Babaei-Aghbolagh:2022leo}.
In this section, we consider a two-parameter deformation of the degenerate $\cE$-model generated by the root-$T\bar{T}$ and $T\bar{T}$ operators and their realization in the four-dimensional theory.

\subsubsection*{Lax pair and extended four-dimensional/two-dimensional system}

As in the previous two cases, we first construct the Lax pair $\cL_{\pm}^{(\la,\gamma)}$ for the two-parameter deformation of the degenerate $\cE$-model.
To achieve this, we utilize the fact that the root-$T\bar{T}$ deformation commutes with the $T\bar{T}$ deformation \cite{Babaei-Aghbolagh:2022uij, Conti:2022egv,Ferko:2022cix, Babaei-Aghbolagh:2022leo}.
This property allows us to construct the deformed Lax pair $\cL_{\pm}^{(\la,\gamma)}$ by applying a dynamical coordinate transformation to the root-$T\bar{T}$ deformed Lax pair (\ref{tt-lax}). 
By following the discussion in Section \ref{sec:dycoord}, the two-parameter deformed Lax pair can be expressed as
\begin{align}\label{two-tt-lax}
    \cL_{\pm}^{(\la,\gamma)}=\frac{1}{2}\bmp\left((\mathcal{P}_{\bmh}^{+}+\mathcal{P}_{\bmh}^{-})({\bm h}^{-1}\partial_{\pm}{\bm h})\right)\pm\frac{1}{2}\,\bmp\left({\bm\cJ}_{\pm}^{(\la,\gamma)}\right)\,,
\end{align}
where the deformed current ${\bm\cJ}_{\pm}^{(\la,\gamma)}$ takes the form
\begin{align}\label{two-dJ}
    {\bm\cJ}_{\pm}^{(\la,\gamma)}=\pm\gamma_{\pm\rho}^{*(\la,\gamma)}\epsilon^{\rho\mu}\, {\bm\cJ}_{\mu}^{(\gamma)}\,.
\end{align}
In the current case, the field-dependent metric $\gamma_{\mu\nu}^{*(\la,\gamma)}$ is obtained by replacing $T_{\mu\nu}^{(\la)}$ with $T_{\mu\nu}^{(\la,\gamma)}$ in the equation (\ref{tt-metric-sol2}) :
\begin{align}\label{two-metric}
\begin{split}
    \gamma_{\pm\pm}^{*(\la,\gamma)}&=\frac{2\la T_{\pm\pm}^{(\la,\gamma)}(1+\la T_{+-}^{(\la,\gamma)}) }{1+2\la T_{+-}^{(\la,\gamma)} +\la^2 \left((T_{+-}^{(\la,\gamma)})^2-T_{++}^{(\la,\gamma)}T_{--}^{(\la,\gamma)}\right) }\,,\\
    \gamma_{+-}^{*(\la,\gamma)}=\gamma_{-+}^{*(\la,\gamma)}&=-\frac{1+2\la T_{+-}^{(\la,\gamma)} +\la^2 \left((T_{+-}^{(\la,\gamma)})^2+T_{++}^{(\la)}T_{--}^{(\la,\gamma)}\right) }{1+2\la T_{+-}^{(\la,\gamma)} +\la^2 \left((T_{+-}^{(\la,\gamma)})^2-T_{++}^{(\la,\gamma}T_{--}^{(\la,\gamma)}\right) }\,,
\end{split}
\end{align}
where $T_{\mu\nu}^{(\la,\gamma)}$ is given in (\ref{em-two}).
In terms of the root-$T\bar{T}$ deformed energy-momentum tensor $T_{\pm\pm}^{(\gamma)}$, the deformed current ${\bm\cJ}_{\pm}^{(\la,\gamma)}$ is rewritten as
\begin{align}\label{two-dJ-2}
   {\bm\cJ}_{\pm}^{(\la,\gamma)}=\frac{(1+2\la x_1^{(\gamma)})\,{\bm\cJ}_{\pm}^{(\gamma)}+2\la T_{\pm\pm}^{(\gamma)}\,{\bm\cJ}_{\mp}^{(\gamma)}}{\sqrt{1+2(2\la) x_1^{(\gamma)}+2(2\la)^2\left((x_1^{(\gamma)})^2-x_2^{(\gamma)}\right)}}\,\,,
\end{align}
where $x_1^{(\gamma)}$ and $x_2^{(\gamma)}$ are 
\begin{align}
     x_1^{(\gamma)}=-\frac{1}{2}\llangle \bmh^{-1}\partial_{+}\bmh ,{\bm\cJ}_{-}^{(\gamma)} \rrangle_{\mathfrak{d}}\,,
     \qquad  x_2^{(\gamma)}=\frac{1}{2}\left(T_{++}^{(\gamma)}T_{--}^{(\gamma)}+(x_1^{(\gamma)})^2\right)\,.
\end{align}
The Lax pair constructed in this way is a solution of the equation of motion (\ref{Lax-eom2}) of the four-dimensional theory, subject to the boundary condition (\ref{Lax-con2}). Furthermore, it satisfies the transformation law (\ref{Lax-con2}) for the same reasons as in the previous two examples.

We next describe the four-dimensional/two-dimensional system by following the discussion in Section \ref{sec:bc-4d2d}.
Since we initially performed the root-$T\bar{T}$ deformation, the extended defect action (\ref{ex-defect}) must be modified by replacing $T^{(\la)}_{\mu\nu}$ with $T^{(\la,\gamma)}_{\mu\nu}$.
We then consider the modified four-dimensional/two-dimensional action:
\begin{align}
    S_{\text{ext}}[A,\bmh]&=S_{\CS}[A]+ S_{\text{defect}}^{(\la,\gamma)}[A,\bmh]\,,\label{4dCS-defect-two}\\
      S_{\text{defect}}^{(\la,\gamma)}[A,\bmh]&= S_{\text{defect}}[A,\bmh]-\la \int_{\Sigma_{\mathfrak{p}}}{\rm det}(T^{(\la,\gamma)}_{\mu\nu})\,dx^+\wedge dx^- \label{defect-action-two}
\end{align}
to ensure a well-defined variation problem. We can verify that the equations of motion for this system are equivalent to the flatness condition of the Lax pair (\ref{two-tt-lax}). Thus the four-dimensional/two-dimensional system, incorporating the Lax pair (\ref{two-tt-lax}), describes a classical integrable field theory.

\subsubsection*{Two parameter deformed action}

Finally, we derive the effective two-dimensional action corresponding to the Lax pair (\ref{tt-lax}). By substituting (\ref{two-tt-lax}) to the modified four-dimensional/two-dimensional action (\ref{4dCS-defect-two}), we obtain the resulting two-dimensional action  
\begin{align}\label{two-ansatz}
    S_{\rm 2d}^{(\la,\gamma)}[\bmh]&=\frac{1}{2}\int_{\Sigma_{\mathfrak{p}}}\biggl(\llangle \bmh^{-1}\partial_{+}\bmh, {\bm\cJ}_{-}^{(\la,\gamma)}\rrangle_{\mathfrak{d}}
    +\frac{1}{2}\varepsilon^{\mu\nu}\llangle \bmh^{-1}\partial_{\mu}\bmh, (\mathcal{P}_{\bmh}^{+}+\mathcal{P}_{\bmh}^{-})(\bmh^{-1}\partial_{\nu}\bmh)\rrangle_{\mathfrak{d}}\biggr)dx^+\wedge dx^-\no\\
    &\qquad\qquad-\frac{1}{2}S_{\mathfrak{d}}^{\text{WZ}}[\bmh]-\la \int_{\Sigma_{\mathfrak{p}}}{\rm det}(T^{(\la,\gamma)}_{\mu\nu}) dx^+\wedge dx^-\,,
\end{align}
where we have used the identity
\begin{align}
    \llangle \bmh^{-1}\partial_{+}\bmh, {\bm\cJ}_{-}^{(\la,\gamma)}\rrangle_{\mathfrak{d}}=\llangle \bmh^{-1}\partial_{-}\bmh, {\bm\cJ}_{+}^{(\la,\gamma)}\rrangle_{\mathfrak{d}}\,.
\end{align}
In the undeformed limit $\la, \gamma\to 0$, the action (\ref{two-ansatz}) reduces to the degenerate $\cE$-model action (\ref{dE-action}).

We now confirm that the two-dimensional action (\ref{two-ansatz}) obtained in this way satisfies the classical flow equations for the $T\bar{T}$ and root-$T\bar{T}$ deformation. To do this, we rewrite the action (\ref{two-ansatz}) in terms of a field-dependent metric, as shown in 
\begin{align}
    S_{\rm 2d}^{(\la,\gamma)}[\bmh]&=\frac{1}{4}\int_{\Sigma_{\mathfrak{p}}}\biggl(\gamma^{*(\la,\gamma)\mu\nu}\llangle \bmh^{-1}\partial_{\mu}\bmh, {\bm\cJ}_{\nu}^{(\gamma)}\rrangle_{\mathfrak{d}}
    +\varepsilon^{\mu\nu}\llangle \bmh^{-1}\partial_{\mu}\bmh, (\mathcal{P}_{\bmh}^{+}+\mathcal{P}_{\bmh}^{-})(\bmh^{-1}\partial_{\nu}\bmh)\rrangle_{\mathfrak{d}}\biggr)dx^+\wedge dx^-\no\\
    &\qquad\qquad-\frac{1}{2}S_{\mathfrak{d}}^{\text{WZ}}[\bmh]-\la \int_{\Sigma_{\mathfrak{p}}}{\rm det}(T^{(\la,\gamma)}_{\mu\nu}) dx^+\wedge dx^-\,.
\end{align}
This makes it possible to use an identity 
\begin{align}
      \frac{d}{d\la}\gamma^{*(\la)\mu\nu}\llangle \bmh^{-1}\partial_{\mu}\bmh,  {\bm\cJ}_{\nu}^{(\gamma)}\rrangle_{\mathfrak{d}}+\frac{4}{\la}\frac{d}{d\la}\left(\la^2\, {\rm det}(T^{(\la,\gamma)}_{\mu\nu})\right)=0
\end{align}
similar to the equation (\ref{TT-id-flow}), which allows us to verify that the deformed action satisfies the classical $T\bar{T}$ flow equation
\begin{align}
    \frac{d}{d\la} S_{\rm 2d}^{(\la,\gamma)}[\bmh]=\textrm{det}(T_{\mu\nu}^{(\la,\gamma)})\,.
\end{align}
From the expression (\ref{two-dJ-2}) of the deformed current, it is straightforward to show that the deformed action (\ref{two-ansatz}) satisfies the root-$T\bar{T}$ flow equation
\begin{align}
    \frac{d}{d\gamma} S_{\rm 2d}^{(\la,\gamma)}[\bmh]=\sqrt{\textrm{det}(\widetilde{T}_{\mu\nu}^{(\la,\gamma)})}\,.
\end{align}
where $\widetilde{T}_{\mu\nu}^{(\la,\gamma)}$ denotes the traceless part of $T_{\mu\nu}^{(\la,\gamma)}$.
Thus, the deformed action (\ref{two-ansatz}) describes the simultaneous deformation of the $T\bar{T}$ and root-$T\bar{T}$ deformations of the degenerate $\cE$-model.

\section{Conclusion and discussion}
	\label{sec:conclusion}

In this study, we examined $T\bar{T}$-type deformations of two-dimensional integrable field theories described by coupling the four-dimensional theory with disorder defects. We can also consider the coupling with a different type of defect, the so-called order defects, which allows the derivation of a different class of two-dimensional integrable field theories, such as the massless Thirring model and the Faddeev-Reshetikhin model, from the four-dimensional theory. The $T\bar{T}$ and the root-$T\bar{T}$ deformations can also be considered in these two-dimensional integrable field theories \cite{Bonelli:2018kik}.

Recently, a novel procedure has been developed to describe $T\bar{T}$ and root-$T\bar{T}$ deformations by introducing auxiliary fields that satisfy specific algebraic equations \cite{Bielli:2024khq,Ferko:2024ali}. Subsequently, in the case of PCM, the auxiliary field sigma models have been derived by starting from a composite system that couples an auxiliary field to the four-dimensional theory \cite{Fukushima:2024nxm}.
It would be interesting to investigate how our construction relates to that in \cite{Fukushima:2024nxm} for the case of the $T\bar{T}$ and the root-$T\bar{T}$ deformations.

It is possible to consider deformations using higher-spin conserved currents similar to the $T\bar{T}$ deformation, and these deformations also preserve integrability \cite{Smirnov:2016lqw}. Recently, using the procedure of introducing an auxiliary field proposed in \cite{Bielli:2024khq}, deformations generated by operators composed of arbitrary higher-spin conserved currents have been intensively studied for the PCM and its variants \cite{Bielli:2024ach,Bielli:2024fnp}. Furthermore, by constructing the associated Lax pairs explicitly, 
we demonstrated the classical integrability of the deformed models. 
Therefore, it would be interesting to investigate whether the higher-spin integrable deformation can be described within the framework of the four-dimensional theory, similar to the $T\bar{T}$-type deformations discussed in our work, once the auxiliary fields are integrated out.
Such an exploration would provide a better understanding of how four-dimensional theory builds on two-dimensional integrable field theories.

An interesting future direction is to understand the results of this paper from the viewpoint of higher-dimensional gravity, since the closed-string sector generally describes deformations of the open-string theory, which contains the four-dimensional Chern–Simons theory.
One natural candidate for such a gravity theory is an appropriate
mixture of the Kodaira-Spencer gravity \cite{Bershadsky:1993cx,Costello:2015xsa}, which couples to the 6d CS theory and 
describes complex structure deformations, and its mirror, the K\"ahler gravity \cite{Bershadsky:1994sr}.
Kodaira-Spencer theory is also closely related to the JT gravity \cite{Post:2022dfi}, whose flat space version has been used to describe $T\bar{T}$-deformations \cite{Babaei-Aghbolagh:2024hti}.
We have seen, however, that our deformations involve a mixture of complex structure deformations and K\"ahler deformations, in particular for $T\bar{T}$-deformations.

From a more fundamental perspective, it would be very important to extend the four-dimensional Chern–Simons setup to the fully quantum level, providing a natural arena to understand the emergence of distinctive features of 
$T\bar{T}$-deformed theories—such as complex energy levels, Hagedorn-type transitions, and, quite intriguingly, regimes with negative specific heat \cite{Barbon:2020amo,Chakraborty:2022xmz}. It would also be interesting to explore whether these instabilities have a counterpart in our four-dimensional bulk description.

\section*{Acknowledgements}

We would like to thank the organizers of the workshop
``Exact techniques and their applications'' (Tropea, Italy, September 2024)
for providing a stimulating environment where this work was presented.

R.T.\ received partial support from the INFN project “Statistical Field Theory (SFT)” and the PRIN (Progetti di Rilevante Interesse Nazionale) Project No.\ 2022ABPBEY, “Understanding Quantum Field Theory through Its Deformations”, funded by the Italian Ministry of University and Research. He also acknowledges the support and hospitality, during the final part of this project, from the School of Mathematics and Physics of the University of Queensland, including an Ethel Raybould Visiting Fellowship.
M.Y.\ was supported in part by World Premier International Research Center Initiative (WPI), MEXT, Japan;
by the JSPS Grant-in-Aid for Scientific Research (Grant No.\ 20H05860, 23K17689, 23K25865); and
by JST, Japan (PRESTO Grant No. JPMJPR225A, Moonshot R\&D Grant No. JPMJMS2061). 

\appendix

\section*{Appendix}

\section{Weyl invariant metric and Beltrami parametrization}\label{Weyl-metric}

Here, we summarize some properties of the Weyl invariant metric.
On a two-dimensional surface, one can define the following Weyl-invariant metric:
\begin{align}
    \gamma^{\mu\nu} :=g^{\mu\nu}\sqrt{\lvert\text{det}(g_{\rho\sigma})\lvert}\,.
\end{align}
By definition, the determinant of $\gamma^{\mu\nu}$ is equal to $-1$:
\begin{align}\label{det-ga}
    \textrm{det}(\gamma^{\mu\nu})=-1\,,
\end{align}
and its inverse $\gamma_{\mu\nu}$ is related to it by
\begin{align}\label{ga-invga}
    \varepsilon^{\mu\rho}\gamma_{\rho\sigma}\varepsilon^{\sigma\nu}=\gamma^{\mu\nu}\,,
\end{align}
where the antisymmetric tensor $\varepsilon^{\mu\nu}$ is normalized as $\varepsilon^{+-}=1$.
To explicitly see this relation, we adopt the Beltrami parameterization of the metric on a two-dimensional surface
\begin{align}
    ds^2=-e^{2\phi}(dx^++\mu dx^-)(dx^-+\bar{\mu}dx^+)\,,
\end{align}
and then the relation (\ref{ga-invga}) indicates
\begin{align}\label{ga-invga-lc}
    \gamma^{++}=-\gamma_{--}\,,\quad \gamma^{--}=-\gamma_{++}\,,\quad \gamma^{+-}=\gamma_{-+}\,.
\end{align}
The components of $\gamma^{\mu\nu}$ and $\gamma_{\mu\nu}$ in terms of the Beltrami differential are
\begin{align}
    \gamma^{\mu\nu}&=
    \begin{pmatrix}
        \gamma^{++}&\gamma^{+-}\\
        \gamma^{-+}&\gamma^{--}
    \end{pmatrix}=-\frac{1}{1-\mu\bar{\mu}}
    \begin{pmatrix}
        -2\mu&1+\mu\bar{\mu}\\
        1+\mu\bar{\mu}&-2\bar{\mu}
    \end{pmatrix}\,,\\
    \gamma_{\mu\nu}&=
    \begin{pmatrix}
        \gamma_{++}&\gamma_{+-}\\
        \gamma_{-+}&\gamma_{--}
    \end{pmatrix}=-\frac{1}{1-\mu\bar{\mu}}
    \begin{pmatrix}
        2\bar{\mu}&1+\mu\bar{\mu}\\
        1+\mu\bar{\mu}&2\mu
    \end{pmatrix}
    \,.
\end{align}
For the flat-space case, the components of $\gamma^{\mu\nu}$ reduce to
\begin{align}
    \gamma^{+-}=\gamma^{-+}=-1\,,\qquad \gamma^{++}=0=\gamma^{--}\,.
\end{align}
In terms of the Beltrami differentials, the projection operators $P^{(\pm)\mu\nu}$ on the surface, defined in (\ref{sheet-proj}), take the form
\begin{align}
\begin{split}
   P^{(+)++}&=\frac{\mu}{1-\mu\bar{\mu}}=P^{(-)++}\,,\qquad  P^{(+)+-}=-\frac{\mu\bar{\mu}}{1-\mu\bar{\mu}}=P^{(-)-+}\,,\\
  P^{(+)-+}&=-\frac{1}{1-\mu\bar{\mu}}=P^{(-)+-}\,,\qquad  P^{(+)--}=\frac{\bar{\mu}}{1-\mu\bar{\mu}}=P^{(-)--}\,,
\end{split}
\end{align}
and $P_{\mu}^{(\pm)\nu}=\gamma_{\mu\rho}P^{(\pm)\rho\nu}$ is given by
\begin{align}\label{proj-bel}
\begin{split}
    P_{+}^{(+)+}&=\frac{1}{1-\mu\bar{\mu}}=P_{-}^{(-)-}\,,\qquad P_{-}^{(+)-}=-\frac{\mu\bar{\mu}}{1-\mu\bar{\mu}}=P_{+}^{(-)+}\,,\\
    P_{-}^{(+)+}&=\frac{\mu}{1-\mu\bar{\mu}}=-P_{-}^{(-)+}\,,\qquad  P_{+}^{(+)-}=-\frac{\bar{\mu}}{1-\mu\bar{\mu}}=-P_{+}^{(-)-}\,.
\end{split}
\end{align}

\section{Collection of technical computations}

In this appendix, we collect technical calculations related to the derivation of several formulas and statements discussed in the main text.

\subsection{Equations of motion for the degenerate \texorpdfstring{$\cE$}{\mathcal{E}}-model}

Here, we derive the equations of motion
\begin{align}
\begin{split}
\label{E-eom}
        &-\partial_{\mu}(P^{(-)\mu\nu}\mathcal{P}^{-}_{\bmh}(\bmh^{-1}\partial_{\nu}\bmh))+\partial_{\mu}(P^{(+)\mu\nu}\mathcal{P}^{+}_{\bmh}(\bmh^{-1}\partial_{\nu}\bmh))\\
    &\quad-[P^{(+)\mu\rho}\mathcal{P}^{+}_{\bmh}(\bmh^{-1}\partial_{\rho}\bmh),P_{\mu}^{(-)\sigma}\mathcal{P}^{-}_{\bmh}(\bmh^{-1}\partial_{\sigma}\bmh)]=0
\end{split}
\end{align}
for the degenerate $\cE$-model on curved space.
To this end, we denote the current by
\begin{align}
    J_{\mu}^{\pm}:=\mathcal{P}_{\bmh}^{\pm}(\bmh^{-1}\partial_{\mu}\bmh)\,,
\end{align}
so that the degenerate $\cE$-model action (\ref{cE-action2}) can be rewritten as
\begin{align}
    S_{\rm 2d}[\bmh]&=\frac{1}{4}\int_{\Sigma_{\mathfrak{p}}}\biggl(\gamma^{\mu\nu}\llangle \bmh^{-1}\partial_{\mu}\bmh, J^+_{\nu}-J^-_{\nu}\rrangle_{\mathfrak{d}}
     +\varepsilon^{\mu\nu}\llangle \bmh^{-1}\partial_{\mu}\bmh,  J^+_{\nu}+J^-_{\nu}\rrangle_{\mathfrak{d}}\biggr)dx^+\wedge dx^-
   \no\\
    &\qquad
   -\frac{1}{2}S_{\mathfrak{d}}^{\text{WZ}}[\bmh]\,.
\end{align}
Consider a variation of the defect field
\begin{align}
    \delta\bmh= \bmh\epsilon\,,\qquad \epsilon\in C^{\infty}(\Sigma_{\mathfrak{p}},\mathfrak{d})\,,
\end{align}
and then the left-invariant current for $\bmh$ varies as 
\begin{align}
    \delta(\bmh^{-1}d_{\Sigma}\bmh)&=d_{\Sigma}\epsilon+[\bmh^{-1}d_{\Sigma}\bmh, \epsilon]\,.
\end{align}
A crucial step in deriving the equations of motion is to compute the variation of the currents $J_{\mu}^{\pm}$.
For this purpose, it is convenient to consider the variation of the gauge transformed currents ${}^{\bmh}J_{\mu}^{\pm}$ which are given by
\begin{align}\label{var-gtcurrent}
    \mathrm{Ad}_{\bmh^{-1}}\delta({}^{\bmh}J_{\mu}^{\pm})=\delta J_{\mu}^{\pm}-\partial_{\mu}\epsilon-[J_{\mu}^{\pm},\epsilon]\,.
\end{align}
Since ${}^{\bmh}J^{\pm}\in \Omega^1(\Sigma_{\mathfrak{p}},\mathfrak{k})$, its variation $\delta({}^{\bmh}J^{\pm})$ also belongs to $\Omega^1(\Sigma_{\mathfrak{p}},\mathfrak{k})$ thus acting the projection operators $\mathcal{P}_{\bmh}^{\pm}$ on both side of (\ref{var-gtcurrent}) gives
\begin{align}\label{wpm-dj}
\begin{split}
    \mathcal{P}_{\bmh}^{\pm}(\delta J_{\mu}^{+})&=\mathcal{P}_{\bmh}^{\pm}\left(\partial_{\mu}\epsilon+[J_{\mu}^+,\epsilon]\right)\,,\\
   \mathcal{P}_{\bmh}^{\pm}(\delta J_{\mu}^{-})&=\mathcal{P}_{\bmh}^{\pm}\left(\partial_{\mu}\epsilon+[J_{\mu}^-,\epsilon]\right)\,.
\end{split}
\end{align}
From the equalities, the variation of $J_{\mu}^{\pm}$ is obtained as
\begin{align}\label{current-var}
    \delta J_{\mu}^{\pm}=\mathcal{P}_{\bmh}^{\pm}(\partial_{\mu}\epsilon+[J_{\mu}^{\pm},\epsilon])\,.
\end{align}
To see this, we introduce the projection operator $\cP_{\bmh}:\mathfrak{d}\to \mathfrak{d}$ by
\begin{align}
    \cP_{\bmh}=\frac{1}{2}(\mathcal{P}_{\bmh}^++\mathcal{P}_{\bmh}^-)\,,
\end{align}
and rewrite (\ref{wpm-dj}) in terms of $\cP_{\bmh}$ as
\begin{align}\label{pdelj}
       \cP_{\bmh}(\delta J_{\mu}^{\pm})&=\cP_{\bmh}\left(\partial_{\mu}\epsilon+[J_{\mu}^{\pm},\epsilon]\right)\,.
\end{align}
Since $\mathcal{E}_{\mathfrak{r}}(\mathcal{P}^{\pm}_{\bmh}(x))=\pm \mathcal{P}^{\pm}_{\bmh}(x)$, the identity (\ref{pdelj}) can also be written in the form
\begin{align}\label{pdelj2}
       \pm \cP_{\bmh}\cE_{\mathfrak{r}}(\delta J_{\mu}^{\pm})&=\cP_{\bmh}\left(\partial_{\mu}\epsilon+[J_{\mu}^{\pm},\epsilon]\right)\,.
\end{align}

For the non-degenerate case, the projection operator $\cP_{\bmh}$ satisfies the identity
\begin{align}\label{epe-id}
    \mathcal{E}_{\mathfrak{r}} \cP_{\bmh}\mathcal{E}_{\mathfrak{r}}=1-\cP_{\bmh}\,.
\end{align}
While this does not hold literally for a degenerate case, it still holds up to a gauge transformation by $F$, we can use this relation for practical computations.
By applying $\cE_{\mathfrak{r}}$ to both sides of (\ref{pdelj2}) and then using the identity (\ref{epe-id}), we obtain
\begin{align}\label{pdelj3}
       \pm \cE_{\mathfrak{r}} \cP_{\bmh}\cE_{\mathfrak{r}}(\delta J_{\mu}^{\pm})&=\pm(1-\cP_{\bmh})(\delta J_{\mu}^{\pm})=\cE_{\mathfrak{r}} \cP_{\bmh}\left(\partial_{\mu}\epsilon+[J_{\mu}^{\pm},\epsilon]\right)\,.
\end{align}
Combining (\ref{pdelj}) and (\ref{pdelj3}) leads to the variation (\ref{current-var}) of $J_{\mu}^{\pm}$.
Finally, using (\ref{current-var}) and the variation of the WZ term 
\begin{align}
    \delta S_{\mathfrak{d}}^{\text{WZ}}[\bmh]&=\int_{\Sigma_{\mathfrak{p}}}\llangle d_{\Sigma}\epsilon,\bmh^{-1}d_{\Sigma}\bmh\rrangle_{\mathfrak{d}}\,,
\end{align}
we can straightforwardly obtain the equations of motion (\ref{E-eom}).

\subsection{Comment on energy-momentum tensor for the non-degenerate case}\label{sec:em-em}

In Remark 2.8 of \cite{Lacroix:2020flf}, the energy-momentum tensor for the (relativistic) non-degenerate $\cE$-models is given as follows:
\begin{align}
    T_{++}&=\frac{1}{4}(T_{tt}+2T_{tx}+T_{xx})=\frac{1}{4}\llangle \cJ_{x} ,(\cE_{\mathfrak{r}}+1)\cJ_{x} \rrangle_{\mathfrak{d}}\,,\label{dem-pp}\\
    T_{--}&=\frac{1}{4}(T_{tt}-2T_{tx}+T_{xx})=\frac{1}{4}\llangle \cJ_{x} ,(\cE_{\mathfrak{r}}-1)\cJ_{x} \rrangle_{\mathfrak{d}}\,,\\
    T_{+-}&=\frac{1}{4}(T_{tt}-T_{xx})=0\,,
\end{align}
where the current $\cJ_{m}$ is defined by
\begin{align}
    \cJ_{x}&=\cP_{\bmh}(\bmh^{-1}\partial_{x}\bmh)+\cE_{\mathfrak{r}}\cP_{\bmh}(\bmh^{-1}\partial_{t}\bmh)\,,\\
    \cJ_{t}&=\cE_{\mathfrak{r}}\cP_{\bmh}(\bmh^{-1}\partial_{x}\bmh)+\cP_{\bmh}(\bmh^{-1}\partial_{t}\bmh)\,.
\end{align}
At first glance, this expression appears to differ from the energy-momentum tensor (\ref{em-dE}) given in the main text. In what follows, we now show explicitly that these two expressions are in fact equivalent in the non-degenerate case.

We focus on the $(++)$ component and begin by rewriting the right-hand side of (\ref{dem-pp}) in the light-cone coordinates.
Since the projection operators $\mathcal{P}_{\bmh}^{\pm}$ satisfy
\begin{align}
    \mathcal{P}_{\bmh}^{\pm}=(1\pm \cE_{\mathfrak{r}})\cP_{\bmh}\,,\label{Whpm_EP}
\end{align}
the current $\cJ_{m}$ can be expressed in terms of $\mathcal{P}_{\bmh}^{\pm}$ as
\begin{align}
     \cJ_{x}&=\mathcal{P}_{\bmh}^{+}(\bmh^{-1}\partial_{+}\bmh) -\mathcal{P}_{\bmh}^{-}(\bmh^{-1}\partial_{-}\bmh)\,,\\
     \cJ_{t}&=\mathcal{P}_{\bmh}^{+}(\bmh^{-1}\partial_{+}\bmh) +\mathcal{P}_{\bmh}^{-}(\bmh^{-1}\partial_{-}\bmh)\,.   
\end{align}
In terms of the projection operators $\mathcal{P}_{\bmh}^{\pm}$, the $(++)$ component (\ref{dem-pp}) of the energy-momentum tensor takes the form
\begin{align}
    T_{++}
    &=\frac{1}{4}\llangle \cJ_{x} ,(\cE_{\mathfrak{r}}+1)\cJ_{x} \rrangle_{\mathfrak{d}}\no\\
    &=\frac{1}{2}\llangle \mathcal{P}_{\bmh}^{+}(\bmh^{-1}\partial_{+}\bmh) ,\mathcal{P}_{\bmh}^{+}(\bmh^{-1}\partial_{+}\bmh)\rrangle_{\mathfrak{d}}-\frac{1}{2}\llangle \mathcal{P}_{\bmh}^{+}(\bmh^{-1}\partial_{+}\bmh) ,\mathcal{P}_{\bmh}^{-}(\bmh^{-1}\partial_{-}\bmh)\rrangle_{\mathfrak{d}}\no\\
    &=\frac{1}{2}\llangle \mathcal{P}_{\bmh}^{+}(\bmh^{-1}\partial_{+}\bmh) ,\mathcal{P}_{\bmh}^{+}(\bmh^{-1}\partial_{+}\bmh)\rrangle_{\mathfrak{d}}\,.
\end{align}
Using the relation (\ref{Whpm_EP}) and taking the transpose of $\mathcal{P}_{\bmh}^{+}$, we have
\begin{align}
    T_{++}&=\frac{1}{2}\llangle \bmh^{-1}\partial_{+}\bmh ,{}^t\cP_{\bmh}(1+\cE_{\mathfrak{r}})(1+\cE_{\mathfrak{r}})\cP_{\bmh}(\bmh^{-1}\partial_{+}\bmh)\rrangle_{\mathfrak{d}}\no\\
    &=\llangle \bmh^{-1}\partial_{+}\bmh ,\cE_{\mathfrak{r}}\cP_{\bmh}(\bmh^{-1}\partial_{+}\bmh)\rrangle_{\mathfrak{d}}\no\\
    &=\frac{1}{2}\llangle \bmh^{-1}\partial_{+}\bmh ,(\mathcal{P}_{\bmh}^{+}-\mathcal{P}_{\bmh}^{-})(\bmh^{-1}\partial_{+}\bmh)\rrangle_{\mathfrak{d}}\,,
\end{align}
where we have used the identity $\cE_{\mathfrak{r}}^2=\cE_{\mathfrak{r}}$ and Proposition 2.3 (i) in  \cite{Lacroix:2020flf}, namely
\begin{align}
    {}^t\cP_{\bmh}=\cE_{\mathfrak{r}}^{-1}\cP_{\bmh}\cE_{\mathfrak{r}}\,.
\end{align}
This precisely matches the energy-momentum tensor given in (\ref{em-dE}).
The $(--)$ component can be shown to be equivalent in the same manner.

\subsection{Gauge invariance of energy-momentum tensor}\label{gaugetr-em}

We show that the energy-momentum tensor 
\begin{align}
    T_{\pm\pm}^{(0)}=-\frac{1}{2}\llangle \bmh^{-1}\partial_{\pm}\bmh ,(\mathcal{P}_{\bmh}^{-}-\mathcal{P}_{\bmh}^{+})(\bmh^{-1}\partial_{\pm}\bmh) \rrangle_{\mathfrak{d}}\,,\qquad T_{+-}^{(0)}=0\label{e-em}
\end{align}
of the degenerate $\cE$-model is invariant under a gauge transformation \begin{align}
    \bmh\to {\bm k}\,\bmh\Delta(f)^{-1}\,,\qquad {\bm k} \in C^{\infty}(\Sigma_{\mathfrak{p}},{\bm K})\,,\quad f \in C^{\infty}(\Sigma,G)\,.
\end{align}
For this purpose, we employ the transformation rule of the projection operators $\mathcal{P}_{\bmh}^{\pm}$ 
\begin{align}\label{W-tr}
    \mathcal{P}^{\pm}_{{\bm k}\bmh\Delta(f)^{-1}}=\text{Ad}_{\Delta(f)}\circ \mathcal{P}_{\bmh}^{\pm}\circ \text{Ad}_{\Delta(f)}^{-1}\,.
\end{align}
This can be verified by direct computation.
Using the transformation formula (\ref{W-tr}), we obtain
\begin{align}
   &\mathcal{P}^{\pm}_{{\bm k}\bmh\Delta(f)^{-1}}(\Delta(f)\bmh^{-1}{\bm k}^{-1}d({\bm k}\,\bmh\Delta(f)^{-1}))\no\\
   &= \mathrm{Ad}_{\Delta(f)}\Bigl(\mathcal{P}^{\pm}_{\bmh}(\bmh^{-1}d\bmh+\mathrm{Ad}_{\bmh^{-1}}({\bm k}^{-1}d{\bm k} ) -\Delta(f)^{-1}d\Delta(f)  )\Bigr)\no\\
   &=\mathrm{Ad}_{\Delta(f)}\Bigl(\mathcal{P}^{\pm}_{\bmh}(\bmh^{-1}d\bmh)\Bigr)-d\Delta(f)\Delta(f)^{-1}\,.
\end{align}
In the second equality we have used $\mathrm{Ad}_{\bmh^{-1}}\left({\bm k}^{-1}d{\bm k} \right)\in  \operatorname{Ker}\,\mathcal{P}_{\bmh}^{\pm}$, and $\mathcal{P}^{\pm}_{\bmh}(\Delta(f)^{-1}d\Delta(f))=\Delta(f)^{-1}d\Delta(f)$ which follows from being $\mathfrak{f}\subset {\rm Im}\mathcal{P}_{\bmh}^{\pm}$.

The energy-momentum tensor (\ref{e-em}) then transforms as
\begin{align}
    &T_{\pm\pm}^{(0)}({\bm k}\,\bmh\Delta(f)^{-1})\no\\
    &=-\frac{1}{2}\llangle {\bm h}^{-1} \partial_{\pm}{\bmh}+\mathrm{Ad}_{\bmh^{-1}}({\bm k}^{-1} \partial_{\pm}{\bm k})-  \Delta(f)^{-1}\partial_{\pm} \Delta(f) 
   ,(\mathcal{P}_{\bmh}^{-}-\mathcal{P}_{\bmh}^{+})(\bmh^{-1}\partial_{\pm}\bmh)\rrangle_{\mathfrak{d}}\,.
\end{align}
Since $\mathfrak{f}^{\bot}=\mathfrak{f}\dot{+}\mathfrak{r}$, $\text{Im}\,\mathcal{P}_{\bmh}^{\pm}=\mathfrak{f}\dot{+} \mathfrak{r}_{\pm}$ and ${\bm k} \in C^{\infty}(\Sigma_{\mathfrak{p}},{\bm K})$, we have
\begin{align}
    &\llangle \mathrm{Ad}_{\bmh^{-1}}({\bm k}^{-1}\partial_{\pm}{\bm k}),(\mathcal{P}_{\bmh}^{-}-\mathcal{P}_{\bmh}^{+})(\bmh^{-1}\partial_{\pm}\bmh) \rrangle_{\mathfrak{d}}=0\,,\\
    & \llangle  \Delta(f)^{-1}\partial_{\pm} \Delta(f)  ,(\mathcal{P}_{\bmh}^{-}-\mathcal{P}_{\bmh}^{+})(\bmh^{-1}\partial_{\pm}\bmh) \rrangle_{\mathfrak{d}}=0\,.
\end{align}
Hence, the energy-momentum tensor (\ref{e-em}) is gauge invariant:
\begin{align}
     T_{\pm\pm}^{(0)}({\bm k}\,\bmh\Delta(f)^{-1})&= T_{\pm\pm}^{(0)}(\bmh)
     \,.
\end{align}

\subsection{Variation of the four-dimensional/two-dimensional action with a field-dependent metric}\label{sec:42-var}

In the main body, we computed the field variations of the four-dimensional action accompanying the Lax pair 
\begin{align}\label{Lax-fm}
    \cL_{\pm}&=\frac{1}{2}\bmp\left((\mathcal{P}_{\bmh}^{+}+\mathcal{P}_{\bmh}^{-})({\bm h}^{-1}\partial_{\pm}{\bm h})\right)+\frac{1}{2}\gamma_{\pm\rho}^{*}({\bm h})\varepsilon^{\rho\mu}\,\bmp\left((\mathcal{P}_{\bmh}^{+}-\mathcal{P}_{\bmh}^{-})({\bm h}^{-1}\partial_{\mu}{\bm h})\right)
\end{align}
involving the field-dependent metric in the cases of root-$T\bar{T}$ and $T\bar{T}$ deformations, respectively.
Here, we present the general formula for the variation of the four-dimensional theory with (\ref{Lax-fm}).

In Section \ref{sec:reduction_curved}, we studied how the four-dimensional/two-dimensional action (\ref{4dCS-defect}) and the Lax pair (\ref{Lax-curve-E1}) corresponding to the degenerate $\cE$-model on a curved space transform under the variation (\ref{var}).
By denoting the variation of the field dependent metric $\gamma_{\mu\nu}^{*}({\bm h})$ by
\begin{align}
   \gamma_{\mu\nu}^{*}({\bm h}')=\gamma_{\mu\nu}^{*}({\bm h})+ \delta\gamma_{\mu\nu}^{*}({\bm h})\,,\qquad \delta\gamma_{\mu\nu}^{*}({\bm h})=\delta\gamma_{\nu\mu}^{*}({\bm h})\,,
\end{align}
The variation of the field-dependent metric $\gamma^{*\mu\nu}$ with upper indices is defined in the same manner as that of a conventional metric:
\begin{align}
    \delta \gamma^{*\pm\pm}=-\delta\gamma_{\mp\mp}^*\,,\qquad \delta\gamma^{*+-}=\delta\gamma_{+-}^{*}\,.
\end{align}
The Lax pair (\ref{Lax-fm}), under the variation (\ref{var}), transforms as
\begin{align}
    {\bm j}^*\cL_{\pm}'&={\bm j}^*\cL_{\pm}+\epsilon \,  {\bm j}^*l_{\pm}\,,
\end{align}
where ${\bm j}^*l_{\pm}={\bm j}^*l_{\pm}^{(0)}+{\bm j}^*l_{\pm}^{(\gamma)}$ is given by
\begin{align}
    {\bm j}^*l_{\pm}^{(0)}&=[{\bm j}^*\cL_{\pm},\sfu]
    +(P_{\pm}^{(+)\mu}\mathcal{P}_{\bmh}^{+}(\partial_{\mu}\sfu)+P_{\pm}^{(-)\mu}\mathcal{P}_{\bmh}^{-}(\partial_{\mu}\sfu))\,,\\
    \epsilon \,  {\bm j}^*l_{\pm}^{(\gamma)}&=\frac{1}{2}\delta\gamma_{\pm\rho}^{*}({\bm h})\varepsilon^{\rho\mu}\,(\mathcal{P}_{\bmh}^{+}-\mathcal{P}_{\bmh}^{-})({\bm h}^{-1}\partial_{\mu}{\bm h})\,.\label{l-var-general}
\end{align}
Here, the variation $l^{(0)}$ takes the same form as (\ref{E-model-var}) and satisfies 
\begin{align}
    \mathrm{Ad}_{\bmh}(d_{\Sigma}\sfu+[{\bm j}^*\cL,\sfu]-{\bm j}^* l^{(0)})\in \Omega^{1}(\Sigma_{\mathfrak{p}},\mathfrak{k})\,.
\end{align}
From the discussion of classical integrability of the two-dimensional effective theory on a curved space in Section \ref{sec:curved}, $l^{(0)}$ does not contribute to the variation of the four-dimensional/two-dimensional action (\ref{4dCS-defect}). Only the variation $l^{(\gamma)}$ associated with the field-dependent metric contributes.
Substituting (\ref{l-var-general}) into the first equality of (\ref{24action-val}), the variation of the 4d–-2d action (\ref{4dCS-defect}) with the Lax pair (\ref{Lax-fm}) is given by
\begin{align}\label{4d2d-var0}
    \delta S_{\text{ext}}[\cL,\bmh]
    &=\frac{i \epsilon}{2\pi}\int_{\Sigma\times C}\omega\wedge\Tr( l^{(\gamma)}\wedge \bar{\partial}\cL)-\frac{\epsilon}{2}\int_{\Sigma_{\mathfrak{p}}}\llangle \mathrm{Ad}_{\bmh}({\bm j}^*l^{(\gamma)}), {}^\bmh({\bm j}^*\cL)\rrangle_{\mathfrak{d}}\no\\
    &\quad+\epsilon\int_{\Sigma_{\mathfrak{p}}}
    \llangle \sfu,d_{\Sigma}({\bm j}^*\cL)+\frac{1}{2}[{\bm j}^*\cL,{\bm j}^*\cL\,]\rrangle_{\mathfrak{d}}\,.
\end{align}

For the same reason discussed in Section \ref{sec:curved}, the first term does not vanish even when the equations of motion $\omega\wedge \bar{\partial}\cL=0$ are imposed, and its integration over $C$ yields
\begin{align}
    \frac{i}{2\pi}\int_{\Sigma\times C}\omega\wedge\Tr( l^{(\gamma)}\wedge \bar{\partial}\cL)=\frac{1}{2}\int_{\Sigma_{\mathfrak{p}}}\llangle {\bm j}^*l^{(\gamma)}, {\bm j}^*\cL\rrangle_{\mathfrak{d}}\,.
\end{align}
Using this identity, the first line in (\ref{4d2d-var0}) simplifies, and by expressing the remaining terms in terms of the undeformed energy-momentum tensor (\ref{em-dE}), the variation (\ref{4d2d-var0}) becomes
\begin{align}
    \delta S_{\text{ext}}[\cL,\bmh]
    &=\frac{1}{2}\epsilon\int_{\Sigma_{\mathfrak{p}}} \left(\delta\gamma^{*++}T_{++}^{(0)}+\delta\gamma^{*--}\,T_{--}^{(0)}\right)dx^+\wedge dx^-\no\\
    &\quad+\frac{1}{2}\epsilon\int_{\Sigma_{\mathfrak{p}}}\delta \gamma^{*+-}\,\llangle \bmh^{-1}\partial_+\bmh, (\mathcal{P}_{\bmh}^{+}-\mathcal{P}_{\bmh}^{-})({\bm h}^{-1}\partial_{-}{\bm h})\rrangle_{\mathfrak{d}}\,dx^+\wedge dx^-\no\\
    &\quad+\epsilon\int_{\Sigma_{\mathfrak{p}}}
    \llangle \sfu,d({\bm j}^*\cL)+\frac{1}{2}[{\bm j}^*\cL,{\bm j}^*\cL]\rrangle_{\mathfrak{d}}\,.
\end{align}

\subsection{Derivation of deformed energy-momentum tensors}\label{sec:TT-em}

Here, we compute the deformed energy-momentum tensors for the degenerate $\cE$-model using dynamical coordinate transformations.

\subsubsection{$T\bar{T}$ deformed case}

We first derive the energy-momentum tensor of the $T\bar{T}$-deformed degenerate $\cE$-model.
To achieve this, we apply the dynamical coordinate transformation to the energy-momentum tensor
\begin{align}\label{em-un}
    T_{\pm'\pm'}^{(0)}=-\frac{1}{2}\llangle \bmh^{-1}\partial_{\pm'}\bmh ,(\mathcal{P}_{\bmh}^{-}-\mathcal{P}_{\bmh}^{+})(\bmh^{-1}\partial_{\pm'}\bmh) \rrangle_{\mathfrak{d}}\,.
\end{align}
The left-invariant current $\bmh^{-1}\partial_{\pm'}\bmh$ transforms as 
\begin{align}\label{tj-j}
        {\bm h}^{-1}\partial_{\pm'}{\bm h}(x')&={\bm h}^{-1}\partial_{\pm}{\bm h}(x)-\la\,T_{\pm'\pm'}^{(0)}(x'){\bm h}^{-1}\partial_{\mp}{\bm h}(x)\,.
\end{align}
Substituting this into (\ref{em-un}) yields 
\begin{align}
\label{Tpcm-eq}
    T^{(0)}_{\pm'\pm'}(x')&=T^{(0)}_{\pm\pm}(x)-2\la  T^{(0)}_{\pm'\pm'}(x')x_1+\la^2 T^{(0)}_{\pm'\pm'}(x')^2 T^{(0)}_{\mp\mp}(x)\,,
\end{align}
where $x_1$ is defined by
\begin{align}
     x_1=\frac{1}{2}\llangle \bmh^{-1}\partial_{+}\bmh ,(\mathcal{P}_{\bmh}^{-}-\mathcal{P}_{\bmh}^{+})(\bmh^{-1}\partial_{-}\bmh) \rrangle_{\mathfrak{d}}\,,
\end{align}
By solving the equation (\ref{Tpcm-eq}) for $ T^{(0)}_{\pm'\pm'}(x')$, one finds
\begin{align}\label{T-Tt}
    T^{(0)}_{\pm'\pm'}(x')&=\frac{1+2\la\,x_1-\sqrt{1+2(2\la) x_1+2(2\la)^2\left(x_1^2-x_2\right)}}{2\la^2T_{\mp\mp}^{(0)}(x)}\,,
\end{align}
where $x_2$ is given by
\begin{align}
   x_2=\frac{1}{2}\left(T_{++}^{(0)}T_{--}^{(0)}+x_1^2\right)(x)\,.
\end{align}
By substituting (\ref{T-Tt}) into (\ref{dT-T0-inv}), we obtain the $T\bar{T}$ deformed energy-momentum tensor
\begin{align}
    T_{\pm\pm}^{(\la)}(x)&=\frac{T_{\pm\pm}^{(0)} }{\sqrt{1+2(2\la) x_1+2(2\la)^2\left(x_1^2-x_2\right)}}\,,\\
    T_{+-}^{(\la)}(x)&=\frac{1+2\la x_1-\sqrt{1+2(2\la) x_1+2(2\la)^2\left(x_1^2-x_2\right)}}{2\la\sqrt{1+2(2\la) x_1+2(2\la)^2\left(x_1^2-x_2\right)}}\,.
\end{align}
Note that the $(\pm\pm)$ components of the deformed energy-momentum tensor can be rewritten in a simplified form
\begin{align}\label{em-un-lc}
    T_{\pm\pm}^{(\la)}=\frac{1}{2}\llangle \bmh^{-1}\partial_{\pm}\bmh ,{\bm\cJ}_{\pm}^{(\la)} \rrangle_{\mathfrak{d}}
\end{align}
by using the deformed current ${\bm\cJ}_{\pm}^{(\la)}$.

\subsubsection{Two-parameter deformed case}

Next, we derive the deformed energy-momentum tensor for the two-parameter deformation case. In this case, the dynamical coordinate transformation is applied in 
deriving the root-$T\bar{T}$ deformed energy-momentum tensor
\begin{align}
    T_{\pm'\pm'}^{(\gamma)}=\frac{1}{2}\llangle \bmh^{-1}\partial_{\pm'}\bmh ,{\bm \cJ}_{\pm'}^{(\gamma)}\rrangle_{\mathfrak{d}}\,.
\end{align}
The entire computation is carried out in exactly the same way as in the standard $T\bar{T}$ deformation. Consequently, the two-parameter deformed energy-momentum tensor $T_{\mu\nu}^{(\la,\gamma)}$ is obtained by replacing $x_1$ and $x_2$ with
\begin{align}
     x_1^{(\gamma)}=-\frac{1}{2}\llangle \bmh^{-1}\partial_{+}\bmh ,{\bm\cJ}_{-}^{(\gamma)} \rrangle_{\mathfrak{d}}\,,
     \qquad  x_2^{(\gamma)}=\frac{1}{2}\left(T_{++}^{(\gamma)}T_{--}^{(\gamma)}+(x_1^{(\gamma)})^2\right)\,,
\end{align}
and we obtain
\begin{align}\label{em-two}
    T_{\pm\pm}^{(\la,\gamma)}(x)&=\frac{T_{\pm\pm}^{(\gamma)} }{\sqrt{1+2(2\la) x_1^{(\gamma)}+2(2\la)^2\left((x_1^{(\gamma)})^2-x_2^{(\gamma)}\right)}}\,,\\
    T_{+-}^{(\la,\gamma)}(x)&=\frac{1+2\la x_1^{(\gamma)}-\sqrt{1+2(2\la) x_1^{(\gamma)}+2(2\la)^2\left((x_1^{(\gamma)})^2-x_2^{(\gamma)}\right)}}{2\la\sqrt{1+2(2\la) x_1^{(\gamma)}+2(2\la)^2\left((x_1^{(\gamma)})^2-x_2^{(\gamma)}\right)}}\,.
\end{align}

\subsection{Variation of \texorpdfstring{$T\bar{T}$}{T\bar{T}} deformed energy-momentum tensor}\label{sec:TT-id}

Here we show the identity
\begin{align}\label{tt-id2}
    -\frac{1}{2}\delta\gamma^{*(\la)\mu\nu}\llangle \bmh^{-1}\partial_{\mu}\bmh, (\mathcal{P}_{\bmh}^{-}-\mathcal{P}_{\bmh}^{+})({\bm h}^{-1}\partial_{\nu}{\bm h})\rrangle_{\mathfrak{d}}+2\la\, \delta \left({\rm det}(T^{(\la)}_{\mu\nu})\right)=0\,,
\end{align}
where the field-dependent metric $\gamma^{*(\la)\mu\nu}$ is 
\begin{align}
\begin{split}
    &\gamma^{*\pm\pm(\la)}=-\frac{2\la T_{\mp\mp}^{(0)}}{ \sqrt{1+2(2\la) x_1+2(2\la)^2\left(x_1^2-x_2\right)}}\,,\\
    &\gamma^{*+-(\la)}=\gamma^{*-+(\la)}=-\frac{1+2\la x_1}{ \sqrt{1+2(2\la) x_1+2(2\la)^2\left(x_1^2-x_2\right)}}\,.
\end{split}
\end{align}
The left-hand side of (\ref{tt-id2}) can be written as
\begin{align}
  ({\rm LHS})&=-2\delta\gamma^{*(\la)+-}x_1+\delta\gamma^{*(\la)++}T^{(0)}_{++}+\delta\gamma^{*(\la)--}T^{(0)}_{--}+2\delta T^{(\la)}_{+-}\no\\
    &=\Bigl[2x_1(1+2\la x_1)-4\la T^{(0)}_{++}T^{(0)}_{--}+\frac{1}{\la}(1+2\la x_1)\Bigr]\delta\left(\frac{1}{ \sqrt{X}}\right)\no\\
    &\quad +\frac{1}{ \sqrt{X}}\delta\Bigl(2\la x_1^2+2x_1-2\la T^{(0)}_{++}T^{(0)}_{--}\Bigr)\no\\
    &=\frac{1}{\la}\left(1+2(2\la) x_1+2(2\la)^2\left(x_1^2-x_2\right)\right)
    \delta\left(\frac{1}{ \sqrt{X}}\right)+\frac{2\delta\Bigl(x_1+2\la (x_1^2-x_2))\Bigr)}{ \sqrt{X}}\no\\
    &=0\,,
\end{align}
where $X=1+2(2\la) x_1+2(2\la)^2\left(x_1^2-x_2\right)$, and we have used the identity ${\rm det}(T^{(\la)}_{\mu\nu})=\la^{-1}\,T_{+-}^{(\la)}$ in the first equality.
Since this result holds without depending on the specific forms of $x_1$ and $x_2$, the same identity holds for the two-parameter deformation case as well, by simply replacing $x_1$ and $x_2$ with $x_1^{(\gamma)}$ and $x_2^{(\gamma)}$, respectively.

\section{Comment on dynamical coordinate transformation for root-\texorpdfstring{$T\bar{T}$}{T\bar{T}} deformation}

One might attempt to interpret the relation (\ref{TTbar-gamma}) as arising from a dynamical coordinate transformation, analogous to the case of $T\bar{T}$ deformations discussed in \cite{Conti:2018tca,Conti:2019dxg}.
In fact, from the field-dependent Weyl-invariant metric (\ref{TTbar-gamma}) for the root-$T\bar{T}$ deformation, the dynamical coordinate transformation discussed in Section \ref{sec:dycoord} can take the formal form:
\begin{align}\label{dcoord-tr-root}
    \begin{pmatrix}
        dx^{+'}\\
        dx^{-'}
    \end{pmatrix}
    =\begin{pmatrix}
        1&\tanh\left(\frac{\gamma}{2}\right) \sqrt{\frac{\Tr j_-j_-}{\Tr j_+j_+}}\\
        \tanh\left(\frac{\gamma}{2}\right)\sqrt{\frac{\Tr j_+j_+}{\Tr j_-j_-}}&1
    \end{pmatrix}
        \begin{pmatrix}
        dx^{+}\\
        dx^{-}
    \end{pmatrix}
    \,,
\end{align}
and its inverse is given by
\begin{align}
    \begin{pmatrix}
        dx^{+}\\
        dx^{-}
    \end{pmatrix}
    =\begin{pmatrix}
        \cosh^2\left(\frac{\gamma}{2}\right)&-\frac{1}{2}\sinh\left(\gamma\right) \sqrt{\frac{\Tr j_-j_-}{\Tr j_+j_+}}\\
       -\frac{1}{2}\sinh\left(\gamma\right) \sqrt{\frac{\Tr j_+j_+}{\Tr j_-j_-}}&   \cosh^2\left(\frac{\gamma}{2}\right)
    \end{pmatrix}
        \begin{pmatrix}
        dx^{+'}\\
        dx^{-'}
    \end{pmatrix}
    \,.
\end{align}
However, the coordinate transformation fails to satisfy the integrability condition
\begin{align}
    &\frac{\partial}{\partial x^+}\left(\frac{\partial x^{+'}}{\partial x^-}\right)=\tanh\left(\frac{\gamma}{2}\right) \frac{\partial}{\partial x^+}\left(\sqrt{\frac{\Tr j_-j_-}{\Tr j_+j_+}} \right)\neq 0\,,
     \qquad \frac{\partial}{\partial x^-}\left(\frac{\partial x^{+'}}{\partial x^+}\right)=0\,.
\end{align}

The breakdown of the integrability condition suggests that the dynamical coordinate transformation (\ref{dcoord-tr-root}) cannot consistently map the equations of motion of the undeformed model to those of the deformed model.
This observation might be consistent with the fact that a variant of flat-space JT gravity theory recently presented in \cite{Babaei-Aghbolagh:2024hti} exhibits a singular limit that removes only the $T\bar{T}$ deformation, making it challenging to construct a gravity theory describing pure root-$T\bar{T}$ deformation.
The construction of such a gravity theory would be an interesting direction for future work.

\bibliographystyle{JHEP}
\bibliography{4dCS}

@article{Chakraborty:2022xmz,
    author = "Chakraborty, Soumangsu and Hashimoto, Akikazu",
    title = "{Comments on the negative specific heat of the $ T\overline{T} $ deformed symmetric product CFT}",
    eprint = "2201.08439",
    archivePrefix = "arXiv",
    primaryClass = "hep-th",
    doi = "10.1007/JHEP03(2022)213",
    journal = "JHEP",
    volume = "03",
    pages = "213",
    year = "2022"
}

@book{nekrassov1996four,
  title     = {Four-dimensional holomorphic theories},
  author    = {Nekrassov, Nikita Alex},
  year      = {1996},
  publisher = {Princeton University},
note = {\url{https://media.scgp.stonybrook.edu/papers/prdiss96.pdf}}
}

@article{Rodriguez:2021tcz,
    author = "Rodr{\'\i}guez, Pablo and Tempo, David and Troncoso, Ricardo",
    title = "{Mapping relativistic to ultra/non-relativistic conformal symmetries in 2D and finite $ \sqrt{T\overline{T}} $ deformations}",
    eprint = "2106.09750",
    archivePrefix = "arXiv",
    primaryClass = "hep-th",
    reportNumber = "CECS-PHY-20/03",
    doi = "10.1007/JHEP11(2021)133",
    journal = "JHEP",
    volume = "11",
    pages = "133",
    year = "2021"
}

@article{Babaei-Aghbolagh:2022uij,
    author = "Babaei-Aghbolagh, H. and Velni, Komeil Babaei and Yekta, Davood Mahdavian and Mohammadzadeh, H.",
    title = "{Emergence of non-linear electrodynamic theories from $T\bar{T}$-like deformations}",
    eprint = "2202.11156",
    archivePrefix = "arXiv",
    primaryClass = "hep-th",
    reportNumber = "IPM/P-2022/13",
    doi = "10.1016/j.physletb.2022.137079",
    journal = "Phys. Lett. B",
    volume = "829",
    pages = "137079",
    year = "2022"
}

@article{Ashwinkumar:2019mtj,
    author = "Ashwinkumar, Meer and Tan, Meng-Chwan",
    title = "{Unifying lattice models, links and quantum geometric Langlands via branes in string theory}",
    eprint = "1910.01134",
    archivePrefix = "arXiv",
    primaryClass = "hep-th",
    doi = "10.4310/ATMP.2020.v24.n7.a1",
    journal = "Adv. Theor. Math. Phys.",
    volume = "24",
    number = "7",
    pages = "1681--1721",
    year = "2020"
}

@article{Lacroix:2021iit,
    author = "Lacroix, Sylvain",
    title = "{Four-dimensional Chern{\textendash}Simons theory and integrable field theories}",
    eprint = "2109.14278",
    archivePrefix = "arXiv",
    primaryClass = "hep-th",
    doi = "10.1088/1751-8121/ac48ed",
    journal = "J. Phys. A",
    volume = "55",
    number = "8",
    pages = "083001",
    year = "2022"
}

@article{Py:2022hoa,
    author = "Py, Victor",
    title = "{$ T\overline{T} $ deformations in curved space from 4D Chern-Simons theory}",
    eprint = "2202.08841",
    archivePrefix = "arXiv",
    primaryClass = "hep-th",
    doi = "10.1007/JHEP08(2022)101",
    journal = "JHEP",
    volume = "08",
    pages = "101",
    year = "2022"
}

@article{Costello:2017dso,
    author = "Costello, Kevin and Witten, Edward and Yamazaki, Masahito",
    title = "{Gauge Theory and Integrability, I}",
    eprint = "1709.09993",
    archivePrefix = "arXiv",
    primaryClass = "hep-th",
    reportNumber = "IPMU17-0136",
    doi = "10.4310/ICCM.2018.v6.n1.a6",
    journal = "ICCM Not.",
    volume = "06",
    number = "1",
    pages = "46--119",
    year = "2018"
}

@article{Costello:2013zra,
    author = "Costello, Kevin",
    title = "{Supersymmetric gauge theory and the Yangian}",
    eprint = "1303.2632",
    archivePrefix = "arXiv",
    primaryClass = "hep-th",
    month = "3",
    year = "2013"
}

@article{Costello:2018gyb,
    author = "Costello, Kevin and Witten, Edward and Yamazaki, Masahito",
    title = "{Gauge Theory and Integrability, II}",
    eprint = "1802.01579",
    archivePrefix = "arXiv",
    primaryClass = "hep-th",
    reportNumber = "IPMU18-0025",
    doi = "10.4310/ICCM.2018.v6.n1.a7",
    journal = "ICCM Not.",
    volume = "06",
    number = "1",
    pages = "120--146",
    year = "2018"
}

@article{Costello:2019tri,
    author = "Costello, Kevin and Yamazaki, Masahito",
    title = "{Gauge Theory And Integrability, III}",
    eprint = "1908.02289",
    archivePrefix = "arXiv",
    primaryClass = "hep-th",
    reportNumber = "IPMU19-0110",
    month = "8",
    year = "2019"
}

@article{Vicedo:2019dej,
    author = "Vicedo, Benoit",
    title = "{Holomorphic Chern-Simons theory and affine Gaudin models}",
    eprint = "1908.07511",
    archivePrefix = "arXiv",
    primaryClass = "hep-th",
    month = "8",
    year = "2019"
}

@article{Costello:2021zcl,
    author = "Costello, Kevin and Gaiotto, Davide and Yagi, Junya",
    title = "{Q-operators are 't Hooft lines}",
    eprint = "2103.01835",
    archivePrefix = "arXiv",
    primaryClass = "hep-th",
    month = "3",
    year = "2021"
}

@article{Costello:2018txb,
    author = "Costello, Kevin and Yagi, Junya",
    title = "{Unification of integrability in supersymmetric gauge theories}",
    eprint = "1810.01970",
    archivePrefix = "arXiv",
    primaryClass = "hep-th",
    doi = "10.4310/ATMP.2020.v24.n8.a1",
    journal = "Adv. Theor. Math. Phys.",
    volume = "24",
    number = "8",
    pages = "1931--2041",
    year = "2020"
}

@inproceedings{Yamazaki:2025yan,
    author = "Yamazaki, Masahito",
    title = "{Gauge Theory and Integrability: An Overview}",
    booktitle = "Proceedings of the International Conference of Basic Science 2025",
    eprint = "2509.07628",
    archivePrefix = "arXiv",
    primaryClass = "hep-th",
    month = "9",
    year = "2025"
}

@article{Post:2022dfi,
    author = "Post, Boris and van der Heijden, Jeremy and Verlinde, Erik",
    title = "{A universe field theory for JT gravity}",
    eprint = "2201.08859",
    archivePrefix = "arXiv",
    primaryClass = "hep-th",
    doi = "10.1007/JHEP05(2022)118",
    journal = "JHEP",
    volume = "05",
    pages = "118",
    year = "2022"
}

@article{Delduc:2019whp,
    author = "Delduc, Francois and Lacroix, Sylvain and Magro, Marc and Vicedo, Benoit",
    title = "{A unifying 2d action for integrable $\sigma$-models from 4d Chern-Simons theory}",
    eprint = "1909.13824",
    archivePrefix = "arXiv",
    primaryClass = "hep-th",
    doi = "10.1007/s11005-020-01268-y",
    journal = "Lett. Math. Phys.",
    volume = "110",
    pages = "1645--1687",
    year = "2020"
}

@article{Ferko:2022cix,
    author = "Ferko, Christian and Sfondrini, Alessandro and Smith, Liam and Tartaglino-Mazzucchelli, Gabriele",
    title = "{Root-$T \bar T$ Deformations in Two-Dimensional Quantum Field Theories}",
    eprint = "2206.10515",
    archivePrefix = "arXiv",
    primaryClass = "hep-th",
    doi = "10.1103/PhysRevLett.129.201604",
    journal = "Phys. Rev. Lett.",
    volume = "129",
    number = "20",
    pages = "201604",
    year = "2022"
}

@article{Borsato:2022tmu,
    author = "Borsato, Riccardo and Ferko, Christian and Sfondrini, Alessandro",
    title = "{On the Classical Integrability of Root-$T \overline{T}$ Flows}",
    eprint = "2209.14274",
    archivePrefix = "arXiv",
    primaryClass = "hep-th",
    month = "9",
    year = "2022"
}

@article{Costello:2015xsa,
    author = "Costello, Kevin and Li, Si",
    title = "{Quantization of open-closed BCOV theory, I}",
    eprint = "1505.06703",
    archivePrefix = "arXiv",
    primaryClass = "hep-th",
    month = "5",
    year = "2015"
}

@article{Costello:2020lpi,
    author = "Costello, Kevin and Stefa\'nski, Bogdan",
    title = "{Chern-Simons Origin of Superstring Integrability}",
    eprint = "2005.03064",
    archivePrefix = "arXiv",
    primaryClass = "hep-th",
    doi = "10.1103/PhysRevLett.125.121602",
    journal = "Phys. Rev. Lett.",
    volume = "125",
    number = "12",
    pages = "121602",
    year = "2020"
}

@article{Bershadsky:1994sr,
    author = "Bershadsky, M. and Sadov, V.",
    title = "{Theory of Kahler gravity}",
    eprint = "hep-th/9410011",
    archivePrefix = "arXiv",
    reportNumber = "HUTP-94-A013",
    doi = "10.1142/S0217751X96002157",
    journal = "Int. J. Mod. Phys. A",
    volume = "11",
    pages = "4689--4730",
    year = "1996"
}

@article{Dubovsky:2017cnj,
    author = "Dubovsky, Sergei and Gorbenko, Victor and Mirbabayi, Mehrdad",
    title = "{Asymptotic fragility, near AdS$_{2}$ holography and $ T\overline{T} $}",
    eprint = "1706.06604",
    archivePrefix = "arXiv",
    primaryClass = "hep-th",
    doi = "10.1007/JHEP09(2017)136",
    journal = "JHEP",
    volume = "09",
    pages = "136",
    year = "2017"
}

@article{Conti:2022egv,
    author = "Conti, Riccardo and Romano, Jacopo and Tateo, Roberto",
    title = "{Metric approach to a $ \mathrm{T}\overline{\mathrm{T}} $-like deformation in arbitrary dimensions}",
    eprint = "2206.03415",
    archivePrefix = "arXiv",
    primaryClass = "hep-th",
    doi = "10.1007/JHEP09(2022)085",
    journal = "JHEP",
    volume = "09",
    pages = "085",
    year = "2022"
}

@article{Conti:2018tca,
    author = "Conti, Riccardo and Negro, Stefano and Tateo, Roberto",
    title = "{The $ \mathrm{T}\overline{\mathrm{T}} $ perturbation and its geometric interpretation}",
    eprint = "1809.09593",
    archivePrefix = "arXiv",
    primaryClass = "hep-th",
    doi = "10.1007/JHEP02(2019)085",
    journal = "JHEP",
    volume = "02",
    pages = "085",
    year = "2019"
}

@article{Smirnov:2016lqw,
    author = "Smirnov, F. A. and Zamolodchikov, A. B.",
    title = "{On space of integrable quantum field theories}",
    eprint = "1608.05499",
    archivePrefix = "arXiv",
    primaryClass = "hep-th",
    doi = "10.1016/j.nuclphysb.2016.12.014",
    journal = "Nucl. Phys. B",
    volume = "915",
    pages = "363--383",
    year = "2017"
}

@article{Bershadsky:1993cx,
    author = "Bershadsky, M. and Cecotti, S. and Ooguri, H. and Vafa, C.",
    title = "{Kodaira-Spencer theory of gravity and exact results for quantum string amplitudes}",
    eprint = "hep-th/9309140",
    archivePrefix = "arXiv",
    reportNumber = "HUTP-93-A025, RIMS-946, SISSA-142-93-EP",
    doi = "10.1007/BF02099774",
    journal = "Commun. Math. Phys.",
    volume = "165",
    pages = "311--428",
    year = "1994"
}

@article{Cavaglia:2016oda,
    author = "Cavagli\`a, Andrea and Negro, Stefano and Sz\'ecs\'enyi, Istv\'an M. and Tateo, Roberto",
    title = "{$T \bar{T}$-deformed 2D Quantum Field Theories}",
    eprint = "1608.05534",
    archivePrefix = "arXiv",
    primaryClass = "hep-th",
    doi = "10.1007/JHEP10(2016)112",
    journal = "JHEP",
    volume = "10",
    pages = "112",
    year = "2016"
}

@article{Chen:2021aid,
    author = "Chen, Bin and Hou, Jue and Tian, Jia",
    title = "{Lax connections in $T\bar{T}$-deformed integrable field theories}",
    eprint = "2102.01470",
    archivePrefix = "arXiv",
    primaryClass = "hep-th",
    doi = "10.1088/1674-1137/ac0ee4",
    journal = "Chin. Phys. C",
    volume = "45",
    number = "9",
    pages = "093112",
    year = "2021"
}

@article{Liniado:2023uoo,
    author = "Liniado, Joaquin and Vicedo, Benoit",
    title = "{Integrable Degenerate $\mathcal{E}$-Models from 4d Chern\textendash{}Simons Theory}",
    eprint = "2301.09583",
    archivePrefix = "arXiv",
    primaryClass = "hep-th",
    doi = "10.1007/s00023-023-01317-x",
    journal = "Annales Henri Poincare",
    volume = "24",
    number = "10",
    pages = "3421--3459",
    year = "2023"
}

@article{Benini:2020skc,
    author = "Benini, Marco and Schenkel, Alexander and Vicedo, Benoit",
    title = "{Homotopical Analysis of 4d Chern-Simons Theory and Integrable Field Theories}",
    eprint = "2008.01829",
    archivePrefix = "arXiv",
    primaryClass = "hep-th",
    doi = "10.1007/s00220-021-04304-7",
    journal = "Commun. Math. Phys.",
    volume = "389",
    number = "3",
    pages = "1417--1443",
    year = "2022"
}

@article{Lacroix:2020flf,
    author = "Lacroix, Sylvain and Vicedo, Benoit",
    title = "{Integrable $\mathcal{E}$-Models, 4d Chern-Simons Theory and Affine Gaudin Models. I.~Lagrangian Aspects}",
    eprint = "2011.13809",
    archivePrefix = "arXiv",
    primaryClass = "hep-th",
    reportNumber = "ZMP-HH/20-22",
    doi = "10.3842/SIGMA.2021.058",
    journal = "SIGMA",
    volume = "17",
    pages = "058",
    year = "2021"
}

@article{Arutyunov:2004yx,
    author = "Arutyunov, Gleb and Frolov, Sergey",
    title = "{Integrable Hamiltonian for classical strings on AdS(5) x S**5}",
    eprint = "hep-th/0411089",
    archivePrefix = "arXiv",
    reportNumber = "AEI-2004-105",
    doi = "10.1088/1126-6708/2005/02/059",
    journal = "JHEP",
    volume = "02",
    pages = "059",
    year = "2005"
}

@article{Klimcik:2019kkf,
    author = "Klim\v{c}\'\i{}k, Ctirad",
    title = "{Dressing cosets and multi-parametric integrable deformations}",
    eprint = "1903.00439",
    archivePrefix = "arXiv",
    primaryClass = "hep-th",
    doi = "10.1007/JHEP07(2019)176",
    journal = "JHEP",
    volume = "07",
    pages = "176",
    year = "2019"
}

@article{Klimcik:2021bqm,
    author = "Klimcik, Ctirad",
    title = "{On Strong Integrability of the Dressing Cosets}",
    eprint = "2107.05607",
    archivePrefix = "arXiv",
    primaryClass = "hep-th",
    doi = "10.1007/s00023-021-01125-1",
    journal = "Annales Henri Poincare",
    volume = "23",
    number = "7",
    pages = "2545--2578",
    year = "2022"
}

@article{Klimcik:1996np,
    author = "Klimcik, C. and Severa, P.",
    title = "{Dressing cosets}",
    eprint = "hep-th/9602162",
    archivePrefix = "arXiv",
    reportNumber = "CERN-TH-96-43",
    doi = "10.1016/0370-2693(96)00669-7",
    journal = "Phys. Lett. B",
    volume = "381",
    pages = "56--61",
    year = "1996"
}

@article{Khan:2022vrx,
    author = "Khan, Ahsan Z.",
    title = "{Holomorphic Surface Defects in Four-Dimensional Chern-Simons Theory}",
    eprint = "2209.07387",
    archivePrefix = "arXiv",
    primaryClass = "hep-th",
    month = "9",
    year = "2022"
}

@article{Gukov:2006jk,
    author = "Gukov, Sergei and Witten, Edward",
    title = "{Gauge Theory, Ramification, And The Geometric Langlands Program}",
    eprint = "hep-th/0612073",
    archivePrefix = "arXiv",
    month = "12",
    year = "2006"
}

@article{Klimcik:1995ux,
    author = "Klimcik, C. and Severa, P.",
    title = "{Dual nonAbelian duality and the Drinfeld double}",
    eprint = "hep-th/9502122",
    archivePrefix = "arXiv",
    reportNumber = "CERN-TH-95-39, CERN-TH-95-039",
    doi = "10.1016/0370-2693(95)00451-P",
    journal = "Phys. Lett. B",
    volume = "351",
    pages = "455--462",
    year = "1995"
}

@article{Klimcik:1996nq,
    author = "Klimcik, C. and Severa, P.",
    title = "{NonAbelian momentum winding exchange}",
    eprint = "hep-th/9605212",
    archivePrefix = "arXiv",
    reportNumber = "CERN-TH-96-142",
    doi = "10.1016/0370-2693(96)00755-1",
    journal = "Phys. Lett. B",
    volume = "383",
    pages = "281--286",
    year = "1996"
}

@article{Klimcik:1995dy,
    author = "Klimcik, C. and Severa, P.",
    title = "{Poisson-Lie T duality and loop groups of Drinfeld doubles}",
    eprint = "hep-th/9512040",
    archivePrefix = "arXiv",
    reportNumber = "CERN-TH-95-330",
    doi = "10.1016/0370-2693(96)00025-1",
    journal = "Phys. Lett. B",
    volume = "372",
    pages = "65--71",
    year = "1996"
}

@article{Babaei-Aghbolagh:2024hti,
    author = "Babaei-Aghbolagh, H. and He, Song and Morone, Tommaso and Ouyang, Hao and Tateo, Roberto",
    title = "{Geometric formulation of generalized root-$T\bar{T}$ deformations}",
    eprint = "2405.03465",
    archivePrefix = "arXiv",
    primaryClass = "hep-th",
    month = "5",
    year = "2024"
}

@article{Conti:2019dxg,
    author = "Conti, Riccardo and Negro, Stefano and Tateo, Roberto",
    title = "{Conserved currents and $\text{T}\bar{\text{T}}_s$ irrelevant deformations of 2D integrable field theories}",
    eprint = "1904.09141",
    archivePrefix = "arXiv",
    primaryClass = "hep-th",
    doi = "10.1007/JHEP11(2019)120",
    journal = "JHEP",
    volume = "11",
    pages = "120",
    year = "2019"
}

@article{Polyakov:1983tt,
    author = "Polyakov, Alexander M. and Wiegmann, P. B.",
    editor = "Stone, M.",
    title = "{Theory of Nonabelian Goldstone Bosons}",
    doi = "10.1016/0370-2693(83)91104-8",
    journal = "Phys. Lett. B",
    volume = "131",
    pages = "121--126",
    year = "1983"
}

@article{Coussaert:1995zp,
    author = "Coussaert, Oliver and Henneaux, Marc and van Driel, Peter",
    title = "{The Asymptotic dynamics of three-dimensional Einstein gravity with a negative cosmological constant}",
    eprint = "gr-qc/9506019",
    archivePrefix = "arXiv",
    reportNumber = "ULB-TH-95-08",
    doi = "10.1088/0264-9381/12/12/012",
    journal = "Class. Quant. Grav.",
    volume = "12",
    pages = "2961--2966",
    year = "1995"
}

@article{Hirano:2024eab,
    author = "Hirano, Shinji and Shigemori, Masaki",
    title = "{Conformal field theory on $ T\overline{T} $-deformed space and correlators from dynamical coordinate transformations}",
    eprint = "2402.08278",
    archivePrefix = "arXiv",
    primaryClass = "hep-th",
    reportNumber = "YITP-24-17",
    doi = "10.1007/JHEP07(2024)190",
    journal = "JHEP",
    volume = "07",
    pages = "190",
    year = "2024"
}

@article{Delduc:2013fga,
    author = "Delduc, Francois and Magro, Marc and Vicedo, Benoit",
    title = "{On classical $q$-deformations of integrable sigma-models}",
    eprint = "1308.3581",
    archivePrefix = "arXiv",
    primaryClass = "hep-th",
    doi = "10.1007/JHEP11(2013)192",
    journal = "JHEP",
    volume = "11",
    pages = "192",
    year = "2013"
}

@article{Kawaguchi:2014qwa,
    author = "Kawaguchi, Io and Matsumoto, Takuya and Yoshida, Kentaroh",
    title = "{Jordanian deformations of the $AdS_5 x S^5$ superstring}",
    eprint = "1401.4855",
    archivePrefix = "arXiv",
    primaryClass = "hep-th",
    reportNumber = "KUNS-2477, ITP-UU-14-05, SPIN-14-05",
    doi = "10.1007/JHEP04(2014)153",
    journal = "JHEP",
    volume = "04",
    pages = "153",
    year = "2014"
}

@article{Hollowood:2014rla,
    author = "Hollowood, Timothy J. and Miramontes, J. Luis and Schmidtt, David M.",
    title = "{Integrable Deformations of Strings on Symmetric Spaces}",
    eprint = "1407.2840",
    archivePrefix = "arXiv",
    primaryClass = "hep-th",
    doi = "10.1007/JHEP11(2014)009",
    journal = "JHEP",
    volume = "11",
    pages = "009",
    year = "2014"
}

@article{Bielli:2024ach,
    author = "Bielli, Daniele and Ferko, Christian and Smith, Liam and Tartaglino-Mazzucchelli, Gabriele",
    title = "{Integrable Higher-Spin Deformations of Sigma Models from Auxiliary Fields}",
    eprint = "2407.16338",
    archivePrefix = "arXiv",
    primaryClass = "hep-th",
    month = "7",
    year = "2024"
}

@article{Ferko:2024ali,
    author = "Ferko, Christian and Smith, Liam",
    title = "{An Infinite Family of Integrable Sigma Models Using Auxiliary Fields}",
    eprint = "2405.05899",
    archivePrefix = "arXiv",
    primaryClass = "hep-th",
    month = "5",
    year = "2024"
}

@article{Bielli:2024khq,
    author = "Bielli, Daniele and Ferko, Christian and Smith, Liam and Tartaglino-Mazzucchelli, Gabriele",
    title = "{T-Duality and $T \overline{T}$-like Deformations of Sigma Models}",
    eprint = "2407.11636",
    archivePrefix = "arXiv",
    primaryClass = "hep-th",
    month = "7",
    year = "2024"
}

@article{Fukushima:2024nxm,
    author = "Fukushima, Osamu and Yoshida, Kentaroh",
    title = "{4D Chern-Simons theory with auxiliary fields}",
    eprint = "2407.02204",
    archivePrefix = "arXiv",
    primaryClass = "hep-th",
    reportNumber = "RIKEN-iTHEMS-Report-24, STUPP-24-270",
    month = "7",
    year = "2024"
}

@article{Coleman:2019dvf,
    author = "Coleman, Evan A. and Aguilera-Damia, Jeremias and Freedman, Daniel Z. and Soni, Ronak M.",
    title = "{$ T\overline{T} $ -deformed actions and (1,1) supersymmetry}",
    eprint = "1906.05439",
    archivePrefix = "arXiv",
    primaryClass = "hep-th",
    doi = "10.1007/JHEP10(2019)080",
    journal = "JHEP",
    volume = "10",
    pages = "080",
    year = "2019"
}

@article{Bielli:2024fnp,
    author = "Bielli, Daniele and Ferko, Christian and Smith, Liam and Tartaglino-Mazzucchelli, Gabriele",
    title = "{Auxiliary Field Sigma Models and Yang-Baxter Deformations}",
    eprint = "2408.09714",
    archivePrefix = "arXiv",
    primaryClass = "hep-th",
    month = "8",
    year = "2024"
}

@article{Bonelli:2018kik,
    author = "Bonelli, Giulio and Doroud, Nima and Zhu, Mengqi",
    title = "{$T \bar{T}$-deformations in closed form}",
    eprint = "1804.10967",
    archivePrefix = "arXiv",
    primaryClass = "hep-th",
    doi = "10.1007/JHEP06(2018)149",
    journal = "JHEP",
    volume = "06",
    pages = "149",
    year = "2018"
}

@article{Babaei-Aghbolagh:2022leo,
    author = "Babaei-Aghbolagh, H. and Babaei Velni, Komeil and Mahdavian Yekta, Davood and Mohammadzadeh, Hosein",
    title = "{Marginal $T\bar{T}$-like deformation and modified Maxwell theories in two dimensions}",
    eprint = "2206.12677",
    archivePrefix = "arXiv",
    primaryClass = "hep-th",
    doi = "10.1103/PhysRevD.106.086022",
    journal = "Phys. Rev. D",
    volume = "106",
    number = "8",
    pages = "086022",
    year = "2022"
}

@article{Barbon:2020amo,
    author = "Barbon, Jos{\'e} L. F. and Rabinovici, E.",
    title = "{Remarks on the thermodynamic stability of $T \bar T$ deformations}",
    eprint = "2004.10138",
    archivePrefix = "arXiv",
    primaryClass = "hep-th",
    doi = "10.1088/1751-8121/ab99ee",
    journal = "J. Phys. A",
    volume = "53",
    number = "42",
    pages = "424001",
    year = "2020"
}

\end{document}